\documentstyle[aps,prl,eqsecnum,psfig,floats,preprint,12pt]{revtex}

\begin{document}

\draft
\title{Phase coherent transmission through \\
       interacting mesoscopic systems}
\author{Gregor Hackenbroich}
\address{Universit\"at GH Essen, 45117 Essen, Germany}

\maketitle

\begin{abstract}
  This is a review of the phase coherent transmission through
  interacting mesoscopic conductors. As a paradigm we study the
  transmission amplitude and the dephasing rate for electron
  transport through a quantum dot in the Coulomb blockade regime.
  We summarize experimental and theoretical work devoted to the {\em
    phase} of the transmission amplitude. It is shown that the
  evolution of the transmission phase may be dominated by
  non--universal features in the short--time dynamics of the quantum
  dot.  The controlled dephasing in Coulomb coupled conductors is
  investigated. Examples comprise a single or multiple quantum dots
  in close vicinity to a quantum point contact. The current through
  the quantum point contact "measures" the state of the dots and
  causes dephasing. The dephasing rate is derived using widely
  different theoretical approaches. The Coulomb coupling between
  mesoscopic conductors may prove useful for future work on electron
  coherence and quantum computing.
\end{abstract}

\newpage
\hfill
{}
\thispagestyle{empty}

\tableofcontents
\newpage

\narrowtext
\setcounter{equation}{0}
\section{Introduction}
\label{Sec1}

Electron transport on mesoscopic length scales has been intensively
studied during the last fifteen years
\cite{Alt91,BeeHou91,Dat95,Imr96}. A variety of new phenomena has
been discovered such as the weak localization, universal conductance
fluctuations, Aharonov--Bohm oscillations in mesoscopic rings, or
persistent currents. The origin of these phenomena is the quantum
mechanical phase coherence of the electronic wave functions. The
degree of coherence can be measured by the phase coherence length
$L_\phi$ which is the typical length on which electrons travel
without loosing their phase coherence. In mesoscopic systems,
$L_\phi$ exceeds the system size, and transport exhibits quantum
interference effects.

How is mesoscopic transport modified by electron--electron
interactions? Interactions generally play a minor role in good
conductors where electron wave functions are delocalized over the
whole system and screening is effective. Such systems, both in the
presence and absence of disorder, have been successfully described
by noninteracting electron models \cite{Bee97}. Interactions only
lead to small corrections, e.g.\ they suppress the phase coherence
of interfering electrons by coupling them to the bath of other
electrons.  In contrast to the transport in weakly interacting
conductors, interactions strongly affect transport in high magnetic
fields and in spatially confined geometries. Important examples are
the fractional quantum Hall effect and the tunneling through small
spatially confined electron islands, known as quantum dots. The
electron interactions in a quantum dot give rise to a characteristic
energy scale called the charging energy.  Similar to the ionization
energy of an atom, the charging energy is the energy necessary to
remove or add a single electron to the quantum dot. The charging
energy in state--of--the--art semiconductor quantum dots typically
exceeds the single--particle level spacing by a factor 10 or more.
Charging effects can drastically alter and nearly suppress the
transport of charge through these systems at temperatures roughly
below 1 K.

\begin{figure}
\hspace*{0cm}
\centerline{\psfig{file=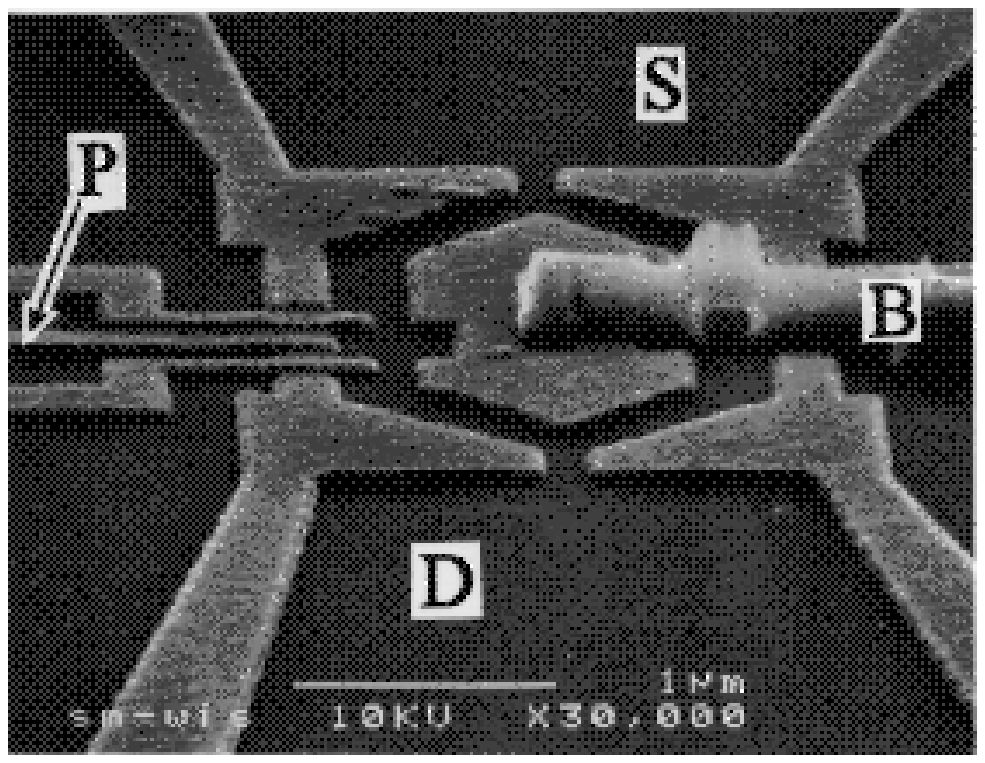,width=9cm,angle=0}}
\caption[]{Electron micrograph of the device used in the
  experiment \protect{\cite{Yac95}}. The brighter regions indicate
  metallic gates; B is a metallic air bridge. Electrons flow between
  the regions labeled S and D through the left or the right arm of
  an Aharonov--Bohm ring. The quantum dot is inserted in the left
  arm.  Taken from Ref.~\protect{\cite{Yac95}}.} \label{yacint}
\end{figure}

In this work we review the phase coherent transmission through {\em
  strongly} interacting systems. As a paradigm, we study the
tunneling through quantum dots. The subject started in 1995 with an
ingenious experiment by Yacoby et al.\ \cite{Yac95}. The authors
utilized a novel interference device with a quantum dot embedded in
one arm of an Aharonov--Bohm (AB) ring (see Fig.~\ref{yacint}). The
measurement of flux--periodic current oscillations through the
device proved for the first time that part of the tunneling current
through a quantum dot is coherent. In a step beyond the
demonstration of coherence, an improved version \cite{Sch97} of the
experiment allowed to measure the phase of the transmission
amplitude through the quantum dot. At sufficiently low temperatures,
this phase is carried by a single resonant many--body state. The
transmission phase displayed a number of unexpected properties. Most
notably, virtually the same transmission phase was found for a whole
sequence of resonant quantum dot states; between neighboring states
the phase displayed a sharp phase slip. These features have been
addressed but not yet fully explained in a large number of
theoretical papers.

The field expanded substantially in 1998 when it was realized that
the coherence of quantum dot states can be controlled by external
means. Controlled decoherence was achieved \cite{Buk98,Spr00} in a
device with a quantum dot that was capacitevly coupled to a quantum
point contact in close vicinity.  The quantum point contact acted as
a measuring apparatus for the number of electrons on the quantum
dot.  Since number and phase are conjugate variables, the
measurement caused the dephasing (decoherence) of electron states in
the quantum dot. The loss of coherence was detected as a suppression
of AB oscillations across the quantum dot.  The experiments
\cite{Buk98,Spr00} for the first time allowed to study fundamental
principles of quantum mechanics such as the number--phase or the
particle--wave duality with solid state devices. New experiments in
this direction, e.g.\ testing the suppression of tunneling due to
frequent measurements, have been proposed \cite{Gur97,Hac98}.

The present work reviews the main theoretical and experimental
results on phase coherent transport through quantum dots. We
emphasize the theoretical developments. The most important formal
tool is scattering theory. We use both the well--known scattering
approach of Landauer to discuss aspects of noninteracting electrons
and many--body scattering theory to address interaction effects. The
scattering methods are supplemented by other techniques, most
prominently by master equation approaches. Most of the theoretical
work was motivated by and must be discussed in connection with the
beautiful experiments \cite{Yac95,Sch97,Buk98,Spr00} on quantum
coherence done by researchers at the Weizmann Institute. An account
of the relevant experimental results, therefore, precedes the
theoretical discussions. It is assumed that the reader is familiar
with ballistic transport in mesoscopic systems. We assume basic
knowledge of the Landauer formula, and of transport through
Aharonov--Bohm rings, quantum point contacts and quantum dots. An
overview about most of these  concepts can be found in Refs.\
\cite{Alt91,BeeHou91,Dat95,Imr96}. For a recent review on transport
through quantum dots in the Coulomb blockade regime see Ref.\
\cite{Kou99}. Reference \cite{Kas93} gives a popular account of quantum
dots and the Coulomb blockade. 

In Sec.~\ref{Sec3} we summarize the first experiments
\cite{Yac95,Sch97} on the phase coherent transmission through
quantum dots in the Coulomb blockade regime. Both experiments aimed
at measuring the phase of the transmission amplitude through a
quantum dot. It turned out, that in the first experiment
\cite{Yac95} this measurement was hampered by peculiar symmetries
resulting from the two--terminal set--up used in that experiment.
Section \ref{Sec4} is devoted to a discussion of these symmetries.
We derive the two--terminal conductance through an AB ring with a
quantum dot. The theory explains the sharp jump of the AB
oscillation pattern found in the vicinity of transmission
resonances. Section \ref{Sec5} deals with the observed phase
similarity of transmission peaks and with the sharp phase jump
between the peaks. We show that these observations may be traced
back to features of the {\em short--time} dynamics in the quantum
dot that completely dominate the transport in the tunneling regime.
In Sec.~\ref{Sec6} we discuss the controlled dephasing of quantum
dot levels due to the coupling to a mesoscopic conductors in the
environment.  Section~\ref{Sec7} deals with new theoretical
proposals to study the dephasing of charge oscillations in coupled
quantum dots. The proposed experiments could provide evidence for a
fundamental quantum effect known as the quantum Zeno effect: the
suppression of quantum transitions due to frequent observations with
a measuring device. We conclude in Sec.~\ref{Sec9}.


\newpage
\setcounter{equation}{0}
\section{Measuring the electron transmission phase}
\label{Sec3}
The phase coherence of electronic wave functions lies at the heart
of mesoscopic interference phenomena like the universal conductance
fluctuations or weak--localization. In most cases, quantum 
interference has been studied with mesoscopic conductors that were
connected to few or many conducting channels. Interaction effects
are usually small in these systems due to good screening of the
electron charge. Charging effects become important when
the conductor is small and only weakly coupled to the external
leads. A paradigm is a quantum dot in the Coulomb blockade regime. 

The idea and the practical implementation of devices for measuring
the electron transmission phase through a quantum dot was
developed in a series of ingenious experiments \cite{Yac95,Sch97} at
the Weizmann Institute. The experiments utilized an Aharonov--Bohm
(AB) ring with a quantum dot in one of its arms. The quantum dot
was in the Coulomb blockade regime. The measurements of the electron
phase revealed a number of unexpected results. The technique for
measuring the electron phase and the main experimental results are
described in this section.\footnote{We do not discuss the experiment
\cite{Buk96} that measured the phase and magnitude of the {\em
reflection coefficient} of a quantum dot.  This experiment
\cite{Buk96} was performed in the integer quantum Hall regime and
used a second quantum dot as an interferometer.}

\subsection{Experiment of Yacoby et al.}
\label{3.1}
\begin{figure}
\hspace*{0cm}
\centerline{\psfig{file=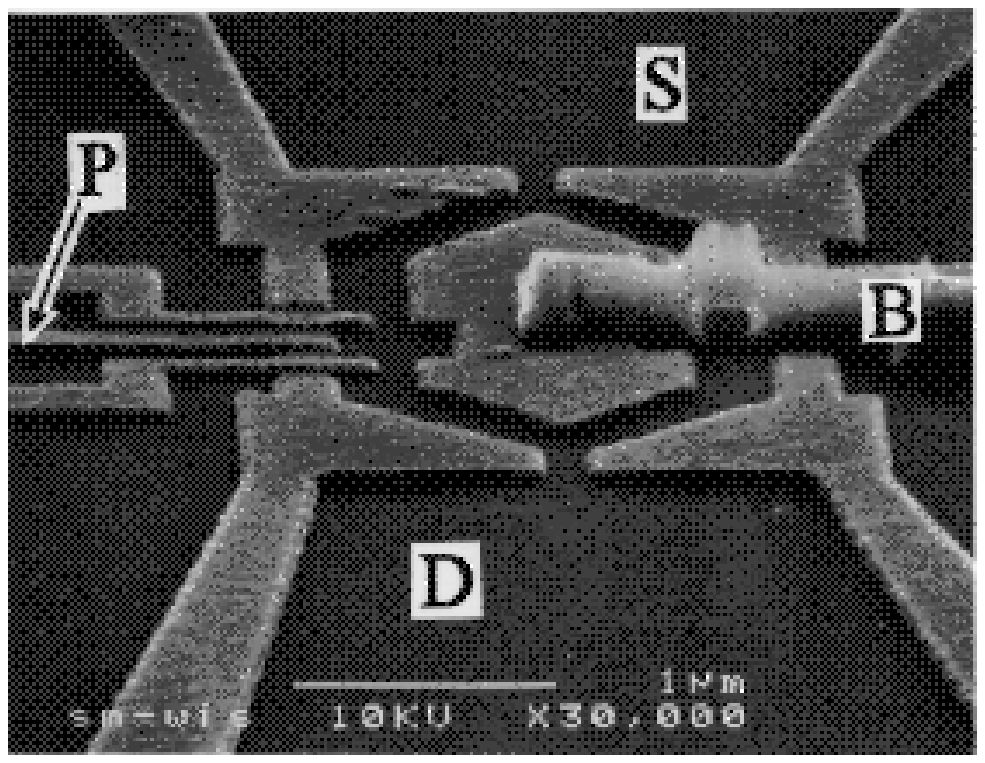,width=9cm,angle=0}}
\caption[]{Electron micrograph of the device used in the
Yacoby--experiment. The brighter regions indicate the metal gates.
Electrons flow between source and drain through the left or the
right arm of the Aharonov--Bohm ring. The quantum dot is inserted in
the left arm. The central metallic island is biased via an air
bridge (B). Taken from Ref.~\protect{\cite{Yac95}}.} \label{yacgraph}
\end{figure}

The first experiment addressing the electron phase through a quantum
dot was realized by Yacoby et al.\ \cite{Yac95}. The experiment
utilized a novel device (see Fig.~\ref{yacgraph}) to measure the phase
evolution through the dot against a fixed reference phase. The
quantum dot was inserted in one arm of an Aharonov--Bohm ring. The
basic idea was to extract the transmission phase through the quantum
dot from the phase of the Aharonov--Bohm current across the ring. 
The device was defined by metallic gates on the top of a GaAs--AlGaAs
heterostructure. The quantum dot was placed in the left arm of the
ring, its conductance could be adjusted by two quantum point contacts.
A third gate (the plunger gate P) controlled the area and the
electrostatic potential at the dot. The dot was about $0.4 \mu m$
$\times$ $0.5 \mu m$ in size. A special lithographic process, invoking
a metallic air bridge (B), enabled to contact the center metal gate
that depleted the electrons underneath the ring's center. Each of the
arms of the Aharonov-Bohm ring supported a few conducting modes. The
ring was connected to two external contacts, source (S) and drain (D),
between which a small voltage was applied.

The elastic mean free path in the two--dimensional electron gas was
about 10 $\mu m$ while the diameter of the Aharonov--Bohm ring was
$\approx 1-1.5$ $\mu m$. The phase coherence length $L_\phi$ was
larger than the ring's circumference. The quantum dot had a
resistance of $1M \Omega$ and a very small capacitance $C \approx
160$ $aF$ corresponding to the charging energy $e^2 /2C \approx 0.5$
$meV$. The dot contained around 200 electrons. Its average
single--particle level spacing was $\Delta \approx 40$ $\mu eV$. The
experiment was performed at a temperature around $100$ $m$K
corresponding to the thermal energy $kT \approx 9$ $\mu eV$. The
intrinsic width $\Gamma$ of the Coulomb peaks was estimated from the
conductance peak height to $\Gamma \approx 0.2$ $\mu eV$. These
scales imply that the quantum dot was in the Coulomb blockade
regime, and that the transmission at each Coulomb peak resulted from
resonant tunneling through a single or few levels of the quantum
dot.

\begin{figure}
\centerline{\psfig{file=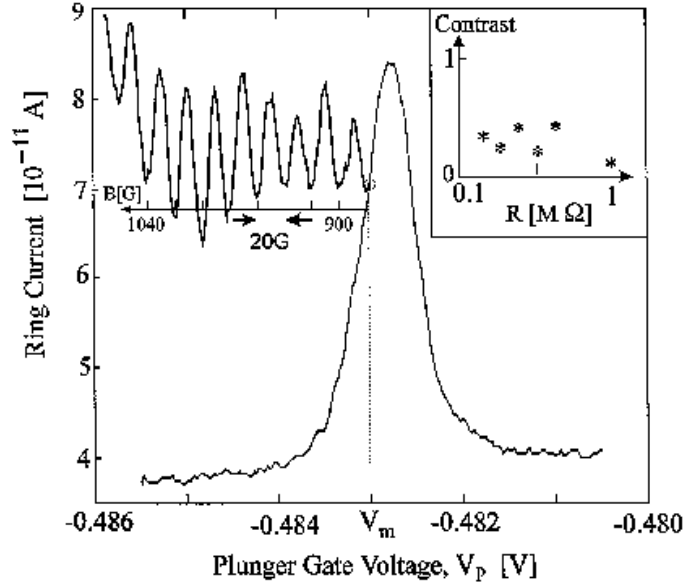,width=9cm,angle=0}}
\caption[]{One of the ring's current peaks as a function of the
plunger gate voltage. The large current of the right arm has been
subtracted. The top left part shows the Aharonov--Bohm
oscillations of the current vs.\ magnetic field at fixed
$V_P = V_m$. Inset: Oscillation contrast defined as peak--to--peak
vs.\ average current through the dot. Taken from Ref.~\protect{\cite{Yac95}}.}
\label{yac1} \end{figure}

The experiment \cite{Yac95} demonstrated for the first time that part
of the tunneling current through a quantum dot is coherent. The
experimental evidence is depicted in Fig.~\ref{yac1}. Shown is the
ring current vs.\ plunger voltage $V_P$ for a fixed small
source--drain voltage. The ring current was obtained by subtracting
from the total current across the ring a large $V_P$--independent
background due to the right arm. The Coulomb blockade in the dot
creates sharp conductance peaks in the ring current for fixed
magnetic field. Fixing the voltage $V_P$ on the side of a current
peak and sweeping the magnetic field, one observes periodic current
oscillations. The period of the oscillations corresponds to one flux
quantum threading the area of the ring, in agreement with the expected
Aharonov--Bohm period. The oscillation contrast, defined as
peak--to--peak current over the average current, was in the
range $0.2$ to $0.4$.

In a next step, the current oscillations were investigated at
different values of the plunger voltage $V_P$. A change in $V_P$ was
expected to modify both the {\em magnitude} and {\em phase} of the
transmission amplitude through the dot. The experiment was motivated
by the idea that the change in the transmission phase would be
directly reflected in a shift of the Aharonov--Bohm oscillations
which, in turn, could be seen experimentally. This one--to--one
correspondence is suggested by the following argument: Suppose the
ring and the leads support only one conducting channel. According to
the Landauer formula, the zero--temperature current between the
leads is proportional to the ring transmission coefficient
$|t(E_F)|^2$ at the Fermi energy $E_F$.  For fully coherent
transport, $t = t_R \exp(2 \pi i \Phi / \Phi_0) + t_L$, where $t_R$
and $t_L$ are the transmission amplitudes through the right and left
arm, respectively, $\Phi$ is the flux through the ring and $\Phi_0$
the flux quantum. This yields the interference term
\begin{equation} 
2 {\rm Re} \{ t_L t_R^* \exp(-2 \pi i \Phi/ \Phi_0)
\}= 2 |t_R| |t_L| \cos(\xi_L -\xi_R - 2 \pi \Phi/ \Phi_0) ,
\label{interf}
\label{3.01} 
\end{equation}
where $\xi_R = {\rm arg} (t_R)$, $\xi_L = {\rm arg} (t_L)$. Any shift in
the phase $\xi_L-\xi_R$ should thus be directly reflected in a shift
of the Aharonov--Bohm oscillations.

The above argument turned out to be incorrect. It neglects multiple
reflections through the ring. The argument fails in particular for a
two--terminal geometry, i.e.\ a ring connected to two external leads.
It was realized \cite{Lev95,Hac96a,Bru96,Yac96,Hac96b} shortly after the
Yacoby--experiment that Onsager symmetries valid for a two--teminal
device restrict the phase of the Aharonov--Bohm oscillations to either
$0$ or $\pi$, spoiling the correspondence between the Aharonov--Bohm
phase and the transmission phase through the quantum dot. A discussion
of this issue is presented in Sec.~\ref{Sec4}. Despite of its failure
for a two--terminal ring, the simple phase--argument catches the
essential idea for measuring the electron phase as it was later
realized in the multiple--terminal device of Schuster et al.\
\cite{Sch97}.

\begin{figure}
\centerline{\psfig{file=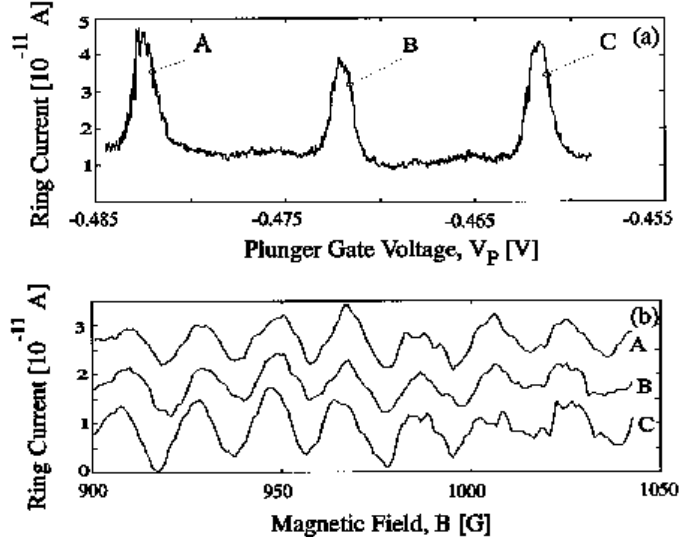,width=9cm,angle=0}}
\caption[]{(a) A series of three Coulomb peaks and (b) the current
oscillations measured at the marked points $A$, $B$, and $C$. All
oscillations are seen to be in phase. The large current of the right
arm has been subtracted. Taken from
Ref.~\protect{\cite{Yac95}}.} \label{yac2} \end{figure}

\begin{figure}
\centerline{\psfig{file=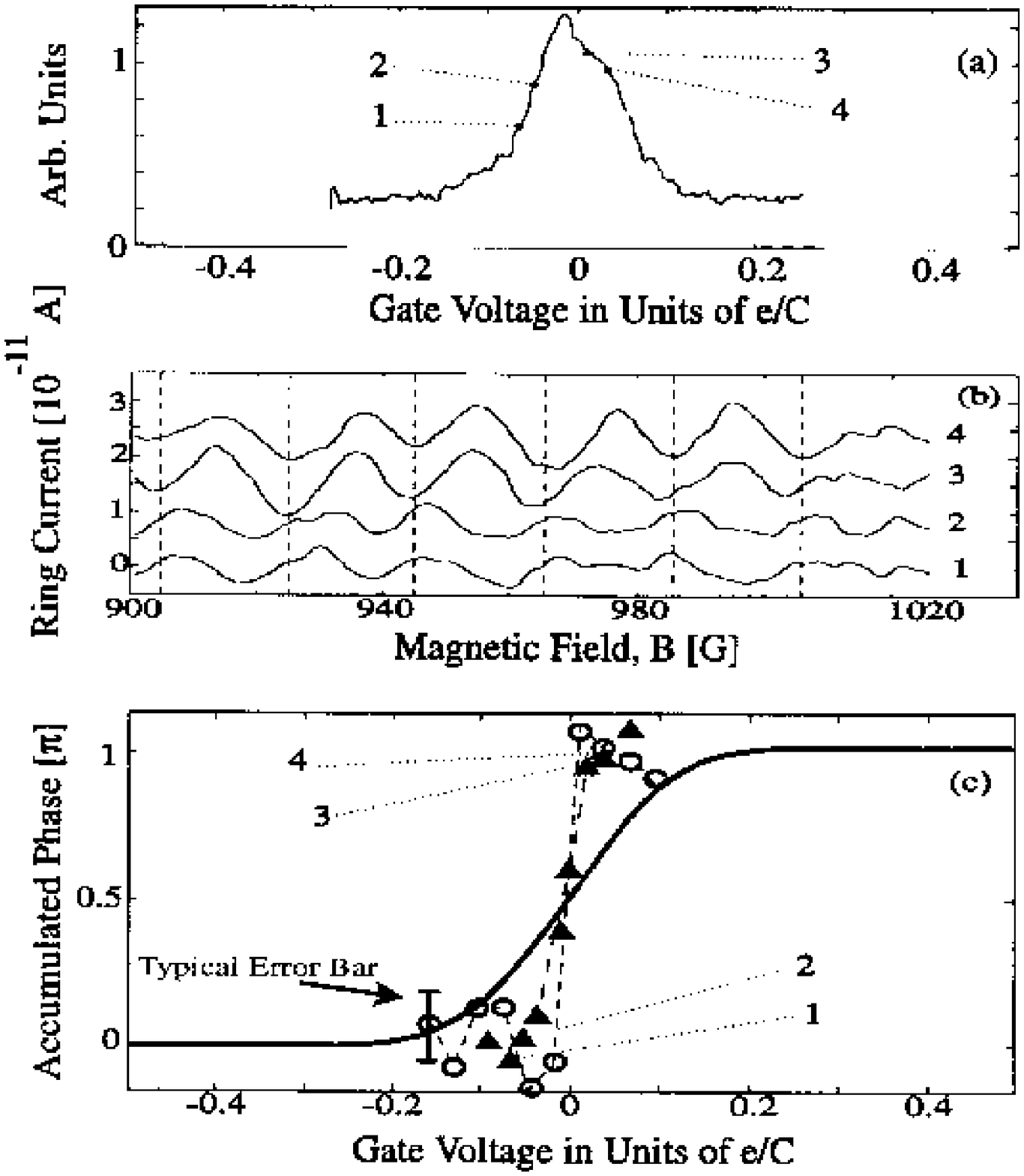,width=9cm,angle=0}}
\caption[]{Evolution of the Aharonov--Bohm phase along a current peak.
(a) Current vs.\ gate voltage at a current peak. (b) A series of
interference patterns taken at the points specified in (a). The
phase jumps between patterns $2$ and $3$. (c) Phase measured at two
peaks (circles and triangles). The broken line is guide to the eye.
The expected behavior of the quantum dot transmission phase in a 1D
resonant tunneling model is shown by the solid line. Taken from
Ref.~\protect{\cite{Yac95}}.} \label{yac3} \end{figure}

Figure \ref{yac2} shows the ring current and the Aharonov--Bohm
oscillations measured at three successive peaks. The oscillations at
similar points (A, B, and C in the figure) have the {\em same}
Aharonov--Bohm phase. This repetition of the phase was found within
the whole sequence of Coulomb peaks (comprising 12 peaks)
investigated in the experiment. The evolution of the Aharonov--Bohm
phase along a single Coulomb peak is displayed in Fig.~\ref{yac3}.
Four different interference patterns taken at the points $1$, $2$,
$3$, and $4$ specified in Fig.~\ref{yac3}(a) are shown in
Fig.~\ref{yac3}(b). The Aharonov--Bohm phase of the patterns shifts
by $\pi$ as one sweeps through the peak. The shift happens rather
abruptly between the points $2$ and $3$. This can be seen in
Fig.~\ref{yac3}(c) which summarizes the phase measurements along a
Coulomb peak. The sharp jump in the measured Aharonov--Bohm phase is
contrasted with the expected phase evolution of the transmission
amplitude assuming resonant tunneling through a single level of the
quantum dot. The latter phase increases smoothly on the scale of the
peak width (which is of order $kT$). Theoretical arguments
\cite{Lev95,Hac96a,Bru96,Yac96,Hac96b} proved that there is no scale
associated with the rather sharp jump seen in the experiment.

\subsection{Experiment of Schuster et al.}
\label{3.2}
\begin{figure}
\vspace*{-0.5cm}
\centerline{\psfig{file=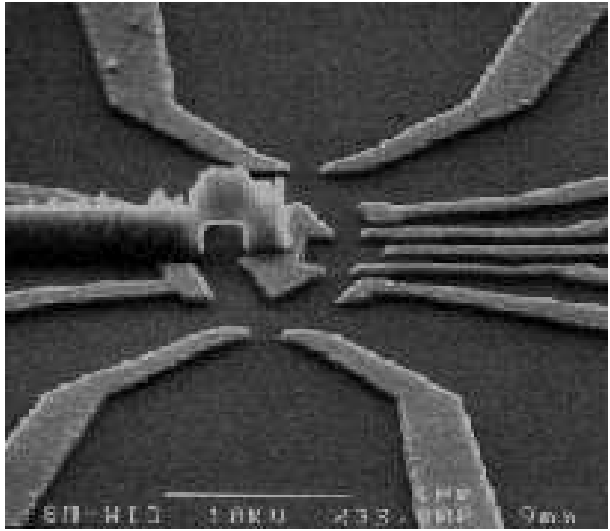,width=8cm,height=7cm,angle=0}}
\vspace*{0.5cm}
\caption[]{Scanning electron micrograph of the double--slit device used
  in the Schuster--experiment \cite{Sch97}. The grey areas are
  metallic gates on the top of the heterostructure. The quantum dot
  is inserted in the right slit. Taken from
  Ref.~\protect{\cite{Sch97}}.} \label{schgraph}
\end{figure}

\begin{figure}
\centerline{\psfig{file=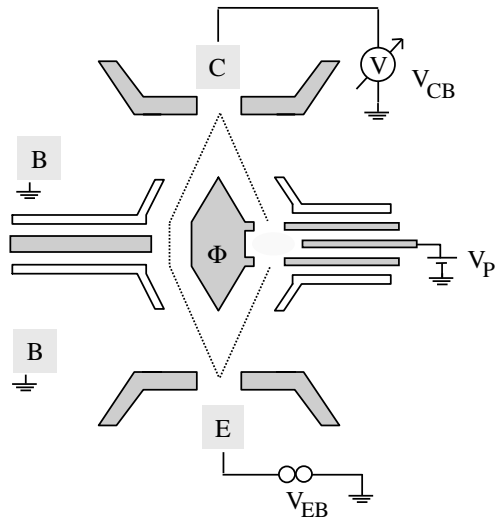,width=7cm,angle=0}}
\caption[]{Schematic of the device structure of the
  Schuster--experiment. An Aharonov--Bohm ring is coupled to an
  emitter (E), a collector (C) and a base region (B). Reflector
  gates reflect diverging electrons towards the collector. The
  quantum dot is defined by the central electrode and the three
  electrodes on the right hand side of the structure. Taken from
  Ref.~\protect{\cite{Sch97}}.}
\label{schschem} \end{figure}

The Aharonov--Bohm phase in a two--terminal device is restricted to
$0$ or $\pi$. No such restriction exists in a multi-terminal probe.
This motivated Schuster et al.\ \cite{Sch97} to perform an
interference experiment similar to the Yacoby--experiment
\cite{Yac95} but with more than two leads connected to the
interferometer. The electron micrograph of the device and a
schematic description of the experiment are shown in
Figs.~\ref{schgraph}, \ref{schschem}. The central element of the
device is an Aharonov--Bohm ring with a quantum dot embedded in its
right arm. Several contacts are connected to ring, namely the
emitter (E), the collector (C) and a base region (B). The base
contacts were held at zero chemical potential. Incorporated in the
structure are additional barriers. They reflect diverging electrons
into the two--slit device and towards the collector. The reflectors
were necessary to enhance the collector signal that could otherwise
not be measured. All contacts were defined by negatively biased
gates on top of the heterostructure.

The quantum dot contained roughly 200 electrons with a mean
single--particle level spacing around $55$ $\mu eV$. The temperature
of the two--dimensional electron gas was $100$ $m$K. The intrinsic
(zero--temperature) width $\Gamma$ of the Coulomb peaks was estimated
to be of the order or even larger than $kT$. This was achieved by
slightly opening the point contacts between the quantum dot and the
ring. The large value of $\Gamma$ corresponds to a quantum dot
resistance larger than the resistance quantum $R_K = h/e^2$ but much
smaller than the dot resistance in the Yacoby--experiment. Working
with a modest quantum dot resistance enabled Schuster et al.\ to
measure the Aharonov--Bohm oscillations not only at the Coulomb peaks
but also in between.

\begin{figure}
\centerline{\psfig{file=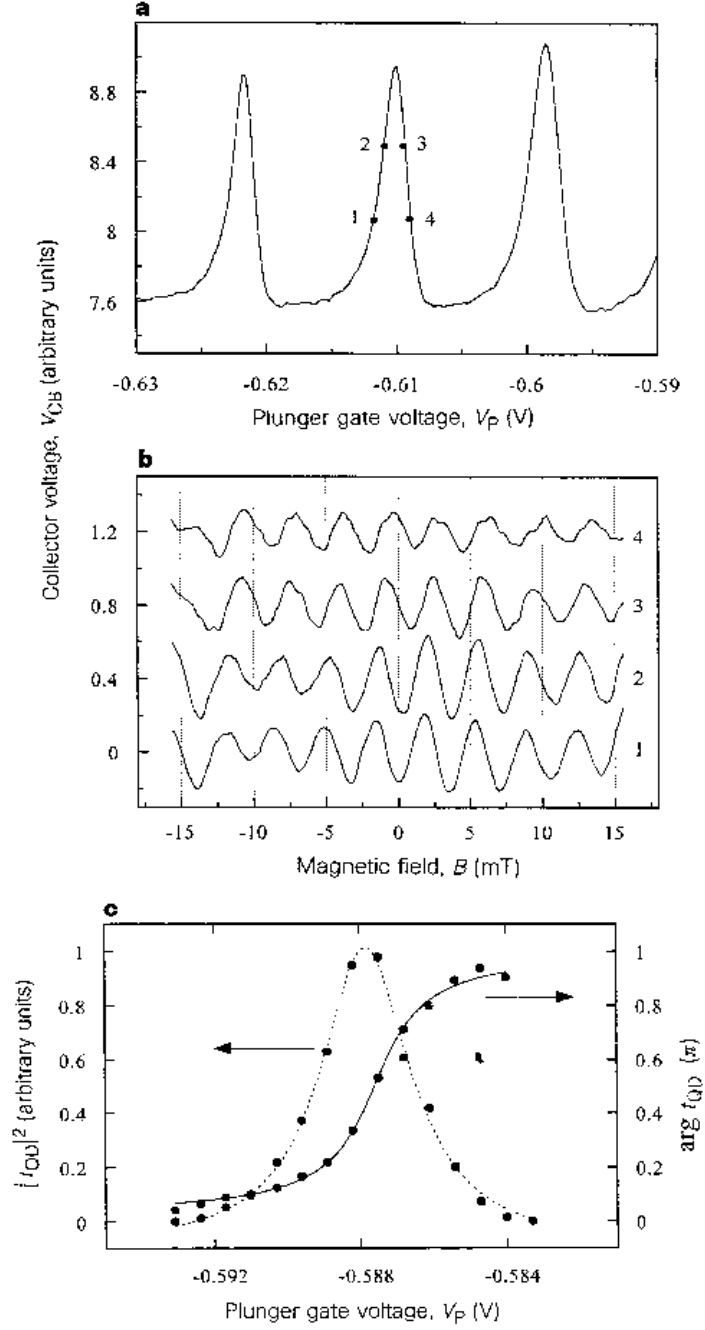,width=9.5cm,angle=0}}
\caption[]{Conductance and phase evolution along a Coulomb peak. (a)
Resonance peaks as a function of the plunger gate voltage. (b) A
series of interference patterns taken at the points specified in
(a). (c) Squared magnitude and phase of the Aharonov--Bohm
oscillations (dots). The solid and dashed line are fits for the
phase and the squared magnitude obtained with a Breit--Wigner model.
Taken from Ref.~\protect{\cite{Sch97}}.} \label{schres1} \end{figure}

Schuster el al.\ investigated the voltage drop $V_{CB}$ between
collector and base for a fixed excitation voltage $V_{EB}$ applied
between the emitter and the base. The Aharonov--Bohm interference in
the transmission coefficient $T_{EC}$ leads to an oscillatory
contribution to $V_{EC}$ from which the Aharonov--Bohm phase is
extracted.  Fig.~\ref{schres1}(a) shows $V_{EC}$ measured as a
function of the gate voltage $V_P$ for a fixed magnetic field. One
observes pronounced resonance peaks as expected for a quantum dot in
the Coulomb blockade regime. When the magnetic field is changed, the
collector signal shows AB oscillations with the expected period
$\Delta B = \Phi_0 / A$ where A is the area of the AB ring. The
observed oscillation patterns measured at the four points 1, 2, 3,
and 4 close to a resonance are shown in Fig.~\ref{schres1}(b). The
oscillation pattern shifts smoothly as one moves through the
resonance. Fig.~\ref{schres1}(c) displays the phase and the squared
magnitude of the AB signal at a resonance peak. The data points are
represented by full circles. The phase shows the expected monotonic
rise by $\pi$ over the width of the resonance.

\begin{figure}
\centerline{\psfig{file=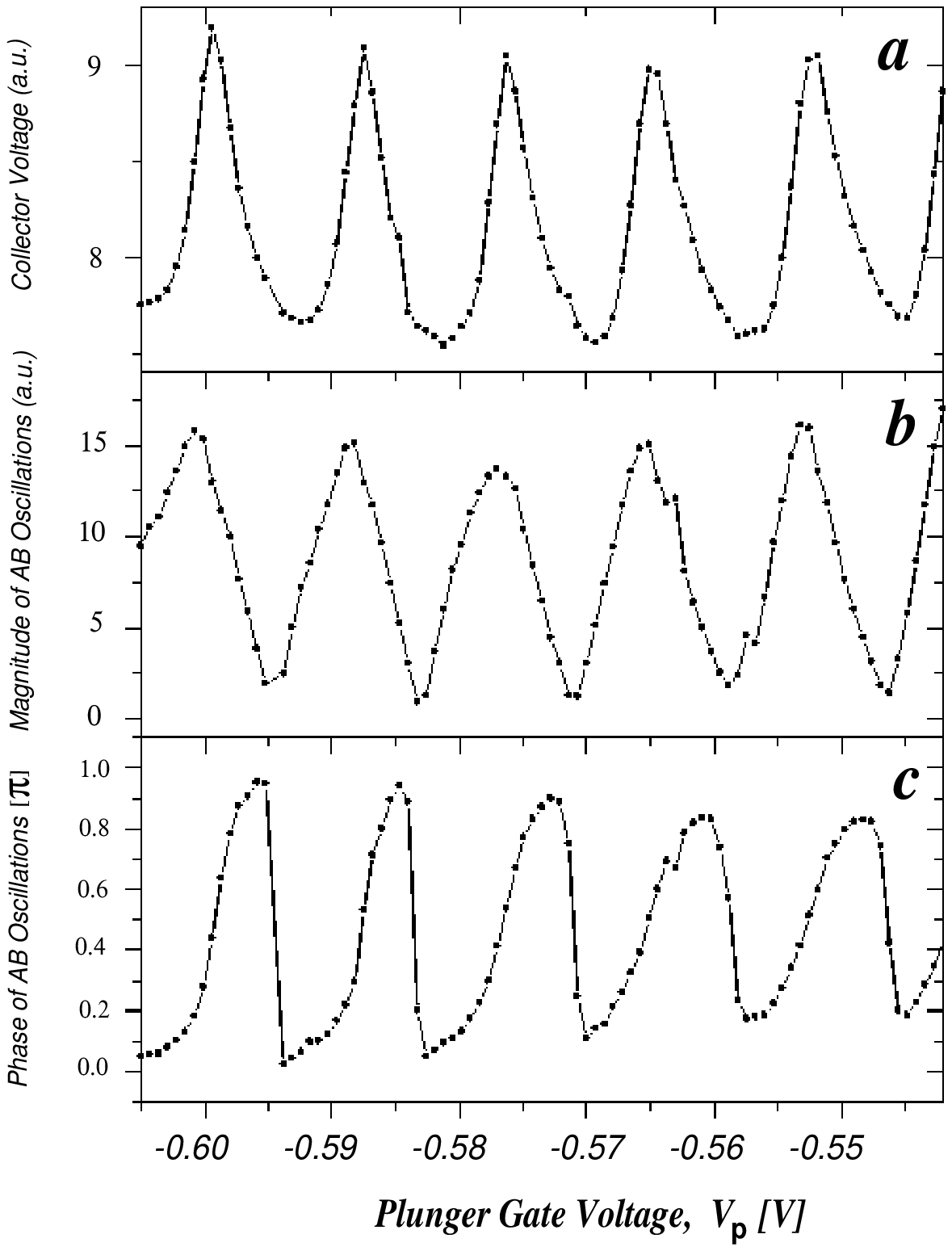,width=8cm,angle=0}}
\caption[]{(a) A series of Coulomb peaks; (b) Magnitude of the 
Aharonov--Bohm oscillations; (c) Phase of the Aharonov--Bohm
oscillations as a function of plunger gate voltage. The solid lines
are guides to the eye. Taken from Ref.~\protect{\cite{Sch97}}.}
\label{schres2} \end{figure}

Schuster et al.\ compared their data with a theoretical model for
resonant transmission. They described the coherent part $t_{\rm QD}$ of
the transmission amplitude through the dot by the Breit--Wigner
ansatz
\begin{equation}
t_{\rm QD} = i C_N {\Gamma_N /2 \over E_F-E_N+i \Gamma_N /2} ,
\label{3.02}
\end{equation}
where $C_N$ is a complex amplitude, $E_F$ the energy of the
electrons transmitted through the device, and $E_N$ and $\Gamma_N$
the energy and the width of the resonance in the quantum dot. Both
the squared magnitude $|t_{\rm QD}|^2$ and the phase ${\rm arg} \,
t_{\rm QD}$ are compared with experimental data in
Fig.~\ref{schres1}(c). Good agreement is found when $\Gamma_N
\approx 4 kT$ is used as a fit parameter.

The collector voltage $V_{CB}$, the magnitude, and the phase of the
AB oscillations measured over a sequence of five peaks are shown in
Fig.~\ref{schres2}. The striking observation is that the phase is
very similar at all peaks. The phase increases roughly by $\pi$ at
each peak. Note that the peaks widen and start to overlap as the
plunger voltage increases. At the same time, the overall variation
of the phase is reduced. The likely origin of these effects is the
electrostatic influence of the plunger on the point contacts at the
quantum dot. They open slightly with increasing plunger voltage. A
striking feature of the data is the sharp phase lapse by $\pi$
between the resonances. The phase lapse occurs when the magnitude of
the AB oscillations vanishes. Unlike the phase jump {\em at
  resonance} found in the Yacoby--experiment, there is a scale
associated with the phase lapse between the peaks. This becomes
evident for increasing $V_P$ when the lapse smoothens and is
resolved experimentally. While Schuster et al.\ were able to model
the phase evolution at resonance, they could not explain the phase
behavior between resonances. The origin of the phase lapse, the
associated energy scale, and the similarity of the phase at
subsequent peaks is addressed in Sec.~\ref{Sec5}.


\newpage
\setcounter{equation}{0}
\section{Phase evolution in a two--terminal device} 
\label{Sec4}
Electrons that tunnel through a quantum dot interact with many other
particles including electrons in the dot and in the surrounding
gates, phonons in the substrate, photons from the environment, etc.
The interactions may spoil or even destroy the phase coherence of
the transmitted electrons. The Yacoby--experiment \cite{Yac95}
demonstrated for the first time that part of the tunnel current is
coherent. As discussed in Sec.\ \ref{Sec3}, the experiment employed
a quantum dot in an AB ring.  Phase coherence was demonstrated by
measuring AB oscillations of the ring current. The oscillations
displayed a rigid phase, the phase taking on only two values $0$ and
$\pi$ with abrupt jumps between these values.  In this section, we
review theoretical explanations
\cite{Lev95,Hac96a,Bru96,Yac96,Hac96b} of this observation. We show
that the phase rigidity is imposed by the two--terminal character of
the Yacoby--experiment (Sec.\ \ref{Sec4.1}). Information about about
the transmission phase through the quantum dot may still be obtained
from the amplitude of the AB oscillations (Sec.\ \ref{Sec4.2}).

\subsection{Phase rigidity}
\label{Sec4.1}
The argument for the phase rigidity was presented by Levy Yeyati and
B\"uttiker \cite{Lev95}. Their argument is based on reciprocity
relations for the transmission amplitude derived by B\"uttiker
\cite{Bue86,Bue88}. These relations state that the two--terminal
conductance is an even function of the flux through the
ring,\footnote{This symmetry was, implicitly, formulated in earlier
  work of Gefen, Imry, and Azbel \cite{Gef84}}
\begin{equation}
G(\Phi)=G(-\Phi).
\label{4.01}
\end{equation}
This relation holds for an arbitrary number of transverse channels
in the leads. We briefly sketch a derivation of Eq.~(\ref{4.01}),
valid for one channel in each lead: Time--reversal invariance
implies $t_{12}(\Phi) = t_{21}(-\Phi)$ for the transmission
amplitude between the two leads labeled $1$ and $2$. The unitarity
of the scattering matrix yields $|t_{12}(\Phi)|^2=
|t_{21}(\Phi)|^2$. Combining these relations, one finds $|t_{12}
(\Phi)|^2= |t_{12} (-\Phi)|^2$.  Substitution in the Landauer
formula yields the claimed symmetry $G(\Phi)=G(-\Phi)$.

The conductance through the AB ring is a periodic function of $\Phi$.
Thus, $G$ can be expanded in a Fourier series
\begin{equation}
G(\Phi) = G_0 + G_1 \cos(2\pi \Phi / \Phi_0+ \delta_1) + \ldots ,
\label{4.02}
\end{equation}
where the dots indicate the higher harmonics. Combining the
expansion (\ref{4.02}) with the symmetry (\ref{4.01}), one finds
that the phase $\delta_1$ can only be $0$ or $\pi$ (or equivalently
an even or odd multiple of $\pi$). In the Yacoby--experiment, all
expansion coefficients are a function of the plunger voltage $V_P$
at the quantum dot. If the phase $\delta_1$ changes upon variation
of $V_P$, this change must be a sharp jump by $\pi$.

The above explanation of the phase rigidity relies on the symmetry
relation (\ref{4.01}). This relation remains valid even in the
presence of inelastic scattering \cite{Lev95}. Inelastic processes
can be included adopting a model suggested by B\"uttiker
\cite{Bue88}: In addition to the two physical leads $1$ and $2$, the
model includes a fictitious third lead that connects the AB ring
with a phase--randomizing reservoir.  The condition that no current
flows through the third lead determines the chemical potential of
the attached reservoir. The two--terminal conductance is found
\cite{Bue88} to be
\begin{equation}
G = 2 {e^2 \over h} \left[ T_{12} +{T_{23}T_{31} \over T_{32} +
T_{31}} \right] ,
\label{4.03}
\end{equation}
where $T_{ij}$ is the transmission coefficient from lead $j$ to lead
$i$. The term $T_{12}$ describes the direct elastic transmission
between leads $1$ and $2$ while the second term on the right hand
side of Eq.~(\ref{4.03}) accounts for the inelastic transmission.
Using the conservation of transmission probability $\sum_n T_{mn}=1$
(for a single channel in each lead), one can write Eq.~(\ref{4.03})
in the form
\begin{equation}
G = 2 {e^2 \over h} \left[ 1-T_{11} + {T_{13} T_{31} \over 1 -T_{33} }
\right] .
\label{4.04}
\end{equation}
The reciprocity relations $T_{mn}(\Phi)=T_{nm}(-\Phi)$ imply that
$G$ is an even function of flux. Consequently, the phase rigidity
persists even in the presence of inelastic scattering. The only
effect of inelastic processes is to reduce the amplitude of the AB
oscillations by decreasing the direct elastic transmission $T_{12}$.

\begin{figure}
\centerline{\psfig{file=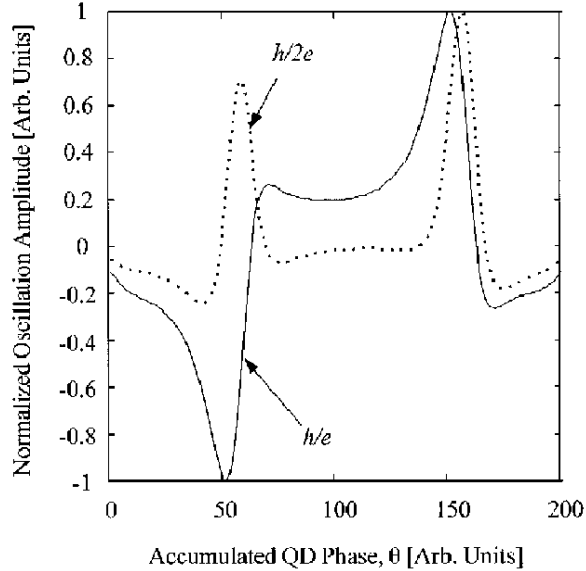,width=8cm,angle=0}}
\caption[]{Amplitude of the AB harmonics vs.\ the phase
  $\theta$. The quantum dot displays transmission peaks around
  $\theta = 70$ and $\theta= 170$. Taken from
  Ref.~\protect{\cite{Yac96}}.}
\label{yactheo1} \end{figure}

\begin{figure}
\centerline{\psfig{file=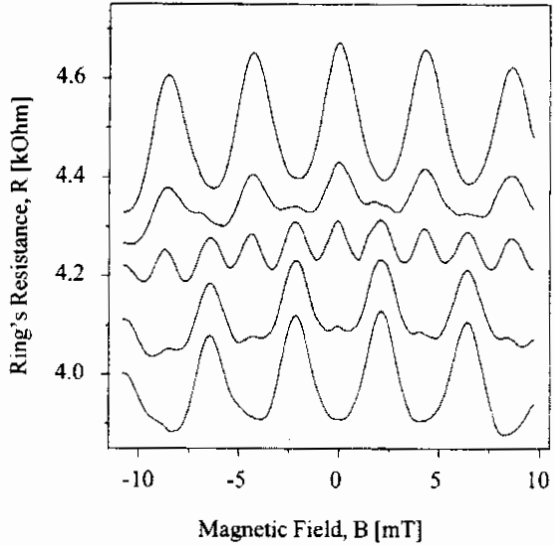,width=8cm,angle=0.4}}
\caption[]{Oscillatory part of the ring resistance as a function of
  the magnetic field. Measured for a two--terminal AB
  ring with an artificial impurity embedded in one of its arms.
  Taken from Ref.~\protect{\cite{Yac96}}.} \label{yactheo2}
\end{figure}

The nature of the AB oscillations close to a phase jump was studied
both theoretically and experimentally by Yacoby, Schuster and
Heiblum \cite{Yac96}. These authors showed that the phase jump
occurs at the point where the amplitude of the $h/e$ periodic AB
oscillations vanishes. At that point, the ring conductance is
dominated by $h/2e$ periodic oscillations. The theoretical analysis
in Ref.~\cite{Yac96} is based on a formula derived by Gefen, Imry
and Azbel \cite{Gef84} for the two--terminal conductance across a
single--channel AB ring with single scatterers in each arm,
\begin{equation}
G = {2 e^2 \over h} \left| {2(A e^{i \phi} + B e^{-i \phi}) \over
D e^{2 i \phi} +E e^{-2i\phi}+C} \right|^2 .
\label{4.05}
\end{equation}
Here $\phi=\pi \Phi / \Phi_0$ is the dimensionless flux through the
ring, and the constants $A$,\ldots,$E$ defined in Ref.~\cite{Gef84}
are functions of the scattering amplitudes of the scatterers in the
ring. The quantum dot in one of the arms is modeled as a
1D--resonant tunneling device. The reflection and transmission
amplitude through the dot can be written in terms of the scattering
amplitudes of the adjacent tunneling barriers and a phase $\theta$
accumulated by the motion in the dot. A variation of the plunger
voltage at the dot is simulated by a variation in $\theta$.

The predicted amplitudes of the $h/e$ and the $h/2e$ harmonics are
shown in Fig.~\ref{yactheo1}. The $h/e$ component grows as a 
conductance peak is approached from the left. Near the peak the $h
/e$ component drops and vanishes for some value of $\theta$. The
current oscillations are then dominated by the $h/2e$ oscillations.
Upon further increase of $\theta$, the $h/e$ component changes sign and
grows again. The change in the sign of the $h/e$ component is
detected experimentally as a jump by $\pi$ in the AB phase.

Jacoby et al.\ \cite{Yac96} corroborated their theoretical arguments
by a measurement of the AB oscillations through a ring with an
artificial impurity. A small gate inserted in one arm of the ring
creates a potential barrier for the electron motion.  The potential
at the gate can be varied by means of a metallic air bridge. The
impurity replaces the quantum dot in the earlier experiment
\cite{Yac95}. The AB oscillations of the ring resistance are
displayed in Fig.~\ref{yactheo2} for increasing impurity strength.
One observes that the $h/e$ component has a rigid phase.  The $h/2e$
oscillations become strong when the $h/e$ oscillations vanish. The
phase jump in the $h/e$ oscillations is thus associated with a
qualitative change in the AB oscillations: The $h/e$ oscillations
vanish and the AB oscillations show $h/2e$ periodicity.

The symmetry (\ref{4.01}) of the two--terminal conductance strictly
holds only in the linear response limit of infinitesimally small
currents and voltages. Bruder et al.\ \cite{Bru96} investigated the
nonlinear response regime using a tunneling Hamiltonian description
of the quantum dot. They expressed the current through the AB ring
in terms of Green functions of the quantum dot. This description
includes charging effects in a non--perturbative way. Bruder et al.\
recover a symmetric conductance in the linear response limit. In the
nonlinear response regime they find derivations from the symmetry
under sign change of the external flux. As a consequence, the phase of
the AB oscillations can change continuously with the voltage at the
quantum dot when a finite voltage difference is applied across the AB
ring.

\subsection{Aharonov--Bohm current}
\label{Sec4.2}
The phase of the AB oscillations in a two--terminal measurement
is rigid and yields little information about the transmission
phase through the quantum dot. The latter phase, however,
strongly affects the amplitude of the AB oscillations as was first
shown by Hackenbroich and Weidenm\"uller \cite{Hac96a,Hac96b}. Their
approach starts from a description of the dot and the AB
ring in terms of a tunneling Hamiltonian. With little modification 
this Hamiltonian also serves as the starting point for the
discussion in Sec.\ \ref{Sec5}. We introduce the model in
Sec.~\ref{Sec4.2.1}. The derivation of the AB amplitude is
sketched in Secs.\ \ref{Sec4.2.2}-\ref{Sec4.2.3}. The calculation
is based on single--particle scattering theory and the Landauer
formula. This restricts the approach to the regime near the
conductance peaks and to temperatures $kT \ll \Delta$, where
$\Delta$ is the single--particle level spacing of the quantum dot. 
 
\subsubsection{Tunneling Hamiltonian}
\label{Sec4.2.1}
\begin{figure}
\centerline{\psfig{file=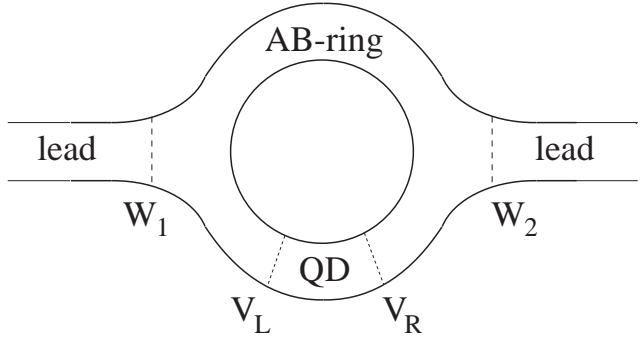,height=4.5cm,angle=0}}
\vspace*{0.2cm}
\caption[]{An AB ring threaded by the magnetic flux $\Phi$
  is connected to two external leads. A quantum dot is embedded in
  one arm of the ring. Tunnel barriers between the dot and the ring
  are modeled by matrix elements $V_{L,R}$. The coupling between the
  ring and the leads is described by the matrix elements $W_{1,2}$.}
\label{qdl1}
\end{figure}

Consider the system schematically represented in Fig.~\ref{qdl1}.  A
conducting ring threaded by the magnetic flux $\Phi$ is connected to
two leads. The ring and the leads may support several transverse
channels. A quantum dot is embedded in one arm of the ring. The
quantum dot is weakly coupled to the ring by potential barriers on
either side of the dot. The quantum dot is in the Coulomb blockade
regime. The electrostatic potential at the dot is controlled by the
gate voltage $V_g$.

The system is described in terms of the tunneling Hamiltonian
\begin{equation} 
H=H_0+H_T,
\label{M1}
\end{equation} 
where $H_0$ describes the isolated subsystems and $H_T$ the
couplings between these subsystems. Explicitly, $H_0$ is given by
\begin{equation} 
H_0  = 
\sum_{mr} \int dE E a_{mE}^{r \dagger} a_{mE}^r + \sum_i \epsilon_i
d_i^{\dagger} d_i + \sum_\lambda {\cal E}_\lambda c_\lambda^{\dagger} 
c_\lambda +U,  
\label{M2} 
\end{equation} 
where $r=1,2$ labels the two leads, $m$ the channels in either lead
and $i$ and $\lambda$ the single--particle states in the AB ring and
the quantum dot, respectively. The respective single--particle
energies are denoted by $E$ (the longitudinal energy in a channel),
$\epsilon_i$ and ${\cal E}_\lambda$. The interaction $U$ has the
form
\begin{equation} 
\label{corr} U = { 1 \over 2} U_0 (\hat{N}^2-\hat{N}),
\end{equation} 
where $\hat{N}=\sum_\lambda c_\lambda^{\dagger} c_\lambda$ is the
total number of electrons on the dot, $U_0=e^2/C$ the charging
energy, and $C$ the total capacitance between the dot and its
surroundings. We use the standard picture of the Coulomb blockade
\cite{Bee91} and assume that the energies ${\cal E}_\lambda$ vary
linearly with the gate voltage 
\begin{equation}
\label{ener} {\cal E}_\lambda = {\cal E}_\lambda^0 + \alpha V_g,
\end{equation} 
where $\alpha$ is some function of the capacitance matrix of the
system. Below we assume that the energies $E_\lambda$ are
non--degenerate.

\vspace*{0.0cm}

The couplings have the form
\begin{equation}
H_T = (\sum_{mri} \int dE W_{mi}^r(E) a_{mE}^{r \dagger} d_i + 
\mbox{H.c.})
  + (\sum_{i p \lambda} V_{\lambda i}^p c_\lambda^{\dagger} d_i +
\mbox{H.c.}),
\label{M3}
\end{equation}
where the matrix elements $W$ couple states in the ring to states of
the leads, $V$ provides the coupling between states in the ring and
the dot, and $p=L,R$ labels either side of the dot. All matrix
elements do not change appreciably on the scale of the charging
energy. This weak energy dependence will be neglected below.

It follows from time--reversal invariance that for vanishing
magnetic flux through the ring, all matrix elements can be
taken real. To account for non--zero flux $\Phi$ through the AB
ring, we attach to each matrix element $V_{\lambda i}^R$ a factor
$\exp(i\phi)$, where $\phi \equiv 2 \pi \Phi/\Phi_0$. We can thus
write
\begin{eqnarray}
\label{M4} & & V_{\lambda i}^L = V_{\lambda i}^{L*} = v_{\lambda i}^L, \\
\label{M5} & & V_{\lambda i}^R \exp(-i \phi) = V_{\lambda i}^{R*} \exp(i \phi) =
v_{\lambda i}^R,
\end{eqnarray}
where $v_{\lambda i}^L$ and $v_{\lambda i}^R$ are real. This
parameterization is adequate whenever the single--particle states
both in the AB ring and in the dot do not change appreciably with
flux, i.e.\ whenever the flux through each arm of the ring and
through the dot is smaller than $\Phi_0$.

\subsubsection{Scattering matrix}
\label{Sec4.2.2}
Near transmission resonances and for temperatures $kT$ much less 
than the single--particle level spacing, scattering through a
quantum dot may be described in a single--particle picture. The
scattering matrix may then be derived explicitely. We first
sketch the caclulation for noninteracting electrons (the limit of
zero charging energy $U_0 \to 0$) and then indicated the changes
resulting from nonzero $U_0$. 

The scattering matrix $S=1-2\pi i T$ at energy $E$ and (dimensionless)
flux $\phi$ is obtained from the Lippmann--Schwinger equation
\begin{equation}
T=H_T+H_T(E-H_0+i \eta)^{-1} T
\label{Lipp}
\end{equation}
for the transition operator $T$. Here, $\eta$ is positive
infinitesimal. Iteration of the Lippmann--Schwinger equation
yields the Born series
\begin{equation}
T = H_T + H_T {1 \over E - H_0 +i \eta} H_T + H_T {1 \over E - H_0 +i
\eta} H_T {1 \over E - H_0 +i \eta} H_T + \cdots .
\label{Born}
\end{equation}
For non--interacting electrons the right--hand--side reduces to a
geometric series which can easily be resumed. For the S--matrix
element $S_{mn}^{rs}$ connecting channel $n$ in lead $s$ with
channel $m$ in lead $r$ one obtains the result
\begin{equation} 
S_{mn}^{rs}(E,\phi)  =  \delta^{rs}
\delta_{mn} -2\pi i [W D_{\rm ring}^{-1} W^{\dagger}]_{mn}^{rs}
- i [\gamma D_{\rm dot}^{-1} \bar{\gamma}^{\dagger}]_{mn}^{rs}, 
\label{Sresult}
\end{equation}
where we used the shorthand $W$ and $V\equiv V^L+V^R$ for the
coupling matrix elements and introduced the propagators
\begin{eqnarray}
\label{Dring} D_{{\rm ring},ik} & = &(E-\epsilon_i)\delta_{ik}+i\pi [ 
W^{\dagger} W]_{ik}, \\
\label{Ddot} D_{{\rm dot},\lambda \mu} & = & (E-{\cal E}_\lambda)
\delta_{\lambda \mu} 
-[V D_{\rm ring}^{-1} V^{\dagger}]_{\lambda \mu}, 
\end{eqnarray}
for the ring and the dot, respectively. The partial width
amplitudes $\gamma_{m \lambda}^s$, $\bar{\gamma}_{m \lambda}^s$
are given by
\begin{eqnarray}
 \label{Ega}  \gamma^r_{m \lambda} & = & \sqrt{ 2 \pi}
  [W (D_{\rm ring})^{-1} V^{\dagger}]^r_{m \lambda}, \\ 
\label{Egb}  \bar{\gamma}^s_{n \mu} & = & \sqrt{ 2 \pi} 
   [W (D_{\rm ring}^{\dagger})^{-1} V^{\dagger}]^s_{n \mu}. 
\end{eqnarray}
The terms on the right--hand--side of Eq.~(\ref{Sresult}) have a
simple physical interpretation. The first term involving the
Kronecker deltas describes the reflection from channel $m$ back into
the same channel. The second term is flux and voltage independent
and accounts for the scattering of elctrons through the free arm of
the ring. The last term describes the scattering of electrons
through the dot. The entire flux and voltage dependence resides in
this term (via both $D_{\rm dot}$ and the partial width amplitudes).
Note that the flux dependence does not reduce to a phase factor
$\exp(i \phi)$ as one naively might have expected. The origin for
the more complicated flux dependence is the multiple scattering of
electrons through the ring which gives rise to higher harmonics of
the current oscillations.  

In the Coulomb blockade regime, the quantum dot is weakly coupled to
the AB ring. The quantum dot resonances are then isolated i.e.\
their spacing is much larger than their width. In this case and
for $E$ close to a resonance, one can approximate $D_{\rm
dot}$ by the diagonal matrix 
\begin{equation} 
\label{Ddiag} (D_{\rm dot})_{\lambda \mu} \equiv
(E-{\cal E}_\lambda- \Delta {\cal E}_\lambda +i \Gamma_\lambda /2)
\delta_{\lambda \mu} . 
\end{equation} 
Here, $\Delta {\cal E}_\lambda$ is the energy shift (of the
resonance position with respect to the bound state energy of the
isolated system), and $\Gamma_\lambda$ the total width of the
resonance with index $\lambda$.  Using the unitarity of $S$, one 
can show that for isolated resonances 
\begin{equation}
\label{Width2}
\Gamma_\lambda = \sum_{mt}|\gamma^t_{m \lambda}|^2= \sum_{mt}
|\bar{\gamma}^t_{m \lambda}|^2.
\end{equation}
Note that because of the oscillatory behavior of the partial width
amplitudes, also $\Gamma_\lambda$ is an oscillating function of
flux. We further note that the $S$-matrix (\ref{Sresult}) has the
well-known symmetry properties
\begin{eqnarray}
\label{unitarity} S(\phi) S^{\dagger} (\phi)  = &  S^{\dagger} (\phi) 
 S(\phi) & =  1 , \\
\label{timerev} S(\phi)S^*(-\phi)  = &  S^*(\phi) S(-\phi) & =  1,
\end{eqnarray}
that follow from current conservation and full time--reversal
invariance (including a reversal of the magnetic field). Both
symmetries together imply that the linear conductance is an even
function of magnetic flux \cite{Bue88} and impose the phase
rigidity as discussed in Sec.\ \ref{Sec4.1}.  

The S--matrix (\ref{Sresult}) for non--interacting electrons may be
generalized to the case of nonzero charging  energy using the
Hartree--Fock approximation. For isolated quantum dot
resonances, one finds \cite{Hac96b} that the scattering matrix
$S(E_F,\phi)$ at the Fermi energy has the matrix elements 
\begin{eqnarray} 
\label{scattm} 
S_{mn}^{rs}(E_F,\phi) & = & \delta^{rs}
\delta_{mn} -2\pi i [W D_{\rm ring}^{-1} W^{\dagger}]_{mn}^{rs}
\nonumber \\ && - i
\sum_{\lambda} \gamma_{m\lambda}^r (E_F-E_\lambda+i \Gamma_\lambda/2)^{-1} 
\bar{\gamma}^{\dagger s}_{n \lambda}, 
\end{eqnarray}     
where $\gamma_{m \lambda}$, $\bar{\gamma}_{n \lambda}$, and
$\Gamma_\lambda$ are defined in Eqs.\ (\ref{Ega}), (\ref{Egb}), and
(\ref{Width2}), respectively. The S--matrix has the same form as in
the non--interacting case (\ref{Sresult}). However, the
single--particle energies ${\cal E}_\lambda$ have been replaced by
the resonance energies $E_\lambda$ which are determined by the set
of self--consistent equations
\begin{eqnarray} 
\label{rener1} E_\lambda = {\cal E}_\lambda + \Delta {\cal E}_\lambda
+  U_0 \sum_{\mu \neq \lambda} \langle n_\mu \rangle, 
\end{eqnarray} 
where the the average occupation probability $\langle n_\lambda
\rangle$ of the level $\lambda$ is given by   
\begin{eqnarray}
  \langle n_\lambda \rangle & = & {1 \over  \pi} 
          \int_{- \infty}^{\infty} dE f(E-E_F) {\Gamma_\lambda \over
|E - E_\lambda + i \Gamma_\lambda /2|^2} .
\label{self1} 
\end{eqnarray} 
Here $f(E-E_F)=[1+\exp((E-E_F)/kT)]^{-1}$ denotes the Fermi
function. The solution to these equations is easiliy found
when the single--particle levels are non--degenerate. Whenever one
level becomes filled, all higher levels are shifted upwards by an
amount equal to the charging energy. In effect, the Hartree--Fock
approach reduces to a picture of isolated resonances with a
`stretched' level spacing due to the Coulomb interaction.

We note that the mean--field approach used here is only valid close
to the resonances and at temperatures $kT$ much less that the
single--particle level spacing $\Delta$ in the quantum dot. The
conductance at higher temperatures $kT \sim \Delta$ can no longer be
reduced to a single--particle scattering problem. The sequential
(incoherent) tunneling through a quantum dot in the regime $kT \sim
\Delta$ has been studied by Beenakker \cite{Bee91} and by Meir et
al.\ \cite{Mei91}. 

\subsubsection{Conductance}
\label{Sec4.2.3}
The dimensionless conductance $g = (h/e^2)G$ of the AB ring with the
quantum dot is obtained from the multi-channel Landauer formula
\begin{equation}
  g=2\int dE \left(-{\partial f \over \partial E}\right)
  \sum_{m,n=1}^N |t_{mn} (E)|^2.
\label{C1}
\end{equation}
Here $t_{mn}(E)=S_{mn}^{12}(E)$ is the transmission amplitude
through the ring for an electron entering the ring via channel $n$
in lead two, and leaving it via channel $m$ in lead one. The
derivative of the Fermi function $f$ is given by $- \big( \partial
f/\partial E \big) = (4 k T)^{-1} \cosh^{-2}((E - E_F)/2 k T)$, and
$E_F$ is the Fermi energy in the leads. A factor two accounts for
the spin degeneracy of the electron.

It is assumed that both the charging energy $U_0$ and $k T$ are much
larger than the resonance widths $\Gamma_\lambda$.  Moreover, $k T$
shall be much smaller than the single--particle level spacing
$\Delta$.  Then an appreciable current can pass the dot only if a
resonance in the dot is close to the Fermi energy, $E_\lambda
\approx E_F$. The contribution of other resonances to the
transmission amplitude can be neglected.  To simplify notation, we
suppress the index $\lambda$ below.  According to
Eq.~(\ref{scattm}), the transmission amplitude has the form
\begin{equation}
t_{mn}=t_{{\rm ring},{mn}}- i {\gamma_{m}^1 \bar{\gamma}_{n}^{2*} \over
        E -E' +i \Gamma /2},
\label{C2}
\end{equation}
where $E'$ is the resonance energy and $t_{{\rm ring},{mn}}=-2i\pi
\sum_{ik} W_{mi}^1 (D^0)_{ik}^{-1} W_{nk}^{2*}$ is the transmission
through the free arm of the ring. This contribution is independent
of $\phi$ and $V_g$ and only weakly dependent on energy (this energy
dependence will be neglected).
 
\begin{figure}
\centerline{\psfig{file=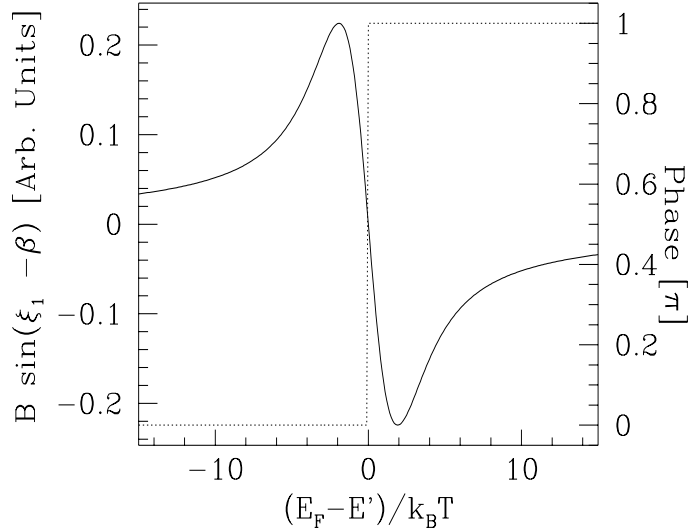,width=9cm,height=7cm,angle=270}}
\vspace*{0.3cm}
\caption[]{ Amplitude of the second term of Eq.~(\ref{C17}) as a
  function of $(E_F-E^\prime)/kT$ for $\xi_1 = \pi/2$. The dashed
  line shows the phase extracted from the sign of $B \sin(\xi_1
  -\beta)$. Taken from Ref.~\protect{\cite{Hac96b}}.}
\label{qdl3}
\end{figure}

We now confine ourselves to a symmetric dot where $v^L_i = v^R_i
\exp(i \chi) = v_i$. Due to time--reversal symmetry, $v_i$ can be
chosen real, and $\chi$ can only take the values $0$ and $\pi$. 
Substituting Eq.~(\ref{C2}) into the Landauer formula, summing
over the channels and integrating over energy, one finds that the
conductance can be written as
\begin{eqnarray} 
g & = &  g_{\rm ring} + [ 1 + \cos(\phi-\chi)] x  {\tilde{\Gamma} 
\over k T} B \sin( \xi -\beta)  \nonumber \\ & & +[1 + \cos(\phi
-\chi)] y { \pi \tilde{\Gamma} \over 2 k T} \cosh^{-2} \left(
E_F-E' \over 2 k T \right),  
\label{C17}
\end{eqnarray}
where $g_{\rm ring}$, $x$, $y$ are positive coefficients, $\xi$ is a
real phase shift, and $\tilde{\Gamma}=(1+\cos[\phi- \chi])^{-1}
\Gamma$ is the flux--independent width. We note that $g$ is an even
function of the magnetic flux $\phi$ as required for a two--terminal
measurement.  Both the amplitude $B$ and the phase $\beta$ are
functions of $(E_F-E^\prime) /(kT)$ (see Fig.~\ref{qdl3}). The phase
$\beta$ may be identified with the transmission phase through the
quantum dot.  It takes the value $\pi/2$ at $E=E'$, while
approaching $0$ for $E \rightarrow -\infty$ and $\pi$ for $E
\rightarrow \infty$, respectively.

Equation (\ref{C17}) allows us to discuss the current through the AB
ring as a function of energy (voltage $V_g$) and magnetic flux. We
refer to the first term on the right hand side of Eq.~(\ref{C17}) as
background term, to the second term as interference term, and to
the third term as resonance term. The background term comprises the
bulk part of the current. This term is independent of both energy
and flux. It is due to transmission of electrons through the free
arm of the AB ring. This term is subtracted in the experimental
analysis from the total current through the device.  The other
terms are smaller than the background term by a factor
$\tilde{\Gamma} / k T \ll 1$. The interference term is due to the
interference of the background amplitude with the transmission
amplitude for passage through the quantum dot. This term displays
resonant behavior and depends explicitly on the geometry of the ring
through the phase shift $\xi$. The amplitude of the resonance term
has the usual temperature dependence known for thermally broadened
resonances \cite{Bee91} of quantum dots directly coupled to leads.
However, in the AB geometry this term is flux--modulated and
contributes to the AB oscillations.

We now focus on the AB phase. As a function of the voltage at the
dot, only the amplitude of the current oscillations is changed. The
AB phase is unaffected unless the interference term changes sign. In
this case, the AB phase jumps by $\pi$. The location of the phase
jump depends on the system specific phase shift $\xi$.
Eq.~(\ref{C17}) predicts that all higher harmonics vanish
identically. This is a consequence of keeping only the resonance
closest to $E_F$ and neglecting the far away resonances. Such
resonances as well as an asymmetric dot produce higher (e.g.\
$\Phi_0/2$ periodic) harmonics which may be observable whenever the
$\Phi_0$ periodic harmonic vanishes.

\vspace*{0.1cm}

Inelastic scattering in the quantum dot destroys the unitarity of
the $S$-matrix. Phenomenologically, this increases the widths of the
transmission resonances $\Gamma$ by the addition of an inelastic
width \cite{Sto85,Bue88} and thereby reduces the amplitudes of both
the interference term and the resonance term. The fundamental
property of the current to be an even function of the applied flux
remains unchanged. As a result, with properly rescaled amplitudes,
Eq.~(\ref{C17}) accounts for the current through the system even in
the presence of inelastic scattering. The basic conclusions
concerning the temperature and flux dependence of the various terms
remain valid.

\newpage
\setcounter{equation}{0}
\section{Transmission phase through a quantum dot} 
\label{Sec5}
The experiment of Yacoby et al.\ \cite{Yac95} proved for the first
time that part of the transmission through a quantum dot in the
Coulomb blockade regime is coherent. The phase of the transmission
amplitude through the dot could not be measured in this experiment due
to the phase rigidity discussed in Sec.~\ref{Sec4}. The problem of
phase rigidity was solved by Schuster et al.\ \cite{Sch97} who
replaced the two--terminal AB ring by a multiple--terminal device.
The AB phase in the Schuster--experiment increased roughly by $\pi$
whenever the gate voltage at the quantum dot was swept through a
transmission resonance, and the profile of the phase evolution was
well described by a Breit--Wigner formula. At the same time, the phase
displayed unexpected properties: (i) It was nearly identical for {\em
  all} resonances investigated in the experiment, and (ii) between
each pair of adjacent resonances the phase sliped roughly by $\pi$ on
a very small energy scale. Both observations (i) and (ii) were totally
unexpected. The theoretical work addressing these observations is
reviewed in this section.
 
\begin{figure} 
\hspace*{0cm} 
\centerline{\psfig{file=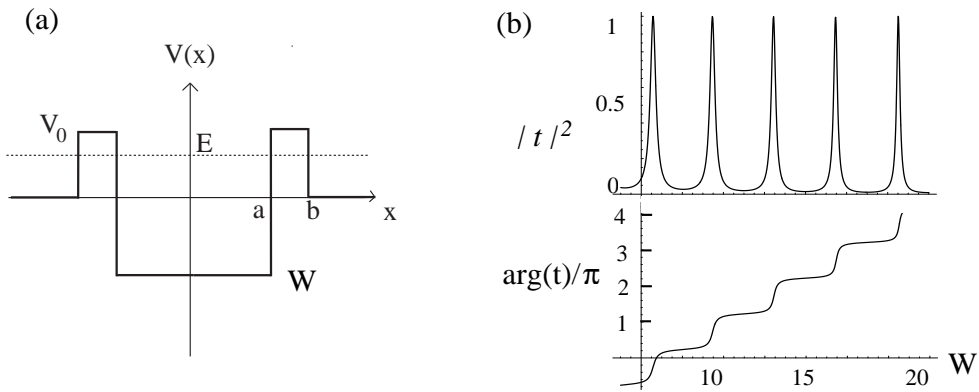,width=13cm,angle=0}} 
\vspace*{0.4cm} 
\caption[]{(a): Double-barrier well potential. (b): Transmission 
  coefficient $|t|^2$ (top) and transmission phase ${\rm arg}(t)$ 
  (bottom) both as a function of $W$. Parameters are $V_0 = 10.0$, 
  $E=2.0$. All energies are given in units of $\hbar^2/2ma^2$.  
} \label{potphase} 
\end{figure}

Our understanding of the transmission phase is largely based on
quantum mechanical scattering of non--interacting particles. The
one--dimensional analogue of a quantum dot is a double--barrier well
as depicted in Fig.~\ref{potphase}(a). The tunneling barriers mimic
negatively biased gates on either side of the dot. A change in the
gate voltage at the quantum dot is modeled by a shift of the potential
in the well. The calculation of the transmission amplitude $t$ through
the double--barrier well is a textbook problem. The transmission
coefficient $|t|^2$ and the transmission phase ${\rm arg} (t)$ are
shown in Fig.~\ref{potphase}(b) as a function of the well potential.
The phase increases smoothly by $\pi$ over the width of a transmission
resonance and stays nearly constant between resonances.  As a result,
the phase of neighboring resonances {\em always differs by $\pi$}. We
refer to such resonances as off--phase resonances below; resonances
with similar transmission phase will be called in--phase resonances.
The appearance of off--phase resonances is not a special property of
the double--barrier quantum well but holds generally for scattering
through time--reversal invariant strictly one--dimensional systems.
In such systems, the wave functions can be chosen real. The wave
functions representing neighboring resonant states differ by one
additional node, and with each node the phase of the transmission
amplitude increases by $\pi$. As a result, the phase evolution in the
one--dimensional model is qualitatively different from the phase
evolution measured in Refs.~\cite{Yac95,Sch97}. Thus two questions
arise: First, how can in--phase resonances occur in the Schuster--
(and Yacoby--) experiment, and why are {\em all} resonances in phase?
Second, why does the phase slip between resonances and why so sharply?
 
Although we will focus on the transmission phase, we note that the
underlying questions go much deeper and concern the nature of
electron transport through quantum dots. Starting with the work of
Jalabert, Stone and Alhassid \cite{Jal92} in the early 1990's, the
theoretical work on quantum dots has almost exclusively addressed
universal aspects described by random matrix theory. Employing
random matrix theory, quantities like the statistical distribution
of the conductance peak heights \cite{Jal92} and parametric
peak--height correlations \cite{Bruu96,Alh96} have been calculated
and tested in various experiments. Because of the enormous success
of these calculations random matrix theory has emerged as a paradigm
for the theoretical description of quantum dots. However, in recent
years a number of experiments showed clear deviations from random
matrix behavior.  Among these are transport experiments on small
vertical quantum dots \cite{Tar96}, conductance peak--height
measurements on lateral quantum dots \cite{Fol96}, and the
Coulomb--blockade interference experiments \cite{Yac95,Sch97}.  We
will show below that the failure of random matrix theory for the
experiments \cite{Yac95,Sch97} may be traced back to features of the
{\em short--time dynamics} in the quantum dot.  Under certain
conditions such {\em non--universal} features can totally dominate
the quantum transport in the tunneling regime and generate strong
deviations from random matrix predictions.

The material covered in this section is organized systematically so
that papers with related theoretical ideas are discussed together.
This ordering displays the underlying physical concepts most clearly.
Sometimes the systematic ordering could only be achieved at the
expense of giving up a strictly chronological order. In Sec.\ 
\ref{Sec5.0} we summarize the scattering theory through a region of
interacting electrons, and derive the transmission phase through a
{\em weakly coupled} quantum dot. This section forms the basis for the
arguments presented in the subsequent sections. In Sec.\ \ref{Sec5.1},
we discuss a general theorem known as the Friedel sum rule. The
Friedel sum rule relates the amount of charge added to a conducting
system to the sum of all scattering phase shifts. We clarify the
relation between the scattering phase shifts that enter in the Friedel
sum rule and the transmission phase measured in a quantum dot
interference device. The phase lapse between resonances is discussed
in Sec.~\ref{Sec5.2}. Mechanisms for in--phase resonances are
summarized in Sec.~\ref{Sec5.3}. We emphasize, that despite of the
arguments for in--phase resonances, non of the theoretical studies
excludes the possibility of out--off--phase resonances.  Further
experimental work is necessary to clarify whether the latter may or
may not be found in quantum dots.
      
\subsection{Transport theory}
\label{Sec5.0}
The mean--field theory for transport through a quantum dot described
in Sec.~\ref{Sec4.2} is only valid near the conductance peaks and at
temperatures much smaller than the mean level spacing $\Delta$. When
these conditions are not met, transport through a quantum dot can no
longer be reduced to a mean--field problem. Several techniques have
been developed to calculate the current through interacting electron
systems. Our presentation follows the approach of Refs.\ 
\cite{Bal99a,Bal99b,Bal99c} which is based on earlier work of Meir,
Wingreen, and Lee \cite{Mei91,Mei92}.

\subsubsection{Generalized Landauer formula}
\label{Sec5.0.1}
We consider an Aharonov--Bohm ring with a quantum dot embedded in
one arm. The arm containing the dot is modeled by the tunneling
Hamiltonian
\begin{eqnarray} 
  H&=& H_{L} + H_{R} +H_{\rm QD} +H_T \ , \\ 
  H_{L(R)} &=& \sum_{k} \epsilon^{L(R)}_k \ {a^{L(R)}_k}^{\dagger} 
  a^{L(R)}_k \ ,  \\ 
  \label{eq:H_dot}
  H_{\rm QD} &=& \sum_{\lambda} \ {\cal E}_\lambda c^{\dagger}_\lambda 
  c_\lambda + \frac{1}{2} U_0 (
  \hat{N}^2 -\hat{N}) \ , \\ 
  H_T &=& \sum_{k,\lambda} \left( V^L_{k,\lambda} a^L_k 
  c^{\dagger}_\lambda +  \mbox{H.c.} \right) + L \leftrightarrow R 
 \ . 
\end{eqnarray} 
Here, $H_{L}$ and $H_R$, respectively, describe the regions to the
left and right of the QD, $H_{\rm QD}$ is the Hamiltonian of the
isolated QD including the charging energy, and $H_T$ represents the
tunneling of electrons in and out of the QD. All energies are
counted from the Fermi level in the leads. For simplicity, we
assumed only one transverse channel in the left and right region.
In the absence of a magnetic field, the matrix elements
$V^{L,R}_{k,\lambda}$ can be chosen real.

An exact formula for the current through the ring can be derived
using the nonequilibrium Keldysh formalism \cite{Mei92}. The current
is expressed in terms of the Fermi function in the leads and local
properties of the interacting region. In general, the formula for
the current includes inelastic scattering, spin flips, and even
scattering processes of several electrons. However, in the linear
response limit and for sufficiently low temperature only elastic
processes are allowed by energy conservation. Then, the conductance
through the Aharonov--Bohm ring reduces to the generalized
Landauer--type formula $G_{\rm ring} = (2 e^2 / h) T_{\rm ring}$,
where the total transmission probability $T_{\rm ring}$ is given by
\begin{eqnarray} 
{T}_{\rm ring} =  \int dE
\left( - {\partial f \over \partial E} \right) \left| t_{0} +
t(E)  \exp[2 \pi i \Phi / \Phi_0] \right|^2 . 
\label{eq:Bal1} 
\end{eqnarray} 
Here $t_{0}$ and $t(E)$ denote the transmission amplitude through
the free arm and the arm with the quantum dot, respectively, $\Phi$
is the magnetic flux through the ring, and $\Phi_0= h/e$ the
elementary flux quantum. Note that we have neglected higher
harmonics.  From the interference term in Eq.\ (\ref{eq:Bal1}) one
can extract the amplitude of transmission through the arm with the
quantum dot
\begin{eqnarray}
\label{eq:tphase}
t_{\rm QD} = \int dE ( -\partial f / \partial E )
t(E).
\end{eqnarray}
We identify $\theta_{\rm QD} \equiv {\rm arg} (t_{\rm QD})$ with the
transmission phase through the quantum dot. The transmission
coefficient through the arm with the dot is given by
\begin{eqnarray}
\label{eq:tcoeff}
{T}_{\rm QD} \equiv \int dE (-\partial f / \partial E ) |t(E)|^2.
\end{eqnarray}
Both $\theta_{\rm QD}$ and $T_{\rm QD}$ can be measured in quantum
dot interference experiments.  The transmission amplitude $t(E)$ can
be expressed in terms of the retarded Green function $G^r_{\lambda
  \mu}$ of the dot
\begin{eqnarray} 
t(E)=\sum_{\lambda ,\mu} V_\lambda^L(E) G^r_{\lambda 
\mu}(E) V_\mu^{R*}(E) \ ,  
\label{eq:Bal1a}
\end{eqnarray} 
where we introduced the tunneling amplitudes $V_\lambda^{L,R} (E) =
[2 \pi \rho^{L,R}(E)]^{1/2} V_{k(E),\lambda}^{L, R}$ with the
density of states $\rho^{L,R}(E)$ in lead $L,R$. To simplify
notation, we will drop the argument $E$ and write $V_\lambda^{L,R}$
below. We emphasize that the Green function $G^r_{\lambda \mu}$ must
be calculated in the presence of interactions and tunneling.

In the limit $t_0 =0$ of vanishing transmission through the free
arm, the ring conductance reduces to the two--terminal conductance
$G_{\rm QD}=(2 e^2/ h) T_{\rm QD}$ through a quantum dot coupled
directly to reservoirs. We derive $G_{\rm QD}$ for the case when the
Green function is diagonal in the single--particle basis of the dot.
Substitution of Eq.~(\ref{eq:Bal1a}) into Eq.~(\ref{eq:Bal1}) then
yields a product of two Green functions which can be evaluated using
the relation
\begin{eqnarray}
G^r_{\lambda \lambda} G^{r*}_{\lambda \lambda} = {1 \over 
\Sigma_{\lambda \lambda} -\Sigma^*_{\lambda \lambda}} 
(G^r_{\lambda \lambda}-G^{r*}_{\lambda \lambda}) =
{{\rm Im} G^r_{\lambda \lambda} \over {\rm Im} \Sigma_{\lambda 
\lambda} },
\label{eq:Bal1b}
\end{eqnarray}
where $\Sigma$ is the self--energy. The diagonal elements
$\Sigma_{\lambda \lambda}= i(\Gamma_\lambda^L+ \Gamma_\lambda^R)/2$
are given in terms of the partial widths $\Gamma_\lambda^{L,R} =
|V_\lambda^{L, R}|^2$ for decay of the state $\lambda$ into the left
and right lead, respectively. The current takes the form
\begin{eqnarray}
\label{eq:Bal1c}
G_{\rm QD} = {2 e^2 \over h} \int dE \left( {\partial f \over \partial E}
\right) \sum_\lambda {\Gamma_\lambda^L \Gamma_\lambda^R \over
\Gamma_\lambda^L +\Gamma_\lambda^R} {\rm Im} G_{\lambda
\lambda}^r(E) .
\end{eqnarray}
This result was first obtained in Refs.\ \cite{Mei91,Mei92}. Note
that $-(1/\pi) {\rm Im} G^r_{\lambda \lambda}(E)$ is the local level
density in the presence of interactions and tunneling.  The above
derivation can be generalized to include the electron spin. In this
case the conductance takes a form similar to Eq.~(\ref{eq:Bal1c}),
but the factor 2 is replaced by an explicit summation over the
electron spin.  The resulting expression has been used to study the
conductance in the Kondo regime \cite{Ng88,Mei92}.

\subsubsection{Retarded Green function}
\label{Sec5.0.2}
To derive the dot Green function we employ the equations--of--motion
method \cite{Mei91}. The derivation starts from the retarded Green
function $G^r_{\lambda \mu}(t)$, defined by
\begin{eqnarray}
\label{eq:Bal10}
G^r_{\lambda \mu}(t) = -i \theta(t) \langle \{ c_\lambda(t), 
c^{\dagger}_\mu(0) \} \rangle,
\end{eqnarray}
where the curly brackets denote the anticommutator. The expectation
value is the thermal average with respect to the Hamiltonian $H$.
The operator $c_\lambda(t)$ is the solution of the Heisenberg
equation
\begin{eqnarray}
\label{eq:Bal11}
i { \partial c_\lambda(t) \over \partial t} = [c_\lambda , H] .
\end{eqnarray}
Differentiating Eq.~(\ref{eq:Bal10}) with respect to $t$ and
substituting Eq.~(\ref{eq:Bal11}) one obtains the equation of motion
\begin{eqnarray}
\label{eq:Bal12}
{\partial \over \partial t} G^r_{\lambda \mu}(t) = -i \delta(t)
\delta_{\lambda \mu} -\theta(t) \langle \{ [c_\lambda,H],
c_\mu^\dagger \} \rangle.
\end{eqnarray}
When $H$ is quadratic in the particle operators, one can express the
right--hand--side of this equation in terms of two-particle Green
functions. This yields a closed set of equations which is readily
solved for the exact Green functions. Since the quantum dot
Hamiltonian $H$ is not quadratic, one generates higher--order Green
functions which must be approximated to obtain a closed set of
equations for $G^r_{\lambda \mu}(t)$. From the solution one finds
the Green function $G^r_{\lambda \mu}(E)$ upon Fourier
transformation. It is illustrative to study $G^r_{\lambda \mu}(E)$
in various limiting cases.

{\bf Isolated dot:} This case is obtained for zero coupling
$V^{L,R}_\lambda =0$. Since no electrons can tunnel into or out of
the quantum dot, the total number of dot electrons $N$ is conserved.
The Fock states of the dot can be separated into classes with $N$
electrons. The grand--canonical expectation value of an operator
$\hat{A}$ can then be expressed in terms of canonical expectation
values
\begin{eqnarray}
\label{eq:Bal13}
\langle \hat{A} \rangle = 
\sum_{N=0}^\infty P_N \langle \hat{A} \rangle_N ,
\end{eqnarray}
where the equilibrium probability $P_N$ to find $N$ electrons on the
dot is given by
\begin{eqnarray}
\label{eq:Bal14}
P_N =  { {\rm tr}_N \exp(-\beta H) \over \sum_{M=0}^\infty {\rm tr}_M
\exp(-\beta H) } .
\end{eqnarray}
The retarded Green function follows upon substitution of $\hat{A} =
-i \theta(t) c_\lambda(t) c_\mu^\dagger (0)$ and $\hat{A} =-i
\theta(t) c_\mu^\dagger (0) c_\lambda(t)$. Fourier transformation
yields the result
\begin{eqnarray} 
\label{eq:Bal15}
  G^r_{\lambda \mu}(E)& = &  \sum_{N=0}^{\infty} 
  P_N \delta_{\lambda \mu} \left[ \frac{1-\langle 
 \hat{n}_\lambda 
    \rangle_N}{E-({\cal E}_\lambda+N \cdot U_0)+i 
   \delta } \right. \nonumber \\ 
 & & \left. \hspace*{1.0cm} + \frac{\langle \hat{n}_\lambda 
\rangle_N}{E-({\cal 
E}_\lambda+(N-1) \cdot U_0 
    +i\delta }\right] \ , 
\end{eqnarray} 
where $\delta \to 0^+$. Note that each level $\lambda$ contributes
twice: The first term on the right--hand--side of
Eq.~(\ref{eq:Bal15}) accounts for {\em electron} propagation through
the level $\lambda$ while the second term describes the propagation
of a {\em hole}. The particle--hole structure originates from the
charging energy which generally assigns the addition or removal of
electrons a different energy. We note that the particle--hole
structure disappears for $U_0 \to 0$.

{\bf Noninteracting electrons:} The limit $U_0 \to 0$ corresponds to
{\em noninteracting electrons}. In this case the equations of motion
close and one readily finds
\begin{eqnarray}
\label{eq:Bal16}
G^r_{\lambda \mu} = [ E -{\cal E} + \Sigma(E)
 ]^{-1}_{\lambda \mu} .
\end{eqnarray}
The self--energy $\Sigma(E) = \Sigma^L(E) + \Sigma^R(E)$ has two
contributions resulting from the coupling to the left and right
lead. They are given by
\begin{eqnarray}
\label{eq:Bal17}
\Sigma^{L,R}_{\lambda \mu}(E) = \sum_k {V^{L,R*}_{k \lambda} 
V^{L,R}_{k \mu}
  \over E-\epsilon_k^{L,R} + i \delta} .
\end{eqnarray}
Since the states in the leads are dense, one can replace the sum in
Eq.~(\ref{eq:Bal17}) by an integral. The imaginary part of this
integral yields the widths $\Gamma_\lambda^{L,R}$ for decay of the
state $\lambda$ into the left and right lead, while the real part
gives an energy shift. For weak coupling $\Gamma^{L,R}_\lambda \ll
\Delta$ and close to the resonances, the Green function can be
approximated by the diagonal terms $G^r_{\lambda \mu} =
\delta_{\lambda \mu} G^r_{\lambda \mu}$ yielding
\begin{eqnarray} 
\label{eq:Bal19}
G^r_{\lambda \mu}(E) = \delta_{\lambda \mu} {1 \over E-{\cal
    E}_\lambda + i \Gamma_\lambda /2} ,
\end{eqnarray}
with the total width $\Gamma_\lambda=\Gamma_\lambda^L+
\Gamma_\lambda^R$. In Sec.\ \ref{Sec4.2} we arrived at the same
result using single--particle scattering theory.

\subsubsection{Weak--coupling limit}
\label{Sec5.0.3}
The weak--coupling regime is characterized by $\Gamma_\lambda \ll
kT, \Delta$. In this regime the Green function can be approximated
by
\begin{eqnarray} 
\label{eq:Bal20}
  G^r_{\lambda \mu}(E)&\approx & 
\sum_{N=0}^{\infty} P_N \delta_{\lambda \mu}
\left[ \frac{1-\langle \hat{n}_\lambda 
    \rangle_N}{E-({\cal E}_\lambda+N \cdot U_0)+i 
   \Gamma_\lambda/2} \right. \nonumber \\ 
 & & \left. \hspace*{1.0cm} + \frac{\langle \hat{n}_\lambda 
\rangle_N}{E-({\cal 
E}_\lambda+(N-1) \cdot U_0 
    +i\Gamma_\lambda /2}\right] \ . 
\end{eqnarray} 
Note that due to the coupling all quantum dot states acquire a
finite width. The widths are identical to the widths obtained in the
non--interacting case.  Corrections arise for $\Gamma_\lambda \geq
kT$ and lead to a normalization of the self--energy. In particular,
the approximation (\ref{eq:Bal20}) fails in the Kondo regime where
higher iterations of the equations of motion become important.

In the weak--coupling limit both the transmission amplitude and the
transmission coefficient can be obtained analytically.  We first
substitute the weak--coupling result for the Green function
(\ref{eq:Bal20}) into Eq.\ (\ref{eq:Bal1a}). The transmission
amplitude (\ref{eq:tphase}) and the transmission coefficient
(\ref{eq:tcoeff}) then reduce to sums of integrals which can be
computed using contour integration. For the transmission amplitude,
the integrals are of the type
\begin{eqnarray}
\label{eq:Bal21}
\int dE \left( {\partial f \over \partial E} \right) \ 
{1 \over E -{\cal E} + i\Gamma/2} & = & {\beta \over 2 \pi i} 
 \psi^{(1)} \left( {1 \over 2}
  + {\beta \Gamma \over 4 \pi} + i {\beta {\cal E} \over 2
    \pi} \right) , 
\end{eqnarray}
and for the transmission coefficient they have the form
\begin{eqnarray}
\label{eq:Bal22}
\int dE \left( {\partial f \over \partial E} \right) {\rm Im} \ 
{1 \over E -{\cal E} + i\Gamma/2} & = & - \pi {\partial \over \partial
  {\cal E}} {\rm Re} \ f({\cal E} + i \Gamma/2) \nonumber \\
& &  -{\beta \over (2
  \pi)^2} \sum_{\sigma = \pm} {\rm Re} \psi^{(1)} \left( {1 \over 2}
  + {\sigma \beta \Gamma \over 4 \pi} + i {\beta {\cal E} \over 2
    \pi} \right) .
\end{eqnarray}
Here $\beta=1/kT$ is the inverse temperature, $f=1/(1+\exp(\beta
E))$ is the Fermi function, and $\psi^{(1)} (z)$ the trigamma
function \cite{Abr72}. Combining results, the transmission amplitude
reduces to a sum over energy levels and occupation numbers,
\begin{eqnarray}
\label{eq:Bal23}
t_{\rm QD} & = & {\beta \over 2 \pi i} \sum_\lambda \sum_{N=0}^\infty
V^L_\lambda V^{R*}_\lambda P_N 
  \left[ [1-\langle \hat{n}_\lambda \rangle ] \psi^{(1)} 
\left( {1 \over 2}  + {\beta \Gamma_\lambda \over 4 \pi} + i 
{\beta ( {\cal E}_\lambda  -U_0 \cdot N) \over 2 \pi} \right)
\nonumber \right.\\ & & \hspace*{2.8cm} \left. + \langle
\hat{n}_\lambda \rangle  \psi^{(1)}  \left( {1 \over 2}  + {\beta
\Gamma_\lambda \over 4 \pi} + i  {\beta ( {\cal E}_\lambda -U_0 \cdot
(N-1)) \over 2 \pi} \right) \right] . 
\end{eqnarray}
The transmission coefficient $T_{\rm QD}$ and the conductance
$G_{\rm QD} = (2 e^2 / h) T_{\rm QD}$ reduce to similar sum. The
result takes a simple form in two limiting cases.  First, in the
cotunneling regime deep in the transmission valley $N$ we have $P_M
\simeq \delta_{M,N}$.  The first term in Eq.\ (\ref{eq:Bal22})
vanishes exponentially, while the trigamma function can be
approximated by its asymptotic expansion \cite{Abr72}. The
conductance then simplifies to
\begin{eqnarray}
\label{eq:Bal24}
G_{\rm QD} = \simeq 2 {e^2 \over h} \sum_\lambda 
\Gamma^L_\lambda
\Gamma^R_\lambda \left[ {\langle \hat{n}_\lambda \rangle_N \over
    ({\cal E}_\lambda + U_0 \cdot (N_1))^2} + {1 - \langle 
\hat{n}_\lambda \rangle_N \over
({\cal E}_\lambda + U_0 \cdot (N))^2} \right].  
\end{eqnarray}
This is the standard result \cite{Ave90} for the conductance in the
cotunneling regime. Second, near the transmission resonances, we can
neglect the terms involving the trigamma function which are smaller
than the Fermi function by a factor $\beta \Gamma_\lambda$. After a
little algebra one finds  
\begin{eqnarray}
\label{eq:Bal25}
G_{\rm QD} \simeq {e^2 \over kT} \sum_\lambda \sum_{N=0}^\infty 
{ \Gamma^L_\lambda \Gamma^R_\lambda \over \Gamma^L_\lambda +
  \Gamma^R_\lambda} P_N [ 1 - \langle \hat{n}_\lambda \rangle_N]
f( {\cal E}_\lambda + U_0 \cdot N ) .  
\end{eqnarray}
This is the well-known result for the conductance in the regime of
sequential tunneling that was obtained independently by Beenakker
\cite{Bee91} using a master equation approach and by Meir, Wingreen,
and Lee \cite{Mei91} using the generalized Landauer formula.

The results derived in this subsection solve the problem of electron
transmission through a weakly coupled dot. They express the
transmission phase in terms of the coupling matrix elements and the
single--particle energies of the quantum dot. These quantities must
be obtained from a dynamical model of the dot and its coupling to
the leads. Various such models are studied below.  Deviations from
the weak--coupling regime are addressed in Secs.~\ref{Sec5.3.4},
\ref{Sec5.3.5}.

\subsection{Friedel sum rule} 
\label{Sec5.1} 
Levy Yeyati and B\"uttiker \cite{Lev95} were the first authors to 
study the origin of the in--phase resonances.  They discussed the 
subject in terms of the Friedel sum rule \cite{Lan61,Har79}. 
The Friedel sum rule is a powerful relation that links the number 
of localized electron states in a conducting region to the sum of 
all scattering phase shifts at the Fermi energy. The sum rule holds 
even in the presence of interactions \cite{Lan61}. There is an 
analogue of the Friedel sum rule in scattering theory. Known as 
Levinson's theorem \cite{New82}, it relates the number of bound 
states in a scattering region to the sum of the scattering phase 
shifts at zero energy. Employing the Friedel sum rule, Levy Yeyati 
and B\"uttiker argued that the addition of a charge $\Delta Q$ to a 
conducting region $\Omega$ should change the phase of the 
transmission amplitude $t$ through that region according to 
\begin{equation}  
\Delta Q /e = \Delta {\rm arg}(t)/\pi .   
\label{5.01} 
\end{equation} 
If the region $\Omega$ is chosen to include only the quantum dot,  
then the addition of one electron to the dot should cause a phase 
increase by $\pi$. Consequently, successive resonances would be 
off--phase by $\pi$ rather that in--phase as in the experiments 
\cite{Yac95,Sch97}. Levy Yeyati and B\"uttiker emphasized, however, 
that the charge $\Delta Q$ entering in Eq.~(\ref{5.01}) is not the 
additional charge on the quantum dot but rather the additional charge 
within the whole coherence volume. This volume includes the quantum 
dot and the AB ring around the quantum dot. A variation of the gate 
voltage at the quantum dot may also modify the electrostatic 
potential in the vicinity of the dot. This could result in the 
addition of extra charge to the AB ring outside the quantum dot. Levy 
Yeyati and B\"uttiker studied a situation where with each electron 
added to the quantum dot, the extra charge $\delta e$ is added to the 
AB ring. In a model calculation for a two--terminal device using the 
values $\delta \sim 0.3$ and $\delta \sim 1$, sequences of two or 
three consecutive in--phase resonances were found. 
 
The connection between the Friedel sum rule and in--phase resonances
was reconsidered and clarified by Lee \cite{Lee99} and Taniguchi and
B\"uttiker \cite{Tan99}. The Friedel sum rule makes a statement
about {\em the sum of all scattering phases}. The sum rule can be
stated in the form \cite{Lan61}
\begin{equation} 
\Delta Q /e = [\Delta \ln {\rm Det} (S)]/2\pi i ,   
\label{5.02} 
\end{equation} 
where $S$ is the scattering matrix. Writing the eigenvalues of 
$S$ in the form $e^{2 i \alpha_i}$, the right hand side represents 
the change in the sum $\alpha  / \pi = \sum_{i=1}^N \alpha_i /  
\pi$ of all eigenphases, where $N$ is the dimension of $S$. Thus,  
\begin{equation} \Delta Q /e = \Delta \alpha /\pi . 
\label{5.02a} 
\end{equation} 
The transmission amplitude $t$, on the other hand, is a matrix 
element of $S$ in a basis of left and right moving scattering 
states. In general, the phase of $t$ is not simply related to the 
sum of the eigenphases $\Delta \alpha \neq \Delta {\rm arg}(t)$. 
The relation (\ref{5.01}) is thus different from the Friedel sum 
rule (\ref{5.02}), (\ref{5.02a}). Generally, the expression 
(\ref{5.01}) is not expected to be correct. 
 
\begin{figure} 
\hspace*{0cm} 
\centerline{\psfig{file=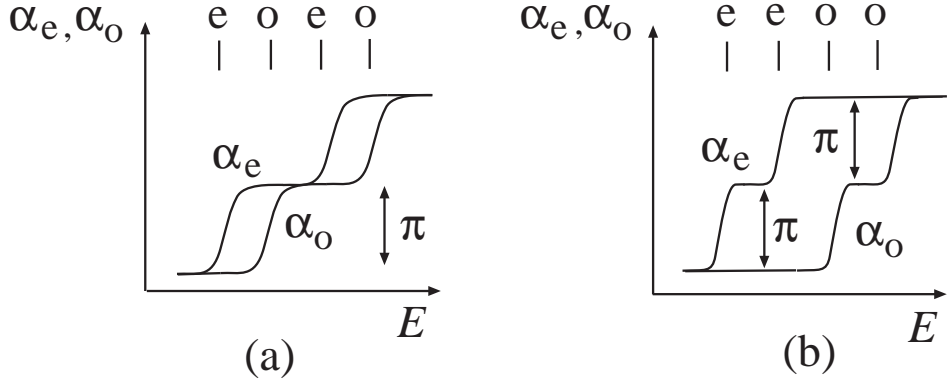,width=12.5cm,height=5.0cm,angle=0}} 
\vspace*{-0.4cm} 
\caption[]{(a) In strictly 1D systems, even and odd resonant levels 
  alternate in energy. (b) In quasi-1D or higher dimensional 
  systems, they do not necessarily alternate, allowing for sequences 
  of resonances with the same parity.} \label{lee} 
\end{figure} 
 
The difference between Eqs.~(\ref{5.01}) and (\ref{5.02a}) can be
explicitely demonstrated in the scattering through a quasi-1D system
\cite{Lee99}. We assume a mirror reflection symmetry $x \to -x$ and
the absence of a magnetic field. The scattering region at $|x|<R$ is
connected to two single--channel leads at $x<-R$ and $x>R$,
respectively. The scattering states can be decomposed into even and
odd scattering states. For $|x|>R$ these states read $\psi_e (x)
=e^{-ik|x|}+e^{2i \alpha_e} e^{ik|x|}$ and $\psi_o (x) = {\rm
  sgn}(x) [e^{-ik|x|}+e^{2i \alpha_o} e^{ik|x|}]$.  Note that there
are two scattering phases $\alpha_e$, $\alpha_o$ corresponding to
two scattering channels, one in each lead. Rather than in a basis of
even and odd scattering states, one can formulate the scattering
problem in a basis of right and left moving scattering states,
$\psi_l(x) = [\psi_e (x) - \psi_o (x)]/2$, $\psi_o(x) = [\psi_e (x)
+ \psi_o (x)]/2$. The scattering matrix in the new basis takes the
form
\begin{equation} S = \left( 
    \begin{array}{cc} 
      r & t^\prime \\ 
      t & r^\prime 
                   \end{array} \right) , 
\label{5.04} 
\end{equation} 
where $t$, $t^\prime$ are the transmission amplitudes and $r$, 
$r^\prime$ the reflection amplitudes from the left and right, 
respectively. From the relations between the basis vectors and 
Eq.~(\ref{5.04}), one obtains the transmission amplitudes 
\begin{equation} 
t = t^\prime = i e^{i \alpha} \sin \beta 
\label{5.05} 
\end{equation} 
in terms of the phases $\alpha \equiv \alpha_e + \alpha_o$ and
$\beta \equiv \alpha_e - \alpha_o$. In true 1D systems, even and odd
resonant states alternate in energy (see Fig.~\ref{lee}(a)) and the
angle $\beta$ is restricted to the range $0<\beta<\pi$. Then both
the total phase $\alpha$ and the transmission phase ${\rm arg} (t)$
change by $\pi$ between neighboring resonances. In quasi-1D systems,
however, even and odd states do not necessarily alternate in energy,
and there may be sequences of states with the same parity (see
Fig.~\ref{lee}(b)). Within such sequences, $\beta$ varies by more
than $\pi$ and $\sin \beta$ changes sign from one resonance to the
next. Neighboring resonances then have the same transmission phase
${\rm arg}(t)$ while the total phase $\alpha$ changes by $\pi$ in
keeping with the Friedel sum rule (\ref{5.02}), (\ref{5.02a}).  Note
that Eq.~(\ref{5.05}) predicts a transmission zero ($\sin \beta =0$)
between resonances with the same parity.  A transmission zero
corresponds to a singular point of the transmission phase. The phase
jumps abruptly by $\pi$ when the system is swept through the
transmission zero.
 
The arguments presented above demonstrate that in--phase resonances
do not violate the Friedel sum rule. They also show that the spatial
dimension of the scattering region is important: While neighboring
resonances are always out--off--phase in strictly 1D systems, both
out--off--phase and in--phase resonances can be found in higher
dimensional systems.

\subsection{Phase lapse} 
\label{Sec5.2} 
 
\subsubsection{Breit--Wigner model} 
\label{5.2.1} 
The sharp phase lapse between resonances was investigated by
Hackenbroich and Weidenm\"uller \cite{Hac97a} for the weak--coupling
regime $\Gamma < kT < \Delta$.  Similar to the calculation for the
two--terminal case, cf.\ Sec.~\ref{Sec4.2}, the multi--terminal
transmission amplitude was obtained using scattering theory and the
diagonal approximation for the dot propagator. In this approximation
the transmission amplitude reduces to a sum of Breit--Wigner
resonances. The internal dynamics of the dot and the coupling to the
leads enter via the resonance energies $E_\lambda$ and the partial
width amplitudes $V^{pm}_ \lambda$ for the decay of the resonance
$\lambda$ into channel $m$ of the lead $p$. Hackenbroich and
Weidenm\"uller did not calculate the resonance parameters from a
dynamical model. They rather {\em assumed} these parameters to be
identical for all resonances and investigated the consequences of
this assumption for the transmission phase: By assumption, all
resonances showed in--phase behavior. At each resonance the phase
increased by $\pi$ on a scale $\sim kT$. Between the resonances, the
phase sliped by $\pi$ on the scale of the intrinsic resonance width
$\Gamma$. The Breit--Wigner model \cite{Hac97a} thus predicts a
phase lapse between resonance peaks on the scale of the intrinsic
resonance width $\Gamma$.
 
We note that the experiment \cite{Sch97} was {\em not} performed in
the weak--coupling regime. Rather, the intrinsic resonance width
$\Gamma$ in the experiment was estimated to be of the order $\Gamma
\sim (3-4) kT$. The Breit--Wigner model of Hackenbroich and
Weidenm\"uller therefore cannot explain the phase lapse observed in
\cite{Sch97}. Moreover, there is theoretical evidence (see Sec.\ 
\ref{Sec5.2.2}) that the diagonal approximation for the quantum dot
propagator fails between the resonances. As a result, the
transmission phase may display an sharp lapse rather than a smooth
decrease on the scale $\Gamma$ as predicted by the Breit--Wigner model.
 
\subsubsection{Transmission zero} 
\label{Sec5.2.2} 
In 1998, a number of authors \cite{Xu98,Deo98,Ryu98,Sun98} studied
models for the transmission through a quantum dot and found abrupt
jumps of the transmission phase. The jumps occured at singular
points where both the real and the imaginary part of the complex
transmission amplitude vanished. Transmission zeros were found in
numerical studies exploring Fano resonances \cite{Xu98,Deo98,Ryu98}
and the interplay of multiple resonances \cite{Sun98}. In Refs.\ 
\cite{Xu98,Deo98,Ryu98} the quantum dot was modeled as a quasi-1D or
2D region of regular shape. In all studies time--reversal invariance
was assumed, the dot was connected to two single--channel leads and
electron-electron interactions were neglected.  In--phase resonances
occurred either due to sequences of states with the same parity
\cite{Xu98,Deo98,Ryu98} or by assuming \cite{Sun98}
state--independent coupling matrix elements to the leads. A more
general approach to the connection between transmission zeros and
in--phase resonances was presented by Lee \cite{Lee99} (see the
discussion after Eq.~(\ref{5.05})). He showed that the transmission
always vanishes between neighboring in--phase resonances, provided
(i) the scattering region is connected to two single--channel leads
and (ii) the system is time--reversal invariant.

Why do the models \cite{Xu98,Deo98,Ryu98,Sun98} display an abrupt
phase jump and not a smooth phase slip on the scale $\Gamma$ as
predicted by the Breit--Wigner model \cite{Hac97a}?  This question
was addressed by Sun and Lin \cite{Sun98} for a quantum dot with
symmetric tunneling barriers. It turned out that the diagonal
approximation for the quantum dot propagator used in
Ref.~\cite{Hac97a} fails for time--reversal invariant systems
coupled to single--channel leads.  Here we present an argument that
proves this failure also for asymmetric barriers.  Consider a
quantum dot that is coupled to two single--channel leads, denoted
$L$, $R$. We assume $kT \ll \Delta$ and include electron--electron
interactions in a mean field way.  The scattering matrix of the
system can be derived with the methods of Sec.~\ref{Sec4.2}. For the
scattering amplitude $S_{mn}$ at energy $E$ one has
\begin{equation} 
S_{mn}(E)=\delta_{mn} - 2 i \pi \sum_{ \mu \nu} W_{m \mu}^* (D^{-1} 
(E))_{\mu \nu} W_{n \nu} , 
\label{5.40} 
\end{equation} 
where $m$, $n$ refer to the physical channels. The indices $\mu$ and 
$\nu$ of the inverse propagator $D_{\mu \nu}$ refer to a complete set 
of quantum dot states, and $W_{m \mu}$ is the coupling matrix element 
between channel $m$ and the dot state $\mu$. The inverse propagator 
$D$ has the form 
\begin{equation} 
D_{\mu \nu}(E) = (E-E_\mu) \delta_{\mu \nu} + i \pi \sum_m W_{m \mu} 
W_{m \nu}^* , 
\label{5.41} 
\end{equation} 
where the sum runs over $m=L,R$ and where $E_\mu$ includes an energy
shift resulting from the coupling to the leads. First consider the
system at zero magnetic field. Then the matrix elements can be
chosen real. To simplify the argument, we restrict ourselves to two
levels $1$, $2$ in the dot. The quantum dot propagator is given by
\begin{equation} 
D^{-1}(E) = {1 \over {\rm det} D(E)} \left( 
    \begin{array}{cc} 
      E-E_2+i \Gamma_{22}/2 & -i \Gamma_{12}/2 \\ 
      -i \Gamma_{21}/2 & E-E_1 + i \Gamma_{11}/2 
                   \end{array} \right) , 
\label{5.42} 
\end{equation} 
with $\Gamma_{\mu \nu} \equiv 2 \pi \sum_m W_{m \mu} W_{m \nu}^*$.
Combining Eqs.~(\ref{5.40}) and (\ref{5.42}), one obtains the
transmission amplitude $t_{LR}(E)$ between left and right lead
\begin{eqnarray}  
t_{LR}(E) & =&  -{2 i \pi \over {\rm det} D(E)} [
\gamma_{11}(E-E_2) +  \gamma_{22}(E-E_1) + \nonumber \\ 
& & {i \over 2} (\gamma_{11} \Gamma_{22} +  
\gamma_{22} \Gamma_{11} -\gamma_{12} \Gamma_{21} -\gamma_{21}  
\Gamma_{12} ) ] , 
\label{5.43} 
\end{eqnarray} 
where $\gamma_{\mu \nu} \equiv W_{L \mu}^* W_{R \nu}$. Substituting
the definitions of $\Gamma_{\mu \nu}$ and $\gamma_{\mu \nu}$ into
Eq.~(\ref{5.43}), one finds that the term in the round brackets on
the right hand side vanishes identically. Therefore, if
$\gamma_{11}$ and $\gamma_{22}$ have the same sign, the transmission
amplitude $t_{LR}$ vanishes for some energy between $E_1$ and $E_2$.
We thus obtain a transmission zero.  Note, that for the above
argument, it is crucial to keep the off-diagonal elements of the
propagator. They were neglected in Ref.~\cite{Hac97a}.
 
It turns out that both the condition of time--reversal invariance
and of single--channel leads are necessary for a transmission zero.
If one of these conditions is relaxed, one generally no longer finds
a vanishing transmission. E.g.\ consider the case that
time--reversal invariance is broken. The matrix elements $W_{m \mu}$
then can take complex values and the term in round brackets in
Eq.~(\ref{5.43}) is no longer guaranteed to vanish. As a
consequence, one generically finds a smooth phase slip and not an
abrupt phase jump between resonances.  However, the phase slip may
take place on an energy scale much smaller that the total width
$\Gamma$. This is suggested by the following observation. In the
experiment, all resonances behave similarly (they display nearly the
same phase and peak height). This suggests that different resonances
have similar coupling matrix elements. For the simple two--level dot
studied here, $\gamma_{\mu \nu} \approx \gamma$ and $\Gamma_{\mu
  \nu} \approx \Gamma$ independent of the level indices $\mu$,
$\nu$. This choice results in a near cancelation of the terms in the
round bracket in Eq.~(\ref{5.43}), and the phase slip may get very
sharp.
 
We finally discuss the effect of electron--electron interactions and
finite temperature on the transmission zeros. Lee \cite{Lee99}
argued that the analysis presented for the noninteracting case
applies equally to interacting systems provided quasiparticle
excitations remain well defined at $E=E_F$. The argument is based on
the Friedel sum rule which is known to hold even in the presence of
interactions. Unfortunately, no numerical or analytical study has
explicitly demonstrated transmission zeros in the presence of
interactions. Finite temperature smears out the transmission zeros.
The sharp phase jump is then replaced by a rapid but continuous lapse
of the phase \cite{Sun98}. The precise energy scale of this lapse
depends on the detailed electron dynamics in the dot. We note that
the phase lapse observed in Ref.~\cite{Sch97} is indeed continuous
and has been resolved experimentally, cf.\ Sec.~\ref{Sec3}.
 
\subsubsection{Disordered dot} 
\label{5.2.3} 
The dots used in the experiments \cite{Yac95,Sch97} were roughly 50
times smaller than the elastic mean free path. Under these
conditions, transport through the dots is ballistic. Most of
the theoretical studies of the transmission phase pertain to the
ballistic regime, and neglected the influence of disorder. Baltin and
Gefen \cite{Bal99b,Bal99d} took a different approach and investigated
the transmission amplitude through a {\em disordered} quantum dot.
They identified a generic mechanism for phase correlations and
formulated an approximate sum rule: According to this rule, the
change in the transmission phase $\Delta \theta$ between two
consecutive transmission valleys is $0$ (mod $2 \pi$).  The frequency
of deviations from this rule is small in $\Delta / U_0$ where
$\Delta$ is the mean single particle level spacing and $U_0$ the
charging energy of the quantum dot. The sum rule pertains to
individual, disorder specific systems. Baltin and Gefen also
calculated a disorder--averaged phase--phase correlation function and
observed an enhancement of correlations with increasing interaction
strength. Below we present the main arguments of the study
\cite{Bal99b}. 

The analysis of Baltin and Gefen proceeds from the scattering theory
in the weak--coupling limit as presented in Sec.~\ref{Sec5.0.3}.
However, the basic argument can be understood from a simple toy
model \cite{Bal99b} based on two assumptions: (i) The
single--particle level spacing is constant ${\cal E}_\lambda =
\lambda \Delta$ and (ii) the product of the coupling matrix elements
is a random variable $V_\lambda^L V_\lambda^{R*} = V
\eta_\lambda$ where $\eta_\lambda$ takes the values $+1$ and $-1$
with equal probability. These simple assumptions model the
fluctuations in the wave functions due to disorder. A more elaborate
random--matrix calculation of the energies and the couplings
confirms the predictions of the toy model described below. Consider
first the noninteracting case ($U_0=0$). The transmission amplitude
then becomes
\begin{eqnarray}
t(E) = V \sum_\lambda {\eta_\lambda \over E -{\cal E}_\lambda + 
i \Gamma_\lambda},
\label{eq:Bal3}
\end{eqnarray}
where the system can be tuned in or out of resonance by the gate
voltage $V_g$ (we assume ${\cal E}_\lambda={\cal E}_\lambda^{(0)} -
e V_g$). At each resonance, the transmission phase increases by $\pi$.
The signs of $\eta_\lambda$ govern the phase evolution between the
resonances. If $\eta_\lambda \cdot \eta_{\lambda+1} >0$ there is a
phase lapse by $\pi$ between the resonances $\lambda$ and $\lambda+1$;
for $\eta_\lambda \cdot \eta_{\lambda+1} <0$ there is no phase lapse.
In strictly one--dimensional systems $\eta_\lambda$ alternate in sign
implying no phase lapse as discussed in Sec.\ \ref{Sec5.1}. For a
disordered noninteracting dot the phase lapses occur at random.

\begin{figure}
\centerline{\psfig{file=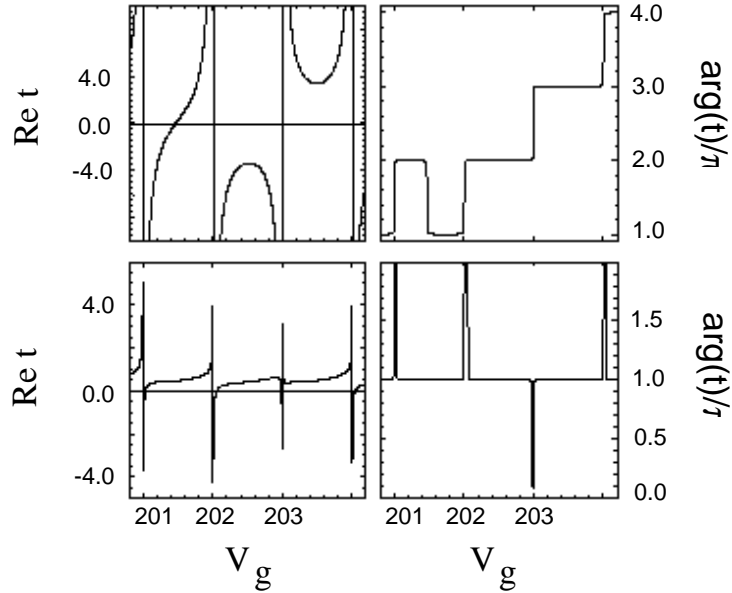,height=8cm,angle=0}}
  \caption[]{Re $t$ (left) and phase (right) for a specific sequence of
    resonances, for $U_0=0$ (upper panel) and $U_0=60\Delta$ (lower
    panel). Taken from
  Ref.~\protect{\cite{Bal99b}}.}
  \label{fig:Bal1}
\end{figure}

The situation is different in the presence of interactions. In the
conductance valleys a large number of random terms $\sim U_0 / \Delta$
contribute to $t(E)$. The terms in the $N$th valley and the $(N+1)$th
valley differ very little from each other (essentially by the
contribution of one level). The transmission in the valleys is
therefore determined by a background that varies little between
neighboring valleys. Figure \ref{fig:Bal1} shows the evolution of
${\rm Re} (t)$ and ${\rm arg}(t)$ for for $kT = \Delta/12$ and
specific series of couplings, both for the noninteracting case $U_0=0$
(upper panel) and $U_0 =60 \Delta$ (lower panel). For interacting
electrons the change in phase $\Delta {\rm arg}(t) =0$ as expected
from the sum rule. Note, however, that the phase evolution as a
function of voltage is different from what has been seen
experimentally: The phase in the disordered dot decreases near the
resonances and not in the valleys between resonances.

\begin{figure}
\centerline{\psfig{file=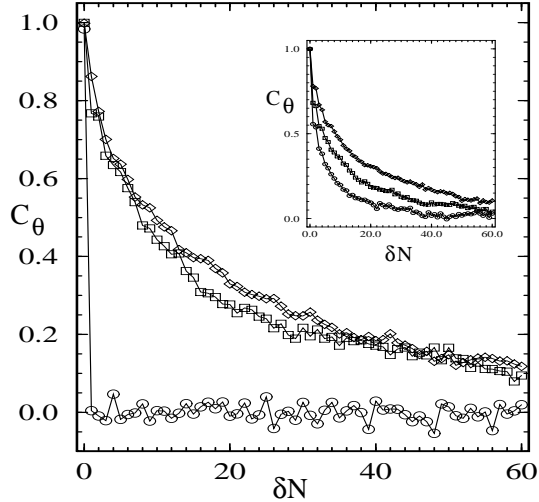,height=8cm,angle=0}}
  \caption[]{Correlation function $C_{\theta}$ vs.\ valley number for
    $kT=\Delta/12$. Interacting electrons (boxes and diamonds,
    $U_0=50\Delta$, $x=\bar{x}=0.5$) show strong correlations,
    non-interacting electrons (circles, $x=\bar{x}=0.25$) do not.
    Toy model statistics (boxes) agrees well with random matrix
    model statistics (diamonds).  Inset: Increasing the interaction
    increases the correlations (upper curve: $U_0=50\Delta$, middle:
    $U_0=25\Delta$, lower: $U_0=12\Delta$). Taken from
  Ref.~\protect{\cite{Bal99b}}.}
  \label{fig:Bal2}
\end{figure}

The crucial influence of the interaction is further illustrated by
the disorder--averaged phase correlation function
\begin{eqnarray}
C_\theta \equiv \langle \cos \theta(x,N) \cos
\theta(\bar{x},N+\delta N) \rangle .
\label{eq:Bal4}
\end{eqnarray}
Here, $x$ and $\bar{x}$ with $0 \leq x,\bar{x} \leq 1$ specify the
voltage in the $N$--th and $(N+\delta N)$--th valley. $C_\theta$
decays slowly on a scale $\delta N \sim U_0/\Delta$. Figure
\ref{fig:Bal2} shows $C_\theta$ vs.\ $\delta N$ for $kT = \Delta
/12$. While there are no correlations for noninteracting electrons,
one observes a slow decay of $C_\theta$ in the presence of
interactions. The decay is slower for stronger interactions as
expected from the arguments given above.

The sum rule found for disordered dots is in accord with the
observations made in the experiment \cite{Sch97}, however, the
evolution of the phase vs.\ gate voltage is in disagreement. We
believe that the latter fact indicates a fundamental difference
between the phase behavior found in disordered (or fully chaotic)
systems and the phase behavior in ballistic dots with mixed
dynamics.

\subsection{In--phase resonances} 
\label{Sec5.3} 
Several mechanisms for peak--correlations and in--phase resonances
have been proposed. These include the effect of finite temperature,
deformations of the dot confining potential, and energy shifts
caused by the coupling between the dot and the leads. We summarize
these mechanisms in Secs.\ \ref{Sec5.3.1}-\ref{Sec5.3.4}. The
mechanisms rely on the assumption that the transmission is dominated
by a few quantum states of the dot.  The {\em dynamical origin} for
this behavior is discussed in Secs.\ \ref{Sec5.3.3}-\ref{Sec5.3.5}.
The investigation suggests that the in--phase resonances result from
features of the short--time dynamics in the dot that completely
dominate the quantum transport in the tunneling regime. Alternative
proposals for peak correlations are reviewed in Sec.~\ref{Sec5.3.6}.

\subsubsection{Finite temperature} 
\label{Sec5.3.1} 
Transport at finite temperatures differs qualitatively from
transport at zero temperature. Close to a conductance peak electrons
tunnel through a single level if $kT \ll \Delta$. In contrast, few
or many levels contribute to each conductance peak if $kT$ is of the
order $\Delta$. Similar sets of levels contribute to neighboring
peaks which generates peak correlations.  Oreg and Gefen
\cite{Ore97} studied the transmission phase correlations resulting
from this effect. More recently, temperature induced correlations of
the conductance peak--height in fully chaotic or diffusive dots have
been calculated \cite{Alh98} using random matrix theory. A
comparison with the peak--height correlations measured for chaotic
dots has been given in Ref.~\cite{Pat98}.

Oreg and Gefen demonstrated in--phase behavior for a simple model
with two levels in the dot. The calculation is an application of the
formalism developed in Sec.~\ref{Sec5.0}. Due to the small number
of levels, all contributions to the transmission amplitude can be
computed explicitly. The starting point is the two--level
Hamiltonian for the quantum dot,
\begin{equation} 
H_{\rm QD} = {\cal E}_a a^\dagger a + {\cal E}_b b^\dagger b + U_0 
a^\dagger a b^\dagger b , 
\label{HD} 
\end{equation} 
where $U_0$ denotes the strength of the electron--electron
interaction.  The dot is coupled via matrix elements $V^L_{a}$,
$V^R_{a}$, $V^L_{b}$, $V^R_{b}$ to two single--channel leads $L$ and
$R$. As in Sec.~\ref{Sec5.0} all energies are counted from the Fermi
energy in the leads. To obtain the transmission amplitude
(\ref{eq:Bal20}) through the dot, we compute the occupation
probabilities of the dot states.  The quantum dot supports four
many--body states. They can be labeled $|0 \rangle$ (no electron in
the dot), $|a \rangle = a^\dagger |0 \rangle$, $|b \rangle =
b^\dagger |0 \rangle$, and $|ab \rangle = a^\dagger b^\dagger |0
\rangle$. The respective energies are $E_0 =0$, $E_a = {\cal E}_a$,
$E_b = {\cal E}_b$, and $E_{ab}= {\cal E}_a + {\cal E}_b +U_0$. The
equilibrium probabilities to find the dot in either of the states
$|i \rangle = |0 \rangle,|a \rangle,|b \rangle,|ab \rangle$ are
given by
\begin{equation} 
P_i = \exp(-\beta E_i) / \sum_i \exp(-\beta E_i). 
\label{Pi} 
\end{equation} 
In terms of the $P_i$, the matrix elements of the retarded Green
function are
\begin{eqnarray}  
{G}_{aa} (E) & = & (P_0 +P_a) {1 \over E 
  -{\cal E}_a + i\Gamma_a /2} \nonumber \\ 
& & +(P_b +P_{ab}) { 1 \over E -({\cal E}_a  +U_0) + i\Gamma_a / 2} , 
\label{Ga} \\ 
{G}_{bb} (E) & = & (P_0 +P_b) {1 \over E 
  -{\cal E}_b + i \Gamma_b /2} \nonumber \\ 
& & +(P_a +P_{ab}) { 1 \over E -({\cal E}_b +U_0) + i \Gamma_b /2}, 
\label{Gb} 
\end{eqnarray} 
where $\Gamma_{a,b}=|V^L_{a,b}|^2+|V^R_{a,b}|^2$. Substitution of
Eqs.~(\ref{Ga}), (\ref{Gb}) into Eq.\ (\ref{eq:Bal23}) yields the
transmission amplitude $t_{\rm QD}$.
 
Two consequtive resonances with similar phase can arise in this
model if one level is significantly stronger coupled to the leads
than the other level. For definiteness, let $a$ be the strongly
coupled level.  At the first resonance, either an electron or a hole
tunnel via level $a$ through the dot while level $b$ is empty. These
processes are represented by the first term on the right hand side
of Eq.~(\ref{Ga}). At the second resonance, again charge is
transfered predominantly via level $a$, this time with level $b$
being occupied. The corresponding processes are described by the
second term on the right hand side of Eq.~(\ref{Ga}). Two
consecutive resonances are then dominated by the same
single--particle level $a$.  This scenario differs qualitatively
from the transmission of independent electrons (which is recovered
in the limit $U_0 \to 0$), where two consecutive resonances are
associated with two different single--particle states.

\begin{figure} 
\centerline{ \hspace*{4cm} \psfig{file=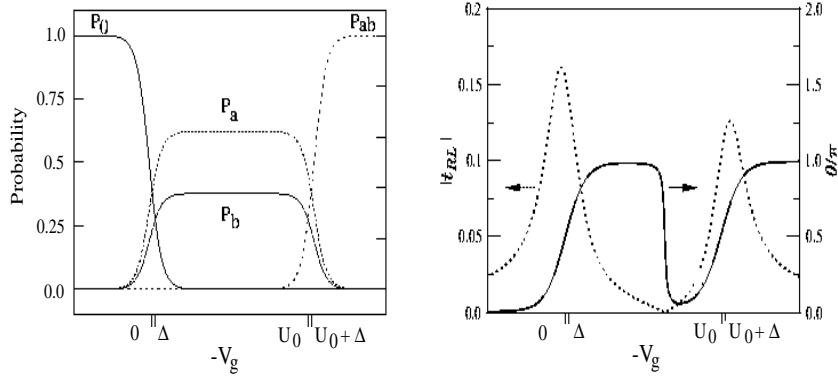,width=16cm,angle=-0.0}}
\vspace*{-3cm}
\caption[]{Left: Occupation probability of the four many--body states 
  as a function of $V_g$. Right: Argument of $t_{RL}$ (solid line) 
  and its magnitude (dashed line) as a function of $V_g$. For the 
  parameters $\Delta = 10 \mu eV$, $kT = 20 \mu eV$, $U_0=500 \mu eV$. 
  Taken from Ref.~\protect{\cite{Ore97}}.} \label{orgef} 
\end{figure} 
 
We use the parameterization ${\cal E}_a  =  V_g$, ${\cal E}_b =  V_g
+\Delta$ of the single--particle energies in terms of the voltage $V_g$
at the dot. The occupation probabilities of the four many body states and
the modulus and phase of $t_{\rm  QD}$ are displayed vs.\ $V_g$ in
Fig.~\ref{orgef}. For both plots, $\Delta =0.5 kT$, and $|V^L_{b}|
=|V^R_{b}|=|0.3 V^L_{a}|=|0.3 V^R_{a}|$. The two resonances are in phase
since both are dominated by the same level $a$. The phase lapse between
resonances takes place on the scale $\Gamma_a$. This is expected since we
neglected the off--diagonal elements of the dot Green function, cf.\
Sec.~\ref{Sec5.2}.
 
\subsubsection{Deformations}  
\label{Sec5.3.2}  
Transport experiments in the Coulomb blockade regime study the
transmission through quantum dots as a function of a gate voltage
$V_g$ applied to the dot. Most theoretical studies of the Coulomb
blockade assumed that $V_g$ exclusively regulates the depth of
potential well in the dot.  Changes in $V_g$ then translate into an
overall shift of the single--particle levels of the dot. This
picture is clearly an oversimplification. In any experimental
realization of a quantum dot, variation of $V_g$ modifies both the
depth and the {\em shape} of the dot confining potential. The
consequences for the Coulomb blockade were first investigated by
Hackenbroich, Heiss, and Weidenm\"uller \cite{Hac97b}. Shape
variations lead to avoided crossings of single--particle levels.
This can generate sequences of conductance resonances which carry
essentially the {\it same} internal wave function. Within such
sequences one may find in--phase resonances and/or strong
correlations of the conductance peak heights. Vallejos, Lewenkopf,
and Mucciolo \cite{Val98} showed that shape deformations may change
the spacing distribution of Coulomb blockade peaks. Stopa
\cite{Stopa98} demonstrated shape deformations in self--consistent
calculations of the dot confining potential. We discuss shape
deformations for a quantum dot with a parabolic confining potential.
Then we consider more realistic confinement potentials with level
repulsion and chaotic classical electron motion.

\begin{figure} 
\vspace*{0.2cm} 
\centerline{\psfig{file=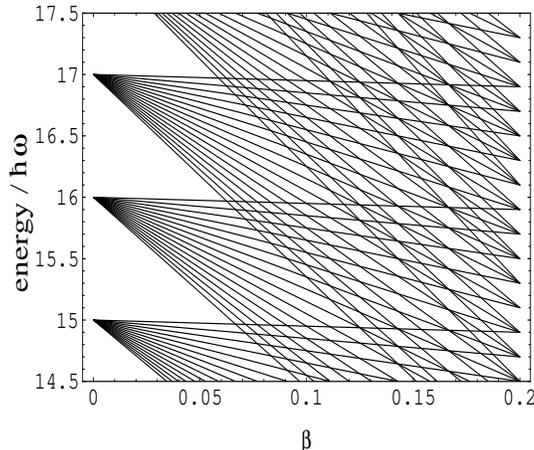,width=7cm,height=6cm,angle=0}} 
\caption[]{ 
  Energy levels of the two-dimensional harmonic oscillator with 
  frequencies $\omega_x=\omega$ and $\omega_y=\omega(1-\beta)$ as 
  functions of $\beta$. Taken from Ref.~\protect{\cite{Hac97b}}.} 
\label{def1} 
\end{figure} 
 
Consider a quantum dot with a parabolic confining potential. The
potential is characterized by two oscillator frequencies $\omega_x$
and $\omega_y$. For simplicity, we keep $\omega_x = \omega$ fixed
while $\omega_y = \omega (1 - \beta)$ depends on the deformation
parameter $\beta$ with $0 \leq \beta < 1$.  Fig.~\ref{def1} shows
part of the single-particle spectrum versus $\beta $. The levels form
a network of intersecting straight lines.  The single--particle state
associate with each line is characterized by two non--negative
integer quantum numbers ($\lambda_x,\lambda_y$). States with small
(large) values of $\lambda_y$ have small (large) negative slopes. We
show below that a pattern similar to Fig.~\ref{def1} may also be
found for the spectrum of deformed non-integrable potentials, with
one difference: The points of intersection disappear and are replaced
by avoided crossings.
 
\begin{figure} 
\centerline{\psfig{file=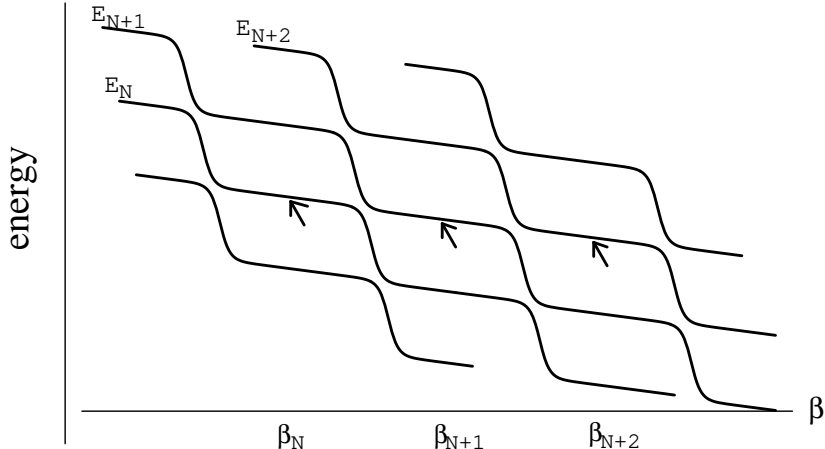,height=6cm,angle=0}} 
\caption[]{ Schematic illustration of the mechanism described in the 
text. At the deformations $\beta_N$, $\beta_{N+1}$, $\beta_{N+2}$ one 
additional level (indicated by the arrows) is occupied. Taken from 
  Ref.~\protect{\cite{Hac97b}}.} 
\label{def2} 
\end{figure}

To illustrate the mechanism for peak correlations, we replace
Fig.~\ref{def1} by the idealized picture shown in Fig.~\ref{def2}: A
set of equally spaced straight lines ($A$-levels) runs nearly
parallel to the $\beta$-axis, a second set ($B$-levels) has large
negative slope. To model the generic case, actual crossings have
been replaced by avoided crossings.  The distance between $B$-lines
is denoted by $\delta \beta_{\rm cross}$.  Wave functions retain
their identity across avoided crossings. Therefore, the wave
function on any $A$-level is nearly independent of $\beta$. A change
of the gate potential and, therefore, of $\beta$ is slow on the
scale of the characteristic times of the quantum dot. Therefore,
each electron in an occupied level follows the deformation $\beta$
adiabatically. At each avoided crossing, the electron wave function
changes from A--type to B--type or vice versa.
 
\vspace*{0.0cm} 
 
Suppose that for some $\beta$ the $N$ lowest single-particle levels
of the dot are occupied and that the highest occupied level is an
$A$-level. What happens as $\beta$ is increased by $\delta \beta_{\rm
cross}$? The last occupied level moves adiabatically down at the
avoided crossing. The associated wave function switches from
A--type to B--type. The $A$-level that was the last occupied level
before the avoided crossing becomes empty. Upon further increase of
$\beta$, this very same A-level can be occupied by another electron
(at deformation $\beta_{N+1}$). This process can repeat itself for a
second, third, etc.\ time {\em if} $\delta \beta_{\rm cross}$
$\approx$ $\delta \beta_e$, the change of $\beta$ needed to pull
another electron into the dot. Then, subsequent conductance peaks
would not be independent, but rather be manifestations of essentially
the same single-particle state. Strong correlations of the
conductance peak heights and the transmission phases are expected
for such sequences.

How does the idealized picture of Figs.\ \ref{def1}, \ref{def2}
change for a more realistic spectrum with level repulsion? As a
generic example consider the Hamiltonian 
\begin{equation}  
\label{spham}  
H={{\bf p}^2\over 2m}+{m\omega^2\over 2} (x^2+(1-\beta 
)^2 y^2)-\rho \hbar \omega L^2 , 
\end{equation} 
where $L$ is the dimensionless $z$-component of the angular momentum
operator. The three-dimensional analogue of $H$ is known as the
Nilsson model and has been quite successful in explaining the spectra
of deformed nuclei \cite{bm}. For $\beta > 0$ and $\rho \neq 0$, $H$
is not integrable and displays level repulsion. 
 
\begin{figure} 
\centerline{\psfig{file=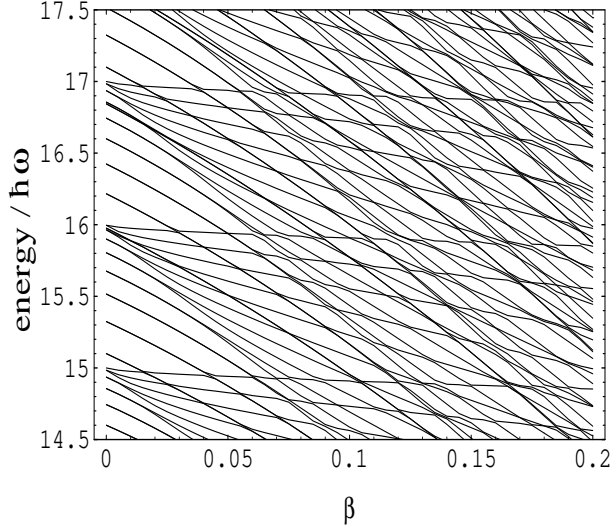,width=8cm,height=7cm,angle=0}} 
\vspace*{0.2cm} 
\caption[]{ 
  Energy levels as a function of $\beta $ of the Hamiltonian 
  (\protect\ref{spham}) for $\rho =0.004$.  Taken from 
  Ref.~\protect{\cite{Hac97b}}.} 
\label{def3} 
\end{figure} 
 
Figure \ref{def3} shows part of the spectrum of $H$ versus $\beta $.
The overall pattern is quite similar to Fig.~\ref{def1}, however all
crossings have been replaced by avoided crossings (mostly not
visible in Fig.~\ref{def3}). One observes a set of nearly flat
lines. Numerical inspection shows that the associated wave functions
retain their identity over a large range of $\beta$. The intensity
of these wave functions is concentrated along periodic orbits which
oscillate in the $x$-direction, with little or no motion in the
$y$-direction. One expects that the bulk part of the current through
the dot is carried by such orbits. The numerical observation of
stable periodic orbits in the Nilsson model of Ref.\ \cite{Hac97b}
provided the first evidence for the dynamical origin of strongly
coupled quantum states. We note that Fig.\ \ref{def3} also reveals
steep levels. The corresponding wave functions change strongly with
deformation as expected for parametric variations in chaotic
systems.

\subsubsection{Integrable dot}  
\label{Sec5.3.3} 
In the preceding sections we presented qualitative arguments for
peak correlations induced either by temperature or deformations. A
synthesis of both approaches was given by Baltin et al.\ 
\cite{Bal99a}. It was shown that a quantum dot of the shape of a
deformed harmonic oscillator can support sequences of up to 30
conductance resonances with similar transmission phase and similar
peak height.  All resonances within such a sequence are dominated by
a single strongly coupled eigenstate of the dot. The study
\cite{Bal99a} pertains to an integrable dot in the regime of
ballistic transport and is therefore restricted in generality and
universality. However, aspects of the study are also relevant for
ballistic dots of more general shape provided these dots support
short orbits that strongly couple to the leads.
 
Consider the tunneling Hamiltonian $H$ defined in Sec.\ \ref{Sec5.0}
and assume that the dot confining potential is an anisotropic
harmonic oscillator potential. The states of the dot can be labeled
by two quantum numbers $\lambda_x$, $\lambda_y$ and the energy
eigenvalues ${\cal E}_\lambda$ for $\lambda =(\lambda_x,\lambda_y)$
are given by
\begin{eqnarray} 
\label{spec} 
{\cal E}_\lambda= &=& \hbar \omega_x (\lambda_x+\frac{1}{2}) + 
\hbar \omega_y(V_g)  
(\lambda_y+\frac{1}{2}) - \alpha V_g +E_0. 
\end{eqnarray} 
To describe the deformation, we assume that the oscillator frequency
$\omega_y(V_g) = \omega_x (1-\gamma (V_g-V_0))$ in the transverse
direction $y$ depends linearly on the gate voltage $V_g$ while the
frequency $\omega_x$ in the longitudinal $x$--direction is held
fixed. The parameter $\alpha$ relates the overall depth of the dot
potential to the gate voltage. The constants $E_0$ and $V_0$ determine
the number of electrons on the dot at zero deformation.  

\begin{figure} 
\centerline{\psfig{file=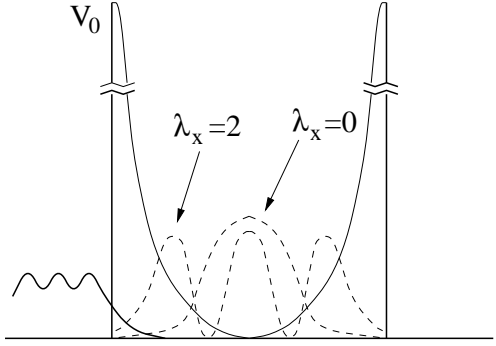,height=4.5cm,angle=270}} 
\vspace*{0.2cm} 
  \caption[]{The thin solid line shows a cross section of the 
    potential in longitudinal direction, the two barriers lying at
    opposite ends. The overlap of the dot wave functions
    (probability shown as dashed lines) with the lead wave function
    (probability shown as a solid line on the left) increases
    strongly with the quantum number $\lambda_x$.  Taken from
    Ref.~\protect{\cite{Bal99a}}.}
\label{over} 
\end{figure}

The matrix elements $V^L_{k \lambda}$, $V^R_{k \lambda}$ for
tunneling from the left and right lead to the quantum dot are given
\cite{duke} by the integrals
\begin{eqnarray} 
\label{tunnel} 
V^{L(R)}_{k,\lambda} &=& \frac{\hbar^2}{2m} \int_{B} dy \left[ 
\psi_k(x,y)^* 
  \frac{\partial \Phi_{\lambda_x,\lambda_y}(x,y)}{\partial x} - 
\Phi_{\lambda_x,\lambda_y} 
  \frac{\partial  \psi_k (x,y)^*}{\partial x} \right]_{x=x_B} \ ,  
\end{eqnarray} 
where $\psi^{L(R)}_k$ denotes the wave function with wave vector $k$
in the left (right) lead , and $\Phi_{\lambda_x,\lambda_y}$ the wave
function in the dot. The integration extends in the $y$--direction
and $x_B$ is arbitrary but must be located within the barrier.  We
restrict ourselves to the case of a single transverse channel in
each lead. The nodes of the wave functions of flat (steep) levels
with large $\lambda_x$ ($\lambda_y$) are predominantly carried by
the x--component (y--component, respectively). Thus, the wave
functions of flat levels extend much further into the barrier region
and have considerably larger matrix elements $V^{L(R)}_{k,\lambda}$
than those of the steep levels. This important property is
illustrated in Fig.~\ref{over}. Due to their stronger coupling, flat
levels will carry larger current that steep levels.
 
According to the scenario discussed in Sec.~\ref{Sec5.3.2} a flat
level of the dot gives rise to a sequence of Coulomb peaks when it
undergoes a single level crossing between every two successive
peaks.  At finite temperature, this condition need not to be met
exactly, but must hold on average for a sufficiently large number of
peaks. Within our model, we can estimate the number $\Delta N$ of
correlated resonances. For $U_0 \gg \Delta$, the distance between
neighboring resonances is $\delta V_g=U_0/\alpha$. The number of
intersection points of a flat level ($\lambda_x \neq 0 \ ,
\lambda_y=0$) with steep levels is found using Eq.~(\ref{spec}). For
the total number of crossings in the voltage range $V_g-V_0$ one
finds
\begin{eqnarray} 
N_c &=& \frac{\lambda_{x}^{2}}{2} \frac{\gamma (V_g-V_0)}{1-\gamma 
(V_g-V_0)}.  
\end{eqnarray} 
The number of steep levels around a given energy increases with with
deformation. This yields an increase of $N_c$, and causes the
divergence for the unphysical case of extreme deformation $\omega_y
\sim 1-\gamma (V_g-V_0) \to 0$. One crossing within the interval
$\delta V_g$ occurs for $(\partial N_c)/(\partial V_g) =
\alpha/U_0$. This condition yields the voltage $V^{*}_g$ where
maximal correlations of the Coulomb peaks are found; $V^{*}_g$ is
used below. The number $\Delta N$ of correlated peaks can be
estimated as the number of resonances for which the flat level
$(\lambda_x,0)$ stays within the energy interval $\Delta$ around
$E_F$. This yields
\begin{eqnarray} 
\label{deltaN} 
\Delta N &\simeq& 2 \sqrt{\lambda_x  
\sqrt{\frac{\alpha}{2\gamma U_0}}} \ , 
\end{eqnarray} 
with the deformation $\omega_y/\omega_x=\sqrt{(\gamma U_0)/(2 \alpha)}\ 
\lambda_x$. The result (\ref{deltaN}) explicitly relates the number of 
correlated levels to our model parameters. We note that $\Delta N$
sets an upper bound for the number of correlated peaks. If $V_{g}$ is
substantially different from $V^{*}_{g}$, the number of level
crossings no longer matches the number of Coulomb peaks. Then the
sequences of correlated peaks are shorter than $\Delta N$ and the
correlations are weaker.  
 
\begin{figure}
\hspace*{0.75cm} \begin{minipage}{7.5cm}
\centerline{\psfig{file=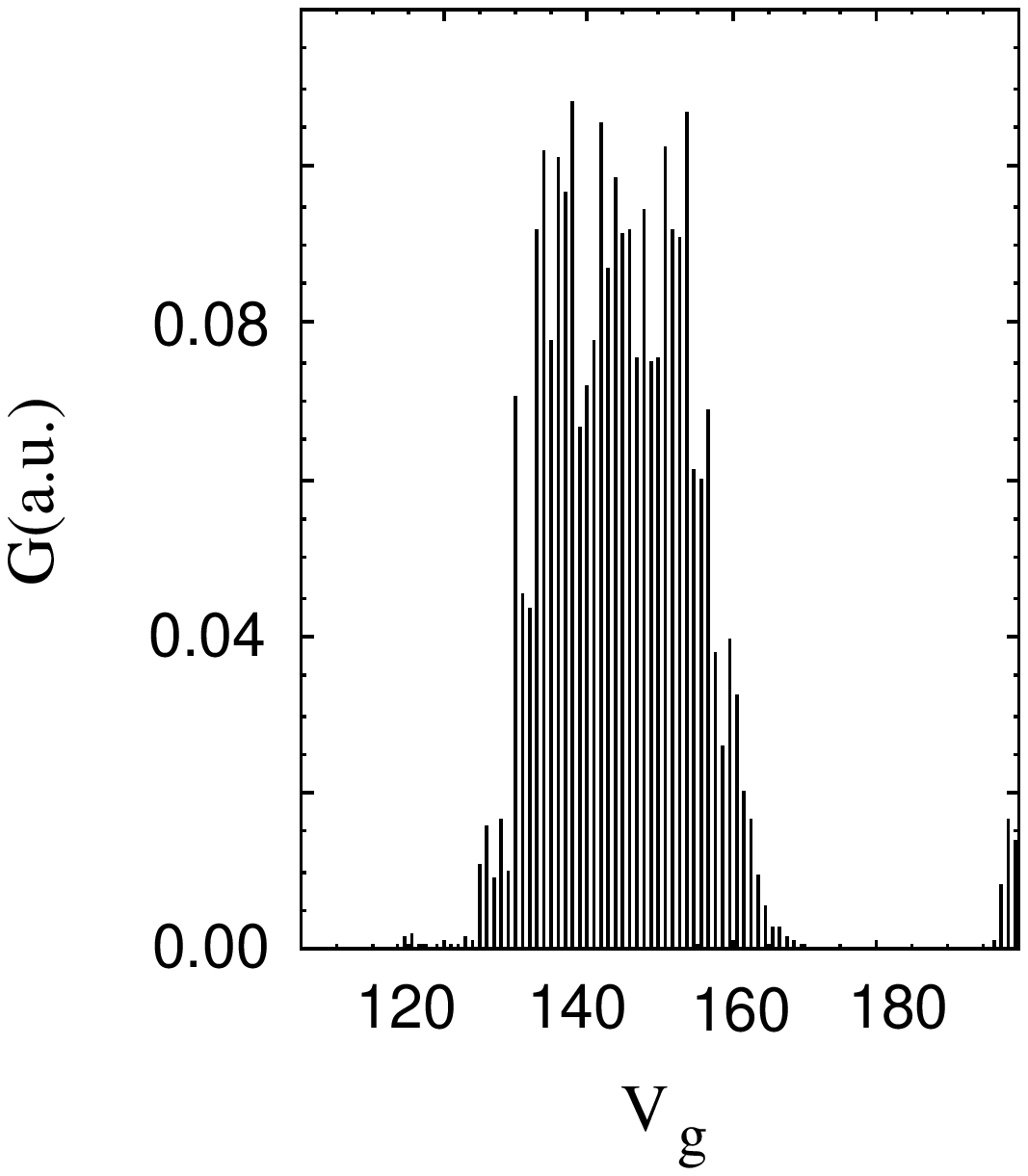,height=7.5cm,width=7.5cm,angle=0}}
\end{minipage}
\begin{minipage}{7.5cm}

  \centerline{\psfig{file=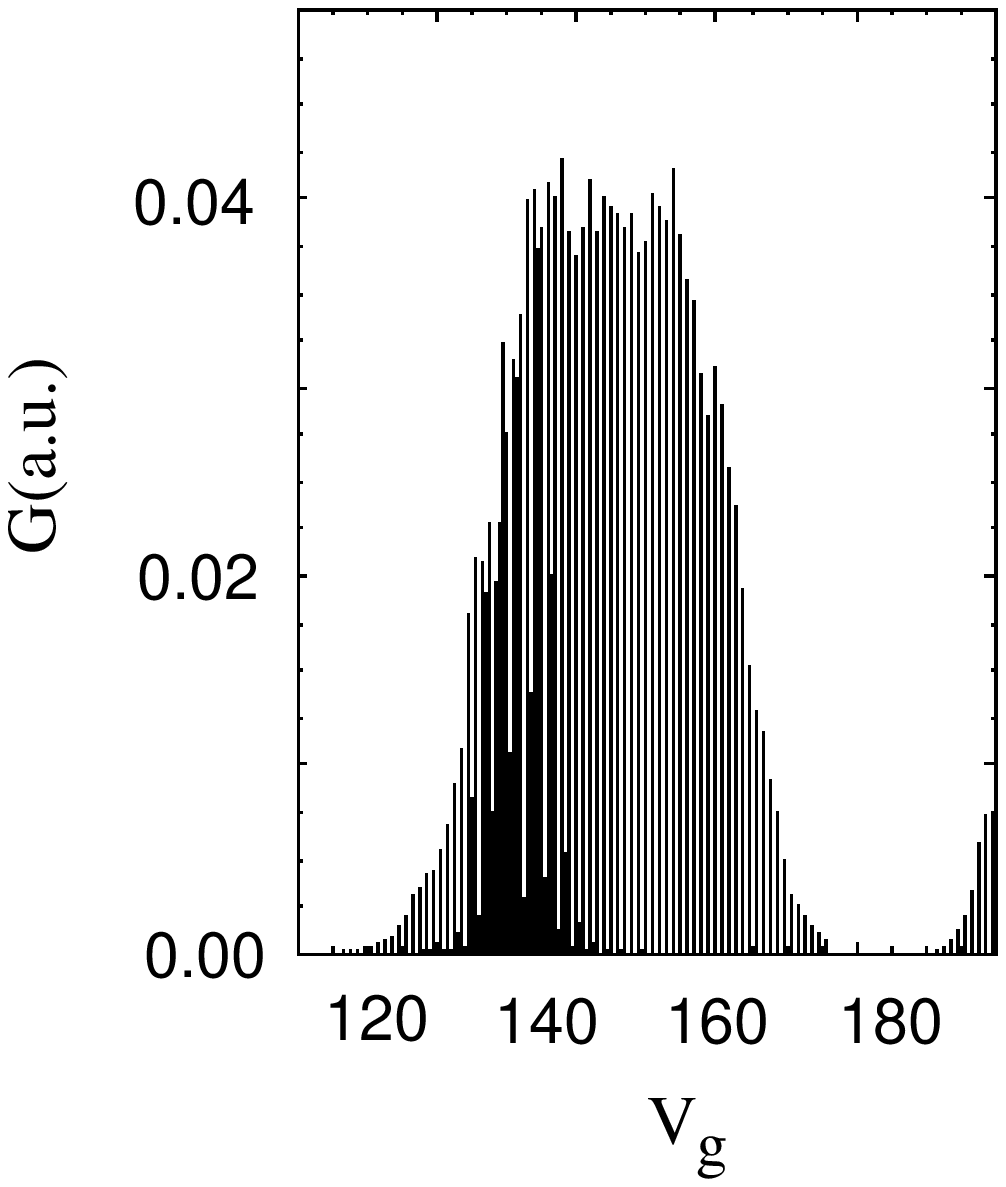,height=7.5cm,width=7.5cm,angle=0}}
\end{minipage}
\caption[]{Conductance $G$ vs.\ gate voltage for (a) $kT=\Delta/5$
(left) and (b) $kT=2\Delta/5$ (right). Taken from
Ref.~\protect{\cite{Bal99a}}.}
\label{conduct}
\end{figure}

The two--terminal conductance $G$ and the transmission phase through
the quantum dot are calculated using the formulas derived in Sec.\ 
\ref{Sec5.0.3}.  In Fig.~\ref{conduct} we show $G$ as a function of
the gate voltage $V_g$ for two values of $kT$. The results are
obtained for $\alpha=1$, $\gamma=0.005$, $E_0=-11$ and $V_0=90$
where the energies and voltages are measured in units of $U_0$.  The
single--particle level spacing is $\Delta=0.03 U_0$ which roughly
corresponds to the experiments \cite{Fol96,Sch97}. About 100 Coulomb
blockade resonances occur in the interval $100 < V_g < 200$.  In
both plots strong peaks with similar peak heights appear whenever a
flat level is close to the Fermi energy. The peak--height
correlations are more pronounced at higher temperature (case (b)).
In the regions $V_g < 130$ and $V_g > 160$, there is no the flat
level close to the Fermi energy and $G$ is small. On the scale of
Fig.~\ref{conduct}, some of the conductance peaks are not visible.

\begin{figure}   
\centerline{\psfig{file=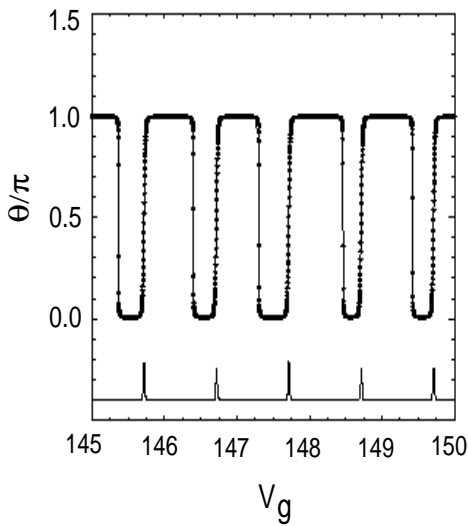,height=7.5cm,width=7.5cm,angle=0}} 
\vspace*{-0.1cm} 
\caption[]{Phase $\theta$ of the transmission amplitude versus gate 
  voltage $V_g$ at $kT=\Delta/5$.  The solid lines at the bottom of
  the plots display the conductance peaks. The flat level
  $\lambda_x=14, \lambda_y=0$ is at or near the Fermi energy. Taken
  from Ref.~\protect{\cite{Bal99a}}.}
\label{phase} 
\end{figure}

The phase $\theta$ of the transmission amplitude is shown vs.\ $V_g$
in Fig.~\ref{phase}. For the parameters chosen, the transmission is
dominated by the flat level $\lambda=(14,0)$; the width of this
level is $\Gamma_\lambda= \Delta/15$. The solid lines at the bottom
of the plot show the conductance peaks and help to identify the
resonance positions. One observes a strikingly similar behavior of
the phase at all resonances. This behavior is found not only within
the $V_g$ interval shown but {\em for the entire interval $130 < V_g
  < 160$ comprising 30 resonances}. The phase increases by $\pi$ at
each resonance and displays a sharp lapse by $\pi$ between adjacent
resonances.  The increase at resonance occurs on the scale $kT$ (we
assumed $kT>\Gamma$) and the phase lapse between resonances on the
scale $\Gamma$ (see the discussion in Sec.~\ref{Sec5.2}).

The identical behavior of $\theta$ at all resonances reflects the
fact that at each resonance, the transmission through the dot is
dominated by a strongly coupled level $\lambda$. Similarly, the
phase lapse between adjacent resonances is caused by the dominant
level: At finite temperature the level $\lambda$ has a finite
probability of being either occupied or empty and, thus, contributes
to both an electron--like and a hole--like cotunneling process. The
contribution of both processes to the transmission amplitude is
\begin{eqnarray} 
  t_{\rm QD}& = & V_{\lambda}^L V_{\lambda}^{R*} \left[ \frac{1-\langle 
\hat{n}_{\lambda} 
    \rangle_N}{E-({\cal E}_{\lambda}+N \cdot U_0)+i \Gamma_{\lambda}/2}  
\right. \nonumber \\ & & \left. \hspace*{1cm} + 
  \frac{\langle \hat{n}_{\lambda} \rangle_N}{E-({\cal E}_{\lambda}+ 
    (N-1) \cdot U_0)+i\Gamma_{\lambda}/2}\right] , 
\label{cotun} 
\end{eqnarray} 
where the first (second) term represents the electron (hole)
contribution, respectively. As the gate voltage $V_g$ scans the
$N^{\rm th}$ conductance valley, the sign of ${\rm Re} (t_{\rm QD})$
reverses, leading to a lapse in the transmission phase.
 
We emphasize that the phase lapse between resonances is a genuine
interaction effect. For vanishing charging energy, the cotunneling
amplitude (\ref{cotun}) would reduce to a single,
temperature--independent term. The phases of the transmission
amplitude in consecutive valleys would not be correlated, and there
would be no systematic phase lapse between resonances. We also note
that the systematic phase slip only occurs at finite temperature. At
zero temperature a flat level contributes to either particle--like or
to hole--like cotunneling and no phase lapse is expected within the
present model.

\subsubsection{Energy shift}  
\label{Sec5.3.4} 
Silvestrov and Imry \cite{Sil99} demonstrated a mechanism for
in--phase behavior that requires neither temperature nor dot
deformations.  They showed that the energy shift resulting from the
coupling between the dot and the reservoirs plays an important role
and may cause in--phase resonances. Energy shifts become relevant
when they are of the order of the mean level spacing. This requires
a coupling of the order $\Gamma \ln (U_0 / \Gamma) \sim 2 \pi
\Delta$. As a result, the scenario of Silvestrov and Imry cannot
explain the phase correlation in the Yacoby--experiment (where
$\Gamma \ll \Delta$) but is possibly relevant for the
strong--coupling regime explored in the Schuster--experiment.  The
number of peaks that may be correlated due to energy shifts is
estimated to be $\sim \Gamma/ (2 \pi \Delta) \ln (U_0 / \Gamma)$.

Slivestrov and Imry illustrated their idea for a quantum dot of
linear dimension $l$ and (dimensionless) confining potential
\begin{equation}
V(x,y)=-4x^2 \left( 1-{x \over l} \right)^2 +\left (y+{x^2 \over 4 l}
\right)^2 \left(1 + 2\left(2 {x \over l}-1\right)^2 \right).
\label{eq:sil1}
\end{equation}
The dynamics in the dot is non--integrable but the potential is
approximately symmetric similarly to the dot used in the
experiments. The dot is connected to a single lead attached at
$x=0$ and extending in negative $x$--direction with a parabolic
confining potential in transverse direction. The Schr\"odinger
equation in the dot and the lead can be solved numerically on a
lattice. The kinetic energy is modeled by a standard nearest
neighbor hopping term. Within the energy interval $1.5 < \epsilon
<4.7$ used below only a single propagating mode exists in the lead.
The energy range corresponds to {\em above--barrier} scattering
through the dot.

\begin{figure}   
\centerline{\psfig{file=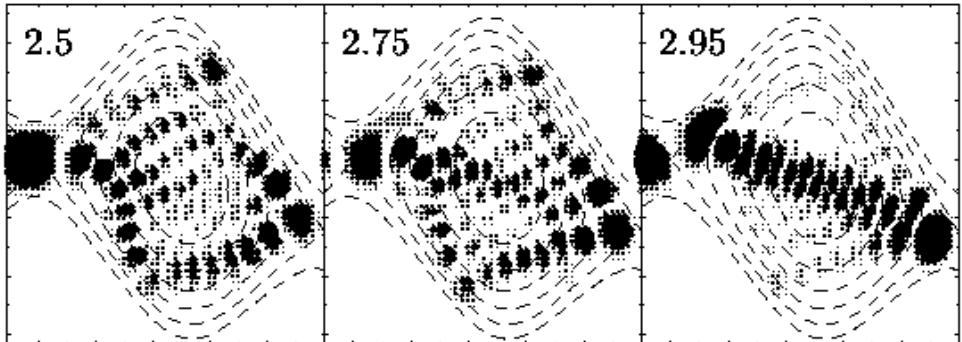,width=13cm,angle=0}} 
\vspace*{-0.1cm} 
\caption[]{Projection of the scattering wave function corresponding
  to a broad level for three different energies. The energies are
  indicated by the numbers in the upper left corners. The broad
  level is the superposition of two contributions which are
  quantized on different classical trajectories. One trajectory
  (left plot) has its classical turning point at the left contact,
  yielding strong coupling to the lead. Taken from
  Ref.~\protect{\cite{Sil99}}.}
\label{fig:sil1} 
\end{figure} 

The numerical solution of the single--particle scattering problem
reveals scattering resonances with very different width. Silvestrov
and Imry extracted the widths by fitting the numerical results with
a sum of Breit--Wigner resonances. This yields broad resonances with
$\Gamma$ of order $(5-7) \Delta$ as well as very narrow resonances
with width much less that $\Delta$.  The projection of the
scattering wave function corresponding to one broad resonance is
shown in Fig.~\ref{fig:sil1}.  It has large probability near the
contact to the lead, and may be understood as the superposition of
two contributions each quantized on a short classical trajectory.

Motivated by their numerical findings for the single--particle
problem, Silvestrov and Imry then investigated many--particle
effects. They studied a model with only one level (denoted $\mu$)
coupled strongly to the lead. The width of this level was assumed to
be $\Gamma \gg \Delta$, the width of all other levels is much
smaller than $\Delta$. Even though one level was strongly
transmitting, the standard charging model was employed to model the
electron--electron interactions. It was assumed that the charging
energy $U_0 \gg \Gamma$. To understand the origin of the energy
shift calculated below, it is useful to decompose the total
occupation number $\hat{N}= \hat{N}^\prime + \hat{n}_\mu$ in two
part, the first part $\hat{N}^\prime =\sum_{\lambda \neq \mu}
\hat{n}_\lambda$ describing the total occupation of the sharp levels
and the part $\hat{n}_\mu$ the occupation of the broad level.  The
charging energy then reads
\begin{equation}
\hat{U} = {1 \over 2} U_0 (\hat{N}^2-\hat{N})={1 \over 2} U_0 
(\hat{N}^{\prime 2} -
\hat{N}^\prime) + U_0 \hat{N}^\prime \hat{n}_\mu .
\label{eq:sil2}
\end{equation}
For fixed $N^\prime$, the charging energy contributes a term $\sim
\hat{n}_\mu$ to the Hamiltonian. This term can be combined with the
single--particle term ${\cal E}_\mu \hat{n}_\mu$ and yields the
effective energy ${\cal E}^{(N^\prime)}_\mu \equiv {\cal E}_\mu+ U_0
\hat{N}^\prime$ which depends on the occupation number $N^\prime$.
Here and below we use the superscript to indicate the number of
electrons in the sharp levels.  For fixed $N^\prime$ and {\em
  vanishing} coupling of the sharp levels, the scattering problem
through the dot reduces to the well-known problem of a single state
interacting with a continuum.  The exact solution of this problem is
known \cite{bm,Sil99}.  The coupling between the state and the
continuum reduces the total energy of the system below its value
found for the decoupled case.  The energy shift introduced by the
coupling is also known exactly \cite{bm,Sil99}.

We now investigate the ground state of the dot as a function of the
gate voltage. Assume that the dot is initially filled with $N$
electrons and that the gate voltage is then tuned beyond the charge
degeneracy point for $N$ and $N+1$ electrons. The states in the dot
are labeled in the order of increasing energy, and it is assumed
that $\mu >N+1$. Neglecting the coupling to the continuum, the
ground state of the dot with $N+1$ electrons is obtained by filling
the lowest $N+1$ levels and leaving the broad level $\mu$ empty.
When the coupling is included, the total energy of this
configuration is given by
\begin{equation}
E_{\rm tot}^{(N+1)} = \sum_{\lambda=1}^{N+1} {\cal E}_\lambda
+ {1 \over 2} U_0 [(N+1)^2-(N+1)] -{\Gamma \over 2 \pi} \ln \left( 
{4 E_F \over {\cal E}_\mu^{(N+1)}} \right),
\label{eq:Sil3}
\end{equation}
where we have omitted the energy of the electrons on the leads
(energies are counted from the Fermi level). The last term on the
right hand side is the energy shift resulting from the coupling
between the dot and the lead. The expression is valid provided
${\cal E}_\mu^{(N+1)} \gg \Gamma$. The corrections for $|{\cal
  E}_{\mu}^{(N+1)}| \leq \Gamma$ can be found in Ref.~\cite{Sil99}.
The energy (\ref{eq:Sil3}) may be compared with the energy
\begin{equation}
E_{\rm tot}^{(N)} = \sum_{\lambda=1}^{N} {\cal E}_\lambda +
{\cal E}_\mu + {1 \over 2}U_0 [(N+1)^2-(N+1)] -{\Gamma \over 2 
\pi} \ln \left( {4
E_F \over {\cal E}_\mu^{(N)}} \right).
\label{eq:Sil4}
\end{equation}
of the configuration where the broad level and $N^\prime=N$ sharp
levels are occupied. In the weak--coupling limit, $E_{\rm
  tot}^{(N+1)}$ is always less than $E_{\rm tot}^{(N)}$, and the
transmission at the $N+1$ Coulomb peak proceed via level $N+1$.
However, when $\Gamma$ is large enough so that
\begin{equation}
{\cal E}_\mu - {\cal E}_{N+1} \leq {\Gamma \over 2 \pi} \ln \left( {
U_0 \over |{\cal E}_\mu^{(N)}| } \right) ,
\label{eq:Sil5}
\end{equation}
the configuration for which level $\mu$ is occupied has lowest
energy. In this case, one finds a broad Coulomb peak caused by the
transmission through the level $\mu$.

When the voltage is further increased, the two functions $E_{\rm
  tot}^{(N)}$ and $E_{\rm tot}^{(N+1)}$ cross. The voltage at the
crossing point is determined by the equation
\begin{equation}
{\cal E}_\mu (V_g) =  U_0 N -{  U_0 \over \exp[2 \pi ({\cal E}_\mu
-{\cal E}_{N+1})/ \Gamma]+1} .
\label{eq:Sil6}
\end{equation}
At the crossing point, the ground state jumps onto the branch
$E_{\rm tot}^{(N+1)}$ and the current--transmitting level $\mu$ is
again empty. At zero temperature, the jump from one electron
configuration to the other is accompanied by a sharp jump by $\pi$
of the transmission phase.  The process of filling the broad level
and jumping to another electron configuration can repeat itself. The
number of consecutive correlated resonances caused by this process
is $\sim \Gamma /(2 \pi \Delta) \ln (U_0 / \Gamma)$. We note that
the analysis described above pertains to the ground state of the
quantum dot. A generalization to finite temperature is required for
a quantitative comparison with the experiment (e.g.\ to explain the
finite scale observed for the phase lapse).
 
\subsubsection{Bouncing--ball tunneling}  
\label{Sec5.3.5} 

In Secs.~\ref{Sec5.3.2}-\ref{Sec5.3.4} we showed that quantum dots
with integrable or mixed classical dynamics may support a subset of
quantum states with exceptionally strong coupling to the leads.
These states are quantized on short periodic orbits that connect the
contacts to the leads. The dynamical origin for strong coupling was
further clarified by Hackenbroich and Mendez \cite{Hac00}. Using a
tunneling Hamiltonian approach, they showed that the relative
coupling strength of the dot states strongly depends on the {\em
  transverse width} of the tunneling region between the dot and the
leads.  Exceptionally strong coupling to few dot states is only
found for sufficiently wide leads, $k a_{\rm eff} >1$. Here, $k$ is
the Fermi wave number and $a_{\rm eff}$ the effective transverse
width of the tunneling barrier. The transport under these conditions
may be termed bouncing--ball tunneling (BBT), as the strongly
conducting states are quantized on classical trajectories bouncing
between the contacts to the leads. A unique fingerprint of BBT is
found in the regime of strong coupling $\Gamma > kT$: Then the peaks
in the tail of a sequence of correlated Coulomb peaks develop a
characteristic line--shape asymmetry. The origin of this asymmetry is
the breaking of the particle--hole symmetry as the strongly coupled
bouncing--ball state moves away from the Fermi energy. 

The crucial role of the transverse barrier width $a_{\rm eff}$ can
be understood from the following simple argument: Confinement in a
lead of width $a_{\rm eff}$ yields the transverse momentum spread
$\hbar / a_{\rm eff}$ for electrons injected in the quantum dot. 
Wide leads therefore result in near normal injection and provide
exceptionally strong coupling to the bouncing--ball states.
In the tunneling Hamiltonian, $a_{\rm eff}$ enters in the the matrix
element $V^l_\lambda$ for tunneling between the lead $l$ and the state
$\lambda$ in the quantum dot. For sufficiently high barriers, 
$V^l_\lambda$ is given by \cite{Nar99}
\begin{eqnarray}
\label{eq_over}
V^l_\lambda = \left( { \hbar^2 \over m^*} \right)   
\left. \int d s \psi_l (s,z) \partial_z \psi_{\lambda}^*(s,z) 
\right|_{z=0} ,
\end{eqnarray}
where the integration is performed along the edge between the
potential barrier and the quantum dot ($\partial_z$ denotes the
derivative normal to the barrier). The wave function $\psi_\lambda$
corresponds to Dirichlet boundary conditions in the dot, while the
barrier tunneling is fully included in the lead wave function
$\psi_l$. The transverse potential in the tunneling region can be
taken quadratic \cite{Nar99} yielding $\psi_l \sim c_l
\exp[-(s-s_l)^2/2 a_{\rm eff}^2]$, where $s$ is the transverse
coordinate, $s_l$ the center of the constriction and $a_{\rm eff}$
its effective width. One can restrict the calculation to the lowest
transverse mode since higher modes are suppressed by the barrier
penetration factor (included in $c_l$).

\begin{figure}
\centerline{\psfig{file=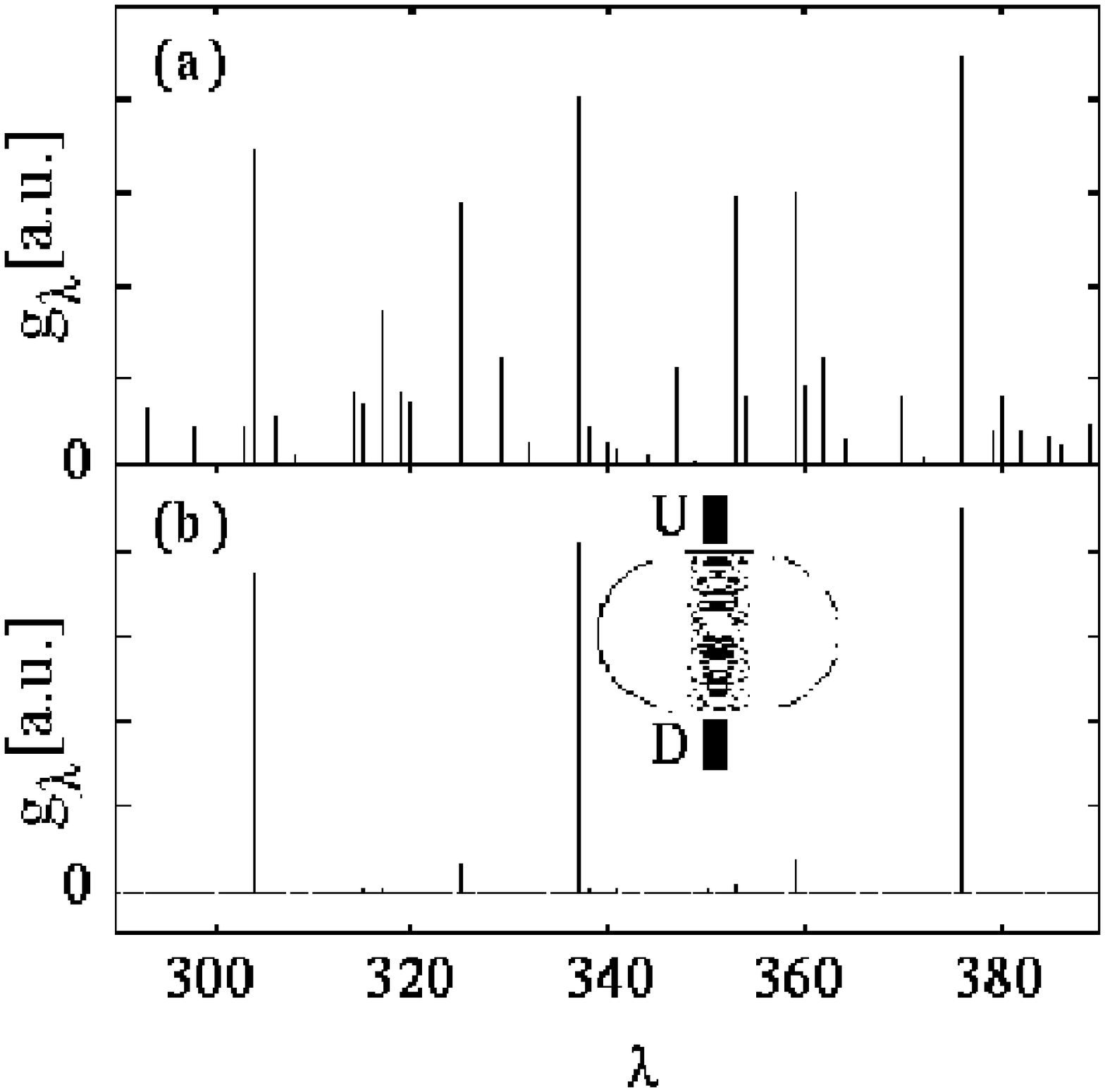,width=8cm,angle=0}} 
\vspace*{0.2cm} 
\caption{Coupling strength $g_\lambda$ for a sequence of 100 states in the
  interval $290 \le \lambda \le 390$. Results are for a quantum dot
  with quadrupolar shape and deformation $\epsilon=0.2$. (a) Narrow
  leads: $k a_{\rm eff} =0.1$. (b) Wide leads: $k a_{\rm eff}=5.0$.
  Here $k=(2 m^* E_\lambda)^{1/2}/\hbar$ is the wave number
  associated with the bouncing--ball state $\lambda=337$. Inset:
  Real--space projection of the state $\lambda=337$.}
\label{ball1}
\end{figure}

The central result of Ref.~\cite{Hac00} is presented in Fig.\ 
\ref{ball1}.  Shown is the coupling strength $g_\lambda \equiv
\Gamma_\lambda^U \Gamma_\lambda^D /(\Gamma_\lambda^U
+\Gamma_\lambda^D)$ over a sequence of 100 quantum states labeled by
the index $\lambda$. Note that $g_\lambda$ is proportional to the
conductance peak height $G_\lambda =(e^2/h) (\pi /2 kT) g_\lambda$
measured in low temperature Coulomb blockade experiments
\cite{Kou99}. The dot is described by a hard--wall confining
potential at the boundary, parameterized in polar coordinates by
$R(\phi) = R_0 [1+\epsilon \cos (2 \phi)]$. Here, $\epsilon$
measures the quadrupolar deformation out of circular shape. A
nonzero value for $\epsilon$ mimics the dots used in
Refs.~\cite{Yac95,Sch97}. As in the experiments, the leads are
attached opposite to each other at the boundary points closest to
the origin (the points with $\phi = \pm \pi/2$). For the value
$\epsilon=0.2$ used for Fig.\ \ref{ball1} the classical dynamics in
the dot is almost completely chaotic {\em except} for two large
islands associated with stable bouncing--ball motion between the
contacts to the leads.

For narrow leads (Fig.\ \ref{ball1}(a)) one observes many peaks with
widely different peak height. In striking contrast, the results for
wide leads (Fig.\ \ref{ball1}(b)) show a few isolated large peaks,
separated by $15-25$ levels with much smaller peak height (not
visible on the scale of Fig.\ \ref{ball1}(b)). All large peaks are
associated with states quantized on stable bouncing--ball orbits.
This is illustrated in the inset for the state $\lambda=337$.  The
height of the small peaks not resolved in Fig.\ \ref{ball1}(b) is
typically two or more orders of magnitude smaller than the maximum
peak height. Such tiny peaks are difficult to resolve in Coulomb
blockade interference experiments. Interference experiments with
wide leads are therefore only sensitive to the strongly coupled
bouncing--ball modes.

We now turn to the calculation of the transmission coefficient $T_{\rm
QD}$ and the phase $\theta_{\rm QD}$. We assume wide leads and $kT
< \Delta$. The case of a weakly coupled dot was studied in Sec.\
\ref{Sec5.0}. Here, we consider a more open dot characterized by
$\Gamma_\mu \sim \Delta$. We identify $\mu$ with the bouncing--ball
state closest to the Fermi energy. All other states $\lambda \neq
\mu$ in the vicinity of $E_F$ have a much smaller width
$\Gamma_\lambda \ll \Delta$.  The Green function for this case may be
obtained using the equations of motion method (see Sec.\
\ref{Sec5.0.2}). It is diagonal up to small off--diagonal corrections
${\cal O} (\sqrt{\Gamma_\lambda \Gamma_\mu} /\Delta)$, and given by
\begin{eqnarray}
\label{GreenGH}
G_{\mu \mu}(E) & = & \sum_{N^\prime =0}^\infty { P_{N^\prime}
\over  E - ({\cal E}_\mu + U_0 \cdot N^\prime) + i 
\Gamma_\mu /2 }.
\end{eqnarray}
Here $N^\prime$ counts the total number of electrons in all dot
levels {\em except} for the level $\mu$ and $P_{N^\prime}$ is the
respective occupation probability. The transmission through the
states $\lambda \neq \mu$ is negligible. Equation (\ref{GreenGH}) is
the generalization of the weak--coupling result derived in Sec.\ 
\ref{Sec5.0.2} to the case of a single strongly conducting quantum
state. The probability $P_{N^\prime} = Z^{-1} \exp[-\Omega
(N^\prime) /kT]$ with $Z=\sum_{N^\prime} \exp[-\Omega (N^\prime)
/kT]$ is related to the thermodynamic potential $\Omega(N^\prime)$
of the dot.  To evaluate $P_{N^\prime}$ we replace
$\Omega_{N^\prime}$ by $\Omega_{N^\prime}^0 + [{\cal E}_\mu +U_0
\cdot N] \langle n_\mu \rangle_{N^\prime}$, where $\Omega_
{N^\prime}^0$ is calculated for the dot with level $\mu$ excluded
from the spectrum, and
\begin{eqnarray} \langle
n_\mu \rangle_{N^\prime} = -{1 \over \pi} \int \! dE  \, {\rm Im} {f(E)
\over E-({\cal E}_\mu + U_0 \cdot N^\prime) +i \Gamma_\mu /2}
\end{eqnarray} 
is the canonical occupation probability of level $\mu$.

\begin{figure}
\centerline{\psfig{file=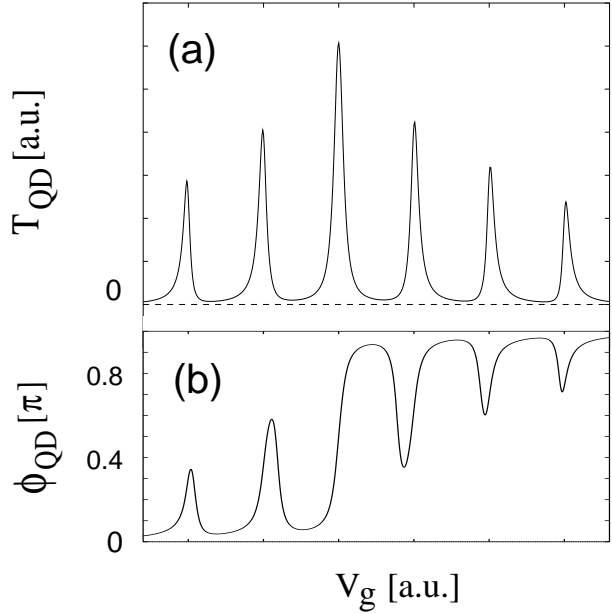,width=8cm,angle=0}} 
\vspace*{0.2cm} 
\caption{(a) Transmission coefficient $T_{\rm QD}$ and (b) phase 
  $\theta_{\rm QD}$ as a function of gate voltage $V_g$ evaluated
  for $kT = 0.2 \Delta$, $\Gamma_\mu=1.5 \Delta$, and $U_0=12
  \Delta$. The peak asymmetry in (a) is a unique fingerprint of
  BBT.}
\label{ball2}
\end{figure}

In Fig.\ \ref{ball2} we show the transmission coefficient $T_{\rm
  QD}$ and the transmission phase $\theta_{\rm QD}$ vs.\ gate
voltage $V_g$ (we assumed ${\cal E}_\mu = {\cal E}^{(0)}_\mu -e
V_g$).  All peaks shown result from transmission through the level
$\mu=337$.  The peaks have comparable height and similar phase in
qualitative agreement with the experiments \cite{Yac95,Sch97}. The
central peak has a Lorentzian shape of width $\Gamma_\mu$.  Note
that the transmission peaks develop a peculiar asymmetry as the
conducting level moves away from the Fermi energy: Each peak to the
left and to the right of the central peak falls off more rapidly on
the side facing the central peak. This pattern extends over the
whole sequence and becomes more pronounced for the peaks in the
tails. The asymmetry results from the breaking of the particle--hole
symmetry in the transmission through the bouncing--ball state and is
a {\em unique fingerprint} of BBT.  Inspection of the data of the
experiment \cite{Sch97} reveals the same asymmetry, providing strong
evidence that the peak correlations in this experiment are due to
BBT.

Note that the magnitude of the conductance peaks in Fig.\
\ref{ball2}(a) decreases in the tails of the sequence. This is in
contrast to the experiment \cite{Sch97} where similar peak--heights
were observed. This discrepancy may have several reasons: First, the
billiard model used here completely neglects deformations of the dot
shape.  As discussed in Sec.~\ref{Sec5.3.2}, \ref{Sec5.3.3}
deformations can enhance peak--height correlations by ``pinning" the
conducting levels close to the Fermi energy. The model further
neglects the electrostatic influence of the plunger gate on the point
contacts defining the tunneling barriers. In most experiment, this
influence is significant: Increasing the plunger voltage, inevitably
opens the dot. To model this effect one may reduce the barrier height
with increasing voltage. In turn, this would result in enhanced peak
heights on one side of a sequence. This voltage induced enhancement
and the suppression by temperature may result in a number of peaks
with similar peak height.

\subsubsection{Other mechanisms}  
\label{Sec5.3.6} 
 
In two recent papers \cite{Wu98,Kan99} Wu et al.\ and Kang proposed
that the in--phase resonances reflect an interference effect in the
AB ring rather than a property of the quantum dot. The authors
investigate the transmission of non--interacting electrons through a
single--channel AB ring. A two--terminal AB ring is studied in
Ref.~\cite{Wu98} and both a two--terminal and a four--terminal ring
in Ref.~\cite{Kan99}. The quantum dot is modeled in both studies as
a symmetric 1D double barrier well (identical barriers on either
side). Full coherence is assumed through the dot and the AB ring.
For an integer value of flux quanta threading the ring, a set of
conductance peaks is found upon variation of the dot potential. All
peaks are in phase. A new set of in--phase peaks appears at
half--integer flux, while the previously found peaks disappear. Each
peak of the new set is located between two peaks of the previous
set. The explanation of the effect is straightforward and follows
from the scattering theory presented in Sec.~\ref{Sec4.2}: There it
was shown that the oscillatory part $G_{\rm AB}$ of the conductance
close to a resonance of the quantum dot is of the form
\begin{equation} 
G_{AB} \propto  1 + \cos(\phi - \chi) , 
\label{GAB} 
\end{equation} 
where $\phi=2 \pi \Phi / \Phi_0$ is the dimensional flux and where 
$\chi$ takes the values $0$ and $\pi$ for even and odd states, 
respectively, of the quantum dot (cf.\ Eq.~(\ref{C17})). As a 
result, only even states contribute to the AB current if an integer 
number of flux quanta are threading the ring. Clearly, all even 
states are in phase. Conversely, for half integer number of flux 
quanta, only odd states contribute and, again, all with the same 
phase. Based on Eq.~(\ref{GAB}) one expects contributions of both 
even and odd states for intermediate values of the flux. Precisely 
this has been reported by Kang \cite{Kan99} for $\Phi= \Phi_0/4$ and 
$\Phi = 3 \Phi_0 / 4$. 
 
The ideas presented in Refs.~\cite{Wu98,Kan99} cannot account for
the experimentally observed in--phase behavior.  First,
Refs.~\cite{Wu98,Kan99} find in--phase resonances only for integer
or half--integer number of flux quanta threading the ring. For
generic values of the flux, neighboring resonances are always out of
phase by $\pi$. Second, the destructive interference responsible for
the in--phase resonances in \cite{Wu98,Kan99} is only complete if
the transport through the AB ring is fully coherent. In the
experiments \cite{Yac95,Sch97}, however, most of the current is
incoherent. The coherent component comprises only about 10 or 20
percent of the total current. The incoherent ring current displays
resonant behavior regardless of the value of the external flux or
the parity of the dot state. Hence, experimentally both even and odd
resonances are observed for all values of flux, and each current
resonance is associated with the charging of the dot by one
additional electron.  The prediction \cite{Wu98,Kan99} of missing
resonances and the addition of the charge $2e$ between neighboring
resonances is in contradiction to the experiments.

\newpage
\setcounter{equation}{0}
\section{Controlled dephasing}
\label{Sec6}

In 1996 Gurvitz \cite{Gur96} realized that mesoscopic electron
devices could be utilized to study and controll decoherence. He 
investigated the resonant tunneling through two capacitevly coupled
quantum dots. Gurvitz showed that the second dot could serve as a
detector for the charge accumulated in the first dot. Solving the
equations of motion of the entire system and then tracing over the
detector variables, Gurvitz obtained the reduced density matrix of
the measured dot. The coupling to the detector dot lead to a
damping of the off--diagonal elements and thus caused decoherence.

Gurvitz' paper has triggered new experimental and theoretical
developments exploring the controlled decoherence of quantum
transport by the environment. Since the destruction of coherence is
not necessarily related to energy relaxation, the new subject is
commonly referred to as controlled dephasing. The first
demonstrations of controlled dephasing were achieved in beautiful
which--path experiments of Buks et al.\ \cite{Buk98} and Sprinzak et
al.\ \cite{Spr00}. The experiments employed the transport through a
single \cite{Buk98} or through two \cite{Spr00} tunnel--coupled
quantum dots. In close proximity and capacitevly coupled to one dot
was a quantum point contact (QPC).  Electrons passing through the
quantum dot interact with electrons in the QPC. This modifies the
transmission through the QPC, so that the QPC serves as a detector
for the charge of the dot. It was found that a current flowing
through the QPC leads to dephasing in the quantum dot. In the
experiment \cite{Buk98} dephasing was detected as a reduction of AB
oscillations using a device with the quantum dot embedded in an AB
ring. The set--up can be viewed as a which--path interferometer, the
archetype of a position measurement apparatus in a double--slit
device.  We discuss the which--path experiments in Sec.\ 
\ref{Sec6.1}.  The theoretical work devoted to dephasing in
mesoscopic electron structures is summarized in Sec.\ \ref{Sec6.2}.

\subsection{Which--path experiments}
\label{Sec6.1}

\subsubsection{Experiment of Buks et al.}
\label{Sec6a.1}
\begin{figure}
\hspace*{0cm}
\centerline{\psfig{file=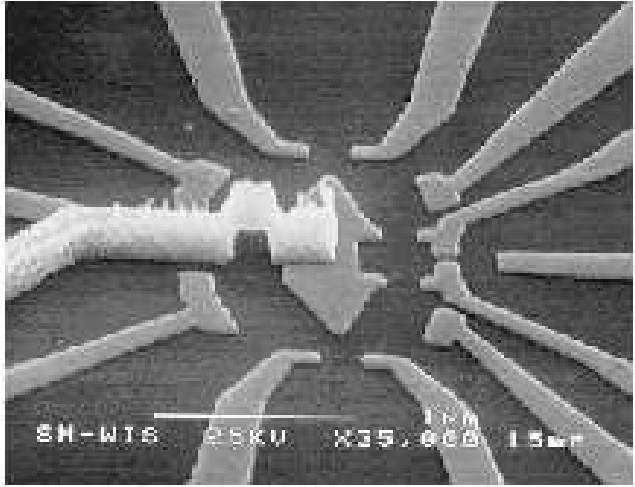,width=11cm,angle=0}}
\centerline{\psfig{file=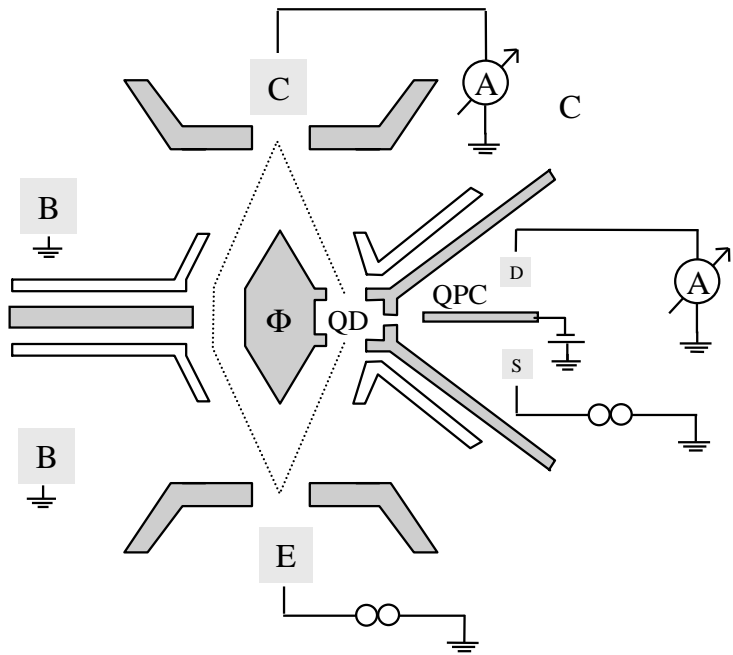,width=11cm,angle=0}}
\caption[]{Top: Scanning electron micrograph of the which--path device. 
  Bright regions indicate metallic gates. A quantum dot is defined
  in the right arm of a multi--terminal AB ring. A
  quantum point contact to the right of the quantum dot serves as a
  which--path detector. Bottom: Schematic description of the
  electrodes and contacts in the which--path experiment. The
  interferometer is composed of three different regions, collector
  C, emitter E, and base regions B. The quantum point contact is to
  the right of the quantum dot. A finite voltage $V_d$ is applied
  across the quantum point contact.  Taken from
  Ref.~\protect{\cite{Buk98}}.} \label{bukschem}
\end{figure}

Fig.~\ref{bukschem} shows the scanning electron micrograph and a
schematic description of the set--up used in the which--path
experiment of Buks et al.\ \cite{Buk98}. Part of the device,
including the multi--terminal AB ring and the quantum dot embedded
in its right arm is identical with the setup used in the
Schuster--experiment (see Sec.\ \ref{Sec3}). The new element is the
QPC to the right of the quantum dot. It is visible as a small
constriction in the two--dimensional electron gas.  The width of the
QPC and hence its transmission coefficient is controlled by the gate
voltage $V_g$. The central gate is contacted with a metallic air
bridge.  This gate at the potential $V_P$ depletes the area between
the two arms of the ring and, at the same time, serves as plunger
gate of the quantum dot. The dot is tuned to the Coulomb blockade
regime with the resistance of the tunneling barriers at either side
of the dot being greater than $h/2e^2$.

\begin{figure}
\vspace{-0.5cm}
\hspace*{-1.5cm}
\centerline{\psfig{file=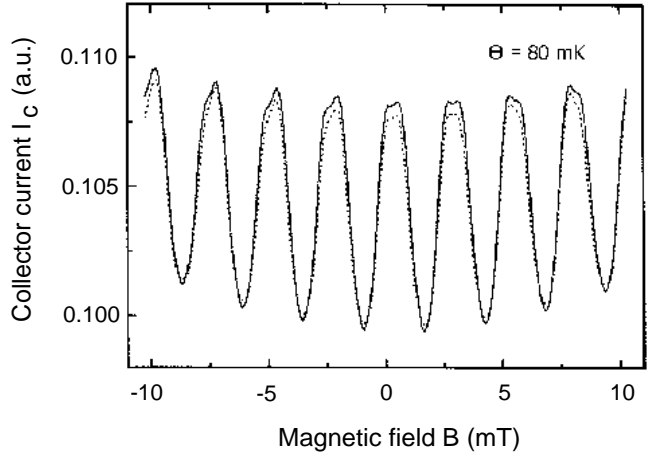,width=14.0cm,angle=0}}
\vspace*{-1.5cm}
\caption[]{AB oscillations of the collector current
  $I_C$. The solid line is measured with QPC drain source voltage
  $V_d=0$ $\mu V$. The dotted line, with reduced visibility, is
  measured with $V_d = 100$ $\mu V$. Taken from
  Ref.~\protect{\cite{Buk98}}.}
\label{bukosc}
\end{figure}

\begin{figure}
\hspace*{0cm}
\centerline{\psfig{file=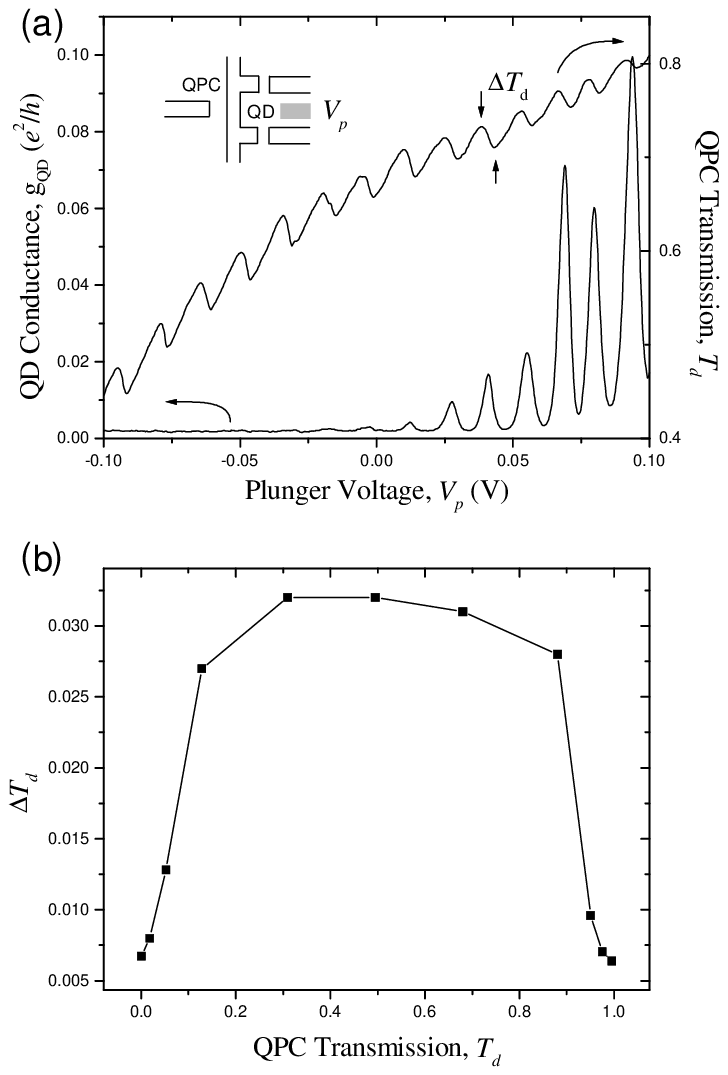,width=8.0cm,angle=0}}
\caption[]{Conduction characteristics of the calibration device
  shown in the inset of (a). (a) Conductance through the quantum dot
  and transmission through the QPC both as a function of the plunger
  gate voltage $V_P$. (b) Induced average change $\Delta {\cal T}_d$
  in the transmission coefficient of the QPC due to adding an
  electron to the quantum dot as a function of ${\cal T}_d$. Each
  data point is obtained by averaging over several Coulomb peaks.
  Taken from Ref.~\protect{\cite{Buk98}}.} \label{bukcal}
\end{figure}

The collector current $I_c$ measured for a fixed a.c.\ emitter
voltage $V_E = 10 \mu V$ displays AB oscillations as depicted in
Fig.~\ref{bukosc}.  The solid line shows the AB oscillations with
vanishing drain--source voltage $V_d =0$ across the QPC. The dotted
line is measured with $V_d = 100 \mu V$. The reduction of the
visibility reflects the dephasing introduced by the which--path
measurement.

To study the dephasing quantitatively as a function of the QPC
characteristics, Buks et al.\ investigated the calibration device
shown in the inset of Fig.\ \ref{bukcal}(a). The device contains a
QD and a QPC similar to these in the which--path interferometer. The
conductance of the QD is scanned through a series of Coulomb peaks
by changing the plunger voltage $V_P$. Due to the proximity of QD
and QPC, the transmission coefficient ${\cal T}_d$ of the QPC is
also affected by the potential of the QD. This gives rise to the
smooth increase of ${\cal T}_d$ with increasing plunger voltage
$V_P$ (see Fig.~\ref{bukcal}(a)). However, whenever a conductance
peak is being scanned and an additional electron is being added to
the QD, ${\cal T}_d$ displays a faster and opposite change with
amplitude $\Delta {\cal T}_d$ on the scale of the peak width. This
decrease of transmission reflects the change in the QD potential due
to the additional electron.  Fig.~\ref{bukcal}(b) shows $\Delta
{\cal T}_d$ averaged over several Coulomb peaks as a function of the
transmission ${\cal T}_d$. The reduced value of $\Delta {\cal T}_d$
near ${\cal T}_d=0$ and ${\cal T}_d=1$ is a consequence of
approaching the conductance plateaus. At the plateaus, changes in
the QPC potential have very little effect on the QPC transmission.

\begin{figure}
\hspace*{-1.0cm}
\centerline{\psfig{file=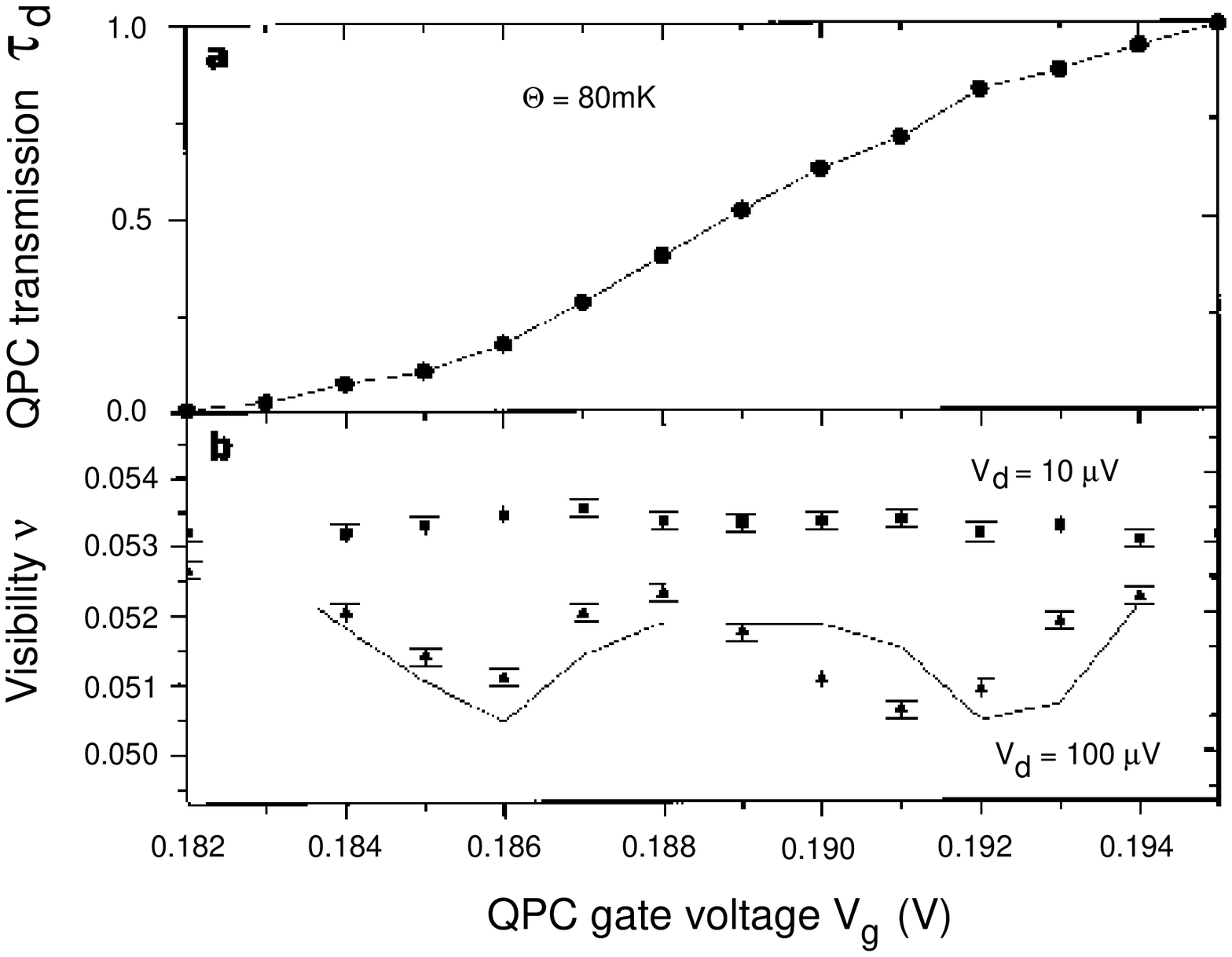,width=15cm,angle=0}}
\vspace*{-0.8cm}
\caption[]{(a) QPC transmission coefficient ${\cal T}_d$ as a
  function of QPC gate voltage $V_g$. (b) Measured visibility of the
  AB oscillations as a function of $V_g$ for two values
  of the QPC drain source voltage $V_d$. The visibility is defined
  as the peak--to--peak signal divided by the average signal. Error
  bars indicate the fluctuations in visibility due to fluctuations
  of the device's properties. Taken from
  Ref.~\protect{\cite{Buk98}}.} \label{bukres}
\end{figure}

\begin{figure}
\hspace*{0cm}
\centerline{\psfig{file=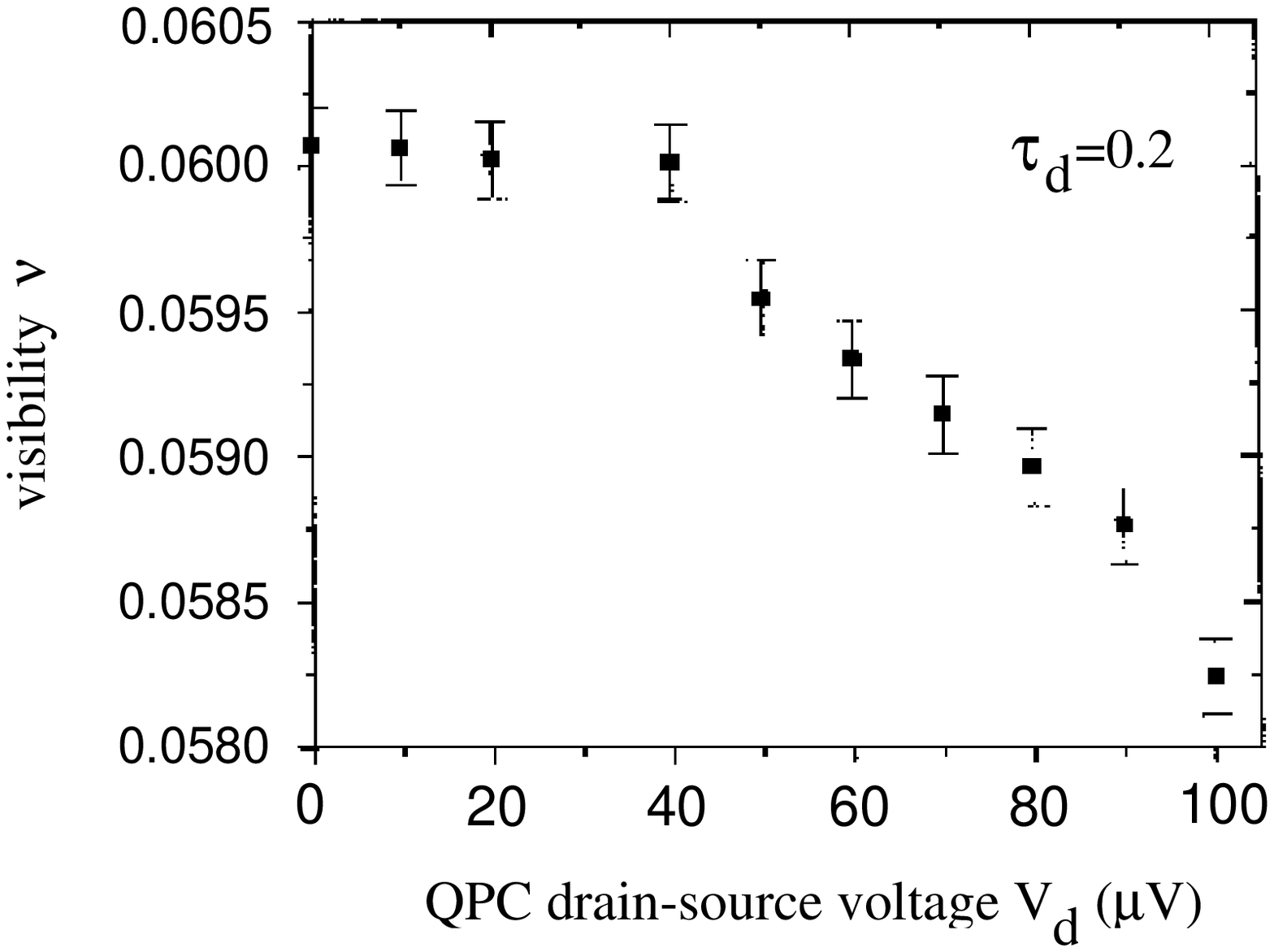,width=9.5cm,angle=0}}
\caption[]{Visibility as a function of QPC drain--source voltage
  $V_d$ for fixed QPC transmission coefficient ${\cal T}_d = 0.2$.
  The visibility decreases linearly for $e V_d > kT$ and saturates
  for low $V_d$. Taken from Ref.~\protect{\cite{Buk98}}.}
\label{bukvolt}
\end{figure}

Fig.~\ref{bukres}(b) summarizes the main result of the experiment.
The visibility of the AB oscillations is shown vs.\ the QPC voltage
$V_g$. The visibility is obtained upon dividing the peak--to--valley
value of $I_C$ by the average $I_C$. The measurement was done when
the QD was tuned to a conductance peak using the central metal
island as a plunger gate. Fig.~\ref{bukres}(a) shows that the
detector transmission increases from $0$ to $1$ as $V_g$ is being
changed. We note that any modulation of the visibility reflects the
dephasing introduced by the detector; a small visibility corresponds
to strong dephasing. For small detector voltage $V_d = 10 \mu V$,
the visibility is practically constant with $V_g$. For $V_d = 100
\mu V$, one observes a pronounced structure: The visibility peaks
near the conductance plateaus ${\cal T}_d = 0,1$, and in between for
${\cal T}_d =0.5$. It turns out that the weak dephasing at these
points is due the small amount of which--path information in the
detector, cf.\ Sec.~\ref{Sec6.2.1}. The detector is inefficient both
at the conductance plateaus and at ${\cal T}_d =0.5$ where the QPC
current is very noisy. The theoretical approaches summarized in Sec.
\ref{Sec6.2} obtain for the visibility the result $\nu = \nu_0
\nu_d$, where $\nu_0$ is the intrinsic visibility of the AB ring and
$\nu_d$ the reduction of the visibility due to the detector. At zero
temperature, $\nu_d$ is calculated to
\begin{equation}
\nu_d = 1-{e V_d \over \pi \Gamma} {(\Delta {\cal T}_d)^2 \over 8 {\cal T}_d (1-
  {\cal T}_d)} ,
\label{nud}
\end{equation}
where $\Gamma$ is the intrinsic (zero--temperature) width of the QD
resonance. The solid line in Fig~\ref{bukres}(b) shows the
prediction of Eq.~(\ref{nud}), where $\Gamma = 0.5 \mu eV$ was used
as a fitting parameter. The quantity $\Delta {\cal T}_d$ was taken
from the calibration device, cf.\ Fig.~\ref{bukcal}.

The dependence of the visibility on the drain source voltage $V_d$
is illustrated in Fig.~\ref{bukvolt}. For $eV_d \gg k T$, the
visibility drops linearly as expected from Eq.~(\ref{nud}). A
deviation from the linear dependence occurs near $e V_d \approx kT$.

\subsubsection{Experiment of Sprinzak et al.}
\label{Sec6a.2}
\begin{figure}
\vspace*{-1cm}
\centerline{\psfig{file=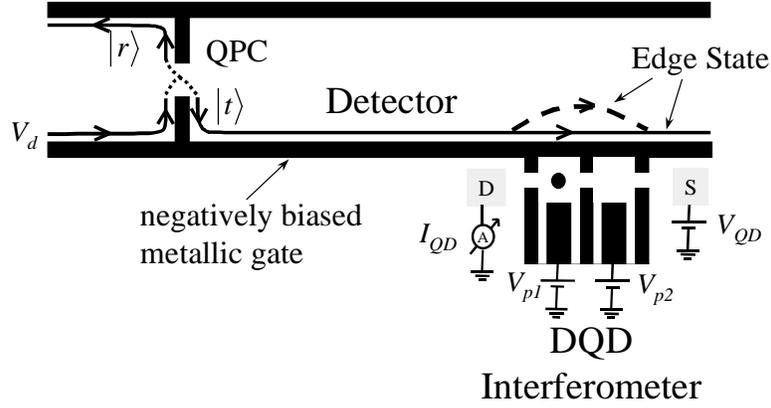,width=11.5cm,angle=0}}
\vspace*{-5.5cm}
\caption[]{Schematic illustration of the DQD interferometer and the
  QPC detector. The device is studied in strong perpendicular
  magnetic field (5--10 Tesla) that imposes chiral electron motion
  near the edges of the sample. The edge states are partly
  transmitted $(| t \rangle)$ and partly reflected $(|r \rangle)$ by
  the QPC. The DQD is weakly coupled to its own leads and is tuned
  to resonance by the two plunger gate voltages $V_{p1}$, $V_{p2}$.
  Taken from Ref.~\protect{\cite{Spr00}}.}
\label{FigSpr1}
\end{figure}

The which--path information in the setup of Buks et al.\ is encoded
in the current flowing through the QPC--detector. Sprinzak et al.\ 
\cite{Spr00} used a different approach and devised a setup in which
the which--path information enters only as a quantum mechanical
phase. The setup is shown in Fig.~\ref{FigSpr1}: A QPC serves as the
detector for the oscillation of charge in a double quantum dot (DQD)
interferometer. The device is subjected to a high magnetic field
corresponding to two filled Landau levels of the two--dimensional
electron gas. Under these conditions most current flows in a chiral
motion along the the edges of the sample. Scattering between the
edge states only occurs at the QPC constriction where the two edges
of the sample are close together. The nearby DQD interferometer
leaves the current through the QPC unaffected but does change the
phase of the transmitted electron mode. The phase of the transmitted
electrons therefore contains information about the state of the DQD.
Note that the DQD is not inserted in a standard two--path
interferometer since the presence of a high magnetic field would
prevent electrons from choosing either path with nearly equal
probability. The experiment, therefore, does not probe the path of
electrons through the interferometer but the state of the
interferometer. In keeping with Ref.\ \cite{Spr00} we still refer to
the setup as a which--path experiment. 

\begin{figure}
\centerline{\psfig{file=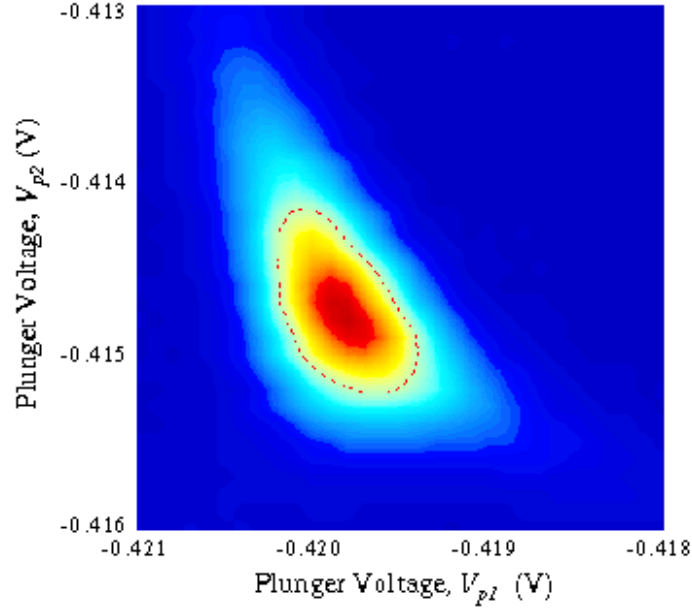,width=10.0cm,angle=0}}
\vspace*{0.2cm}
\caption[]{View of one Coulomb peak as a function of the two plunger
  gate voltages $V_{p1}$ and $V_{p2}$. The dashed line is a contour
  drawn at half maximum of the peak height. The area enclosed by
  this contour is being used as a measure of the peak width.
  Taken from Ref.~\protect{\cite{Spr00}}.}
\label{FigSpr2}
\end{figure}

The dephasing rate of the interferometer can be extracted from the
conductance through the DQD. The DQD is being tuned to resonance by
means of the two plunger gate voltages $V_{p1}$ and $V_{p2}$.  For
the two quantum dots in series, resonances are found when resonant
levels in both dots are degenerate. The resulting Coulomb peaks are
located on a hexagonal lattice and each peak has the width $2
\Gamma$ where $\Gamma$ is the intrinsic resonance width of a single
dot \cite{Vaa95,Kou99}. In the presence of the detector one expects
to observe a reduction of the peak height and a broadening of the
peak width to $2(\Gamma+\hbar/t_d)$. A single resonance peak with
its contour at half maximum is shown in Fig.\ \ref{FigSpr2}.  Due to
the asymmetry of the peak shape, the area in this contour is taken
as a measure of the dephasing rate.

\begin{figure}
\centerline{\psfig{file=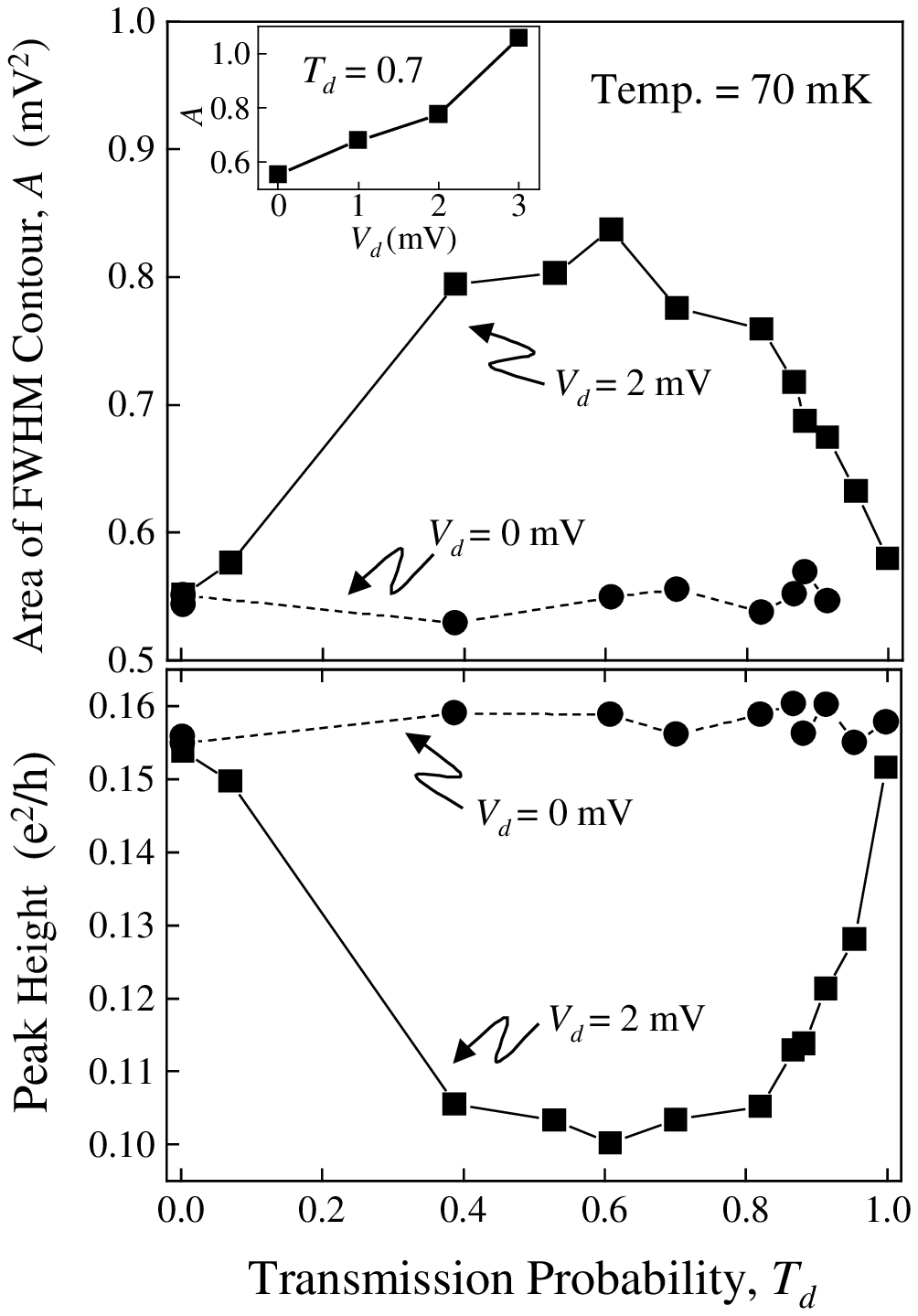,width=7cm,angle=0}}
\vspace*{0.2cm}
\caption[]{Top: The area of the contour at half peak height as a
  function of the transmission probability ${\cal T}_d$ for two
  values of the applied bias $V_d =0$ and $2 mV$. The dependence
  agrees qualitatively with the expected ${\cal T}_d (1-{\cal
    T}_d)$. Inset: The dependence of the contour area on the applied
  voltage $V_d$ for ${\cal T}_d=0.7$. Bottom: The peak height (in
  units
  of $e^2/h$) as a function of the transmission probability ${\cal
    T}_d$ for two values of applied bias $V_d=0$ and $2 mV$. Taken
  from Ref.~\protect{\cite{Spr00}}.}
\label{FigSpr3}
\end{figure}

Figure \ref{FigSpr3} shows the dependence of the contour area and
the peak height as a function of the QPC transmission probability
${\cal T}_d$. Both plots were obtained for a fixed QPC voltage
$V_d$.The measured area (Fig.~\ref{FigSpr3}a) yields a dephasing
rate proportional ${\cal T}_d (1-{\cal T}_d)$. Similarly, the peak
height (Fig.~\ref{FigSpr3}b) has an inverse dependence on that
expression. Note that the result for the dephasing rate differs from
the rate $\propto (\Delta {\cal T}_d)^2/[{\cal T}_d (1-{\cal T}_d]$
found in the experiment of Ref.\ \cite{Buk98}. We show in Sec.\ 
\ref{Sec6.2} that the difference is due the different nature of the
which--path information: In the present device this information
resides in the transmission phase while in the case of the
experiment \cite{Buk98} it is encoded in the transmission
probability of the QPC. The inset of Fig.\ \ref{FigSpr3} shows that
the contour area and hence the dephasing rate grow nearly linearly
with the detector voltage $V_d$.

The measurements show that a phase change in the detector leads to
dephasing of the interferometer even though no interference
experiment is being performed in the detector. The question arises
onto whether the detector must be phase coherent in order to dephase
the interferometer. To investigate this question, Sprinzak et al.\ 
inserted an artificial dephaser (a floating Ohmic contact) in the
path of the transmitted electrons before they reached the DQD. Even
though phase coherence was destroyed, dephasing persisted (however,
a small decrease of dephasing was observed and traced back to the
finite capacitance of the Ohmic contact which effectively eliminated
high frequency components of the shot noise).

\subsection{Theoretical approaches}
\label{Sec6.2}
In the past, experimental investigations of decoherence (dephasing)
have been executed with photons, cooled atoms, and neutrons. The
which--path experiments of Refs.\ \cite{Buk98,Spr00} for the first
time allowed to investigate controlled dephasing with mesoscopic
electron structures.  Widely different methods have been used
\cite{Ale97,Lev97,Gur97,Buk98,Hac98,Sto98,Bue99,Bue99a,Spr00} to
investigate the problem theoretically. Conceptually, some of the
authors investigated the influence of the interferometer on the
state of the detector \cite{Ale97,Buk98,Spr00} while other authors
\cite{Lev97,Gur97,Hac98,Sto98} analyzed how the detector affects the
interferometer. Both descriptions lead to equivalent results as is
known since the work of Stern et al.\ \cite{Ste90}.

In Sec.\ \ref{Sec6.2.1} we present a simple argument for the
dephasing rate in the which--path interferometer. The argument
displays the intimate link between the dephasing rate and the
efficiency with which the QPC detector measures the state of the
interferometer.  Rigorous derivations of the dephasing rate and the
suppression of AB oscillations in a which--path interferometer are
given in Sec.\ \ref{Sec6.2.2}-\ref{Sec6.2.5}.

\subsubsection{Heuristic argument}
\label{Sec6.2.1}
A qualitative argument for the dephasing of quantum dot states due
to the coupling to a QPC detector was given by Aleiner et al.\ 
\cite{Ale97} and Buks et al.\ \cite{Buk98}. Their argument relates
the dephasing rate to the which--path information obtained in the
detector. The original argument only considers the information
contained in the current through the QPC. Below we generalize the
argument to include, in addition, the information encoded in the
phase of the transmission amplitude through the QPC.

An electron in the quantum dot interacts electrostatically with
electrons in the QPC--detector. The interaction generally modifies
both the modulus and the phase of the transmission amplitude through
the QPC.  The change in the transmission probability induced by the
presence of one electron on the quantum dot is denoted by $\Delta
{\cal T}_d$; the corresponding change of the transmission phase is
$\Delta \phi$.  In experiment, these changes cannot be detected with
certainty due to current and phase fluctuations. The current
fluctuations are due to the well--known quantum shot noise (for
reviews on shot noise see Refs.\ \cite{Jon97,Rez98,Bla99}): Let $N=
(2 e V_d /h) t_{\cal T}$ be the number of electrons that probe the
QPC biased with the voltage $V_d$ during a time $t_{\cal T}$.  Then
the number of transmitted electrons $N_{\cal T}$ is a binomial
random variable with the expectation value $\langle N_{\cal T}
\rangle = N {\cal T}_d$ and the standard deviation $\delta N_{\cal
  T} = \sqrt{N{\cal T}_d(1-{\cal T}_d)}$.  From the uncertainty
$\delta N_{\cal T}$ one finds the uncertainty $\delta {\cal T}_d =
\sqrt{{\cal T}_d(1-{\cal T}_d)/N}$ in ${\cal T}_d$. Detection of the
electron on the quantum dot requires that $\delta {\cal T}_d$ is
less than or equal to $\Delta {\cal T}_d$.  The equality yields the
dephasing rate
\begin{equation}
{ 1 \over t_{\cal T}} \propto {e V_d \over h} {(\Delta {\cal
    T}_d)^2  \over {\cal T}_d  (1-{\cal T}_d) }  
\label{decoh1}
\end{equation}
reflecting the which--path information encoded in the detector
current.  A similar reasoning applies to the transmission phase
$\phi$: The uncertainty in $\phi$ is given by $\delta \phi \sim 1/
\delta N_\phi $ where $\delta N_\phi$ is the uncertainty in the
number of electrons transmitted in the time $t_\phi$. The condition
that $\delta \phi$ equals the phase change $\Delta \phi$ yields the
dephasing rate
\begin{equation}
{ 1 \over t_\phi} \propto {e V_d \over h} {\cal T}_d  
(1-{\cal T}_d) (\Delta \phi)^2 .
\label{decoh2}
\end{equation}
Adding the dephasing rates due to the change in the current and the
phase, one finds that the total dephasing rate has the form
\begin{equation}
{ 1 \over t_d} = C_{\cal T} { e V_d \over h} {(\Delta {\cal
    T}_d)^2  \over {\cal T}_d  (1-{\cal T}_d) } + C_\phi
  {e V_d \over h} {\cal T}_d  
(1-{\cal T}_d) (\Delta \phi)^2 ,
\label{decoh3}
\end{equation}
where $C_{\cal T}$ and $C_\phi$ are coefficients of order $1$. The
rigorous calculations presented in Sec.\ \ref{Sec6.2} support the
simple estimate (\ref{decoh3}) and yield expressions for the
coefficients $C_{\cal T}$ and $C_\phi$.

\subsubsection{Green function}
\label{Sec6.2.2}
\begin{figure}
\hspace*{0cm}
\centerline{\psfig{file=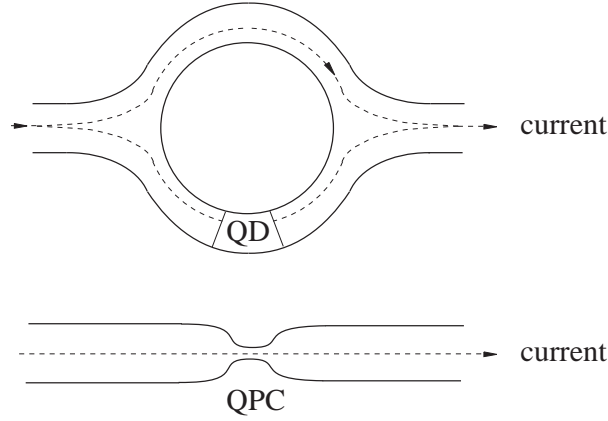,width=8cm,angle=0}}
\vspace*{0.1cm}
\caption[]{Schematic view of the which--path interferometer. The
  quantum dot is embedded in the lower arm of an AB ring. A QPC in the
  vicinity of the ring acts as a which--path detector.}
\label{ab-ring}
\end{figure}

Controlled dephasing and the suppression of AB oscillations due to
the interaction with a QPC have been investigated by Aleiner,
Wingreen, and Meir \cite{Ale97}.  The model consists of a QD in the
Coulomb blockade regime which is embedded in an AB ring and
capcitively coupled to a QPC (Fig.\ \ref{ab-ring}). It is assumed
that $kT$ is much smaller than the single--particle level spacing
$\Delta$.  At a transmission peak two charging states of the dot
with $N$ and $N+1$ electron are degenerate.  Electrons pass through
a single level in the dot. The amplitude $t_{\rm QD}(\epsilon)$ can
be expressed in terms of the retarded Green function of this level,
\begin{equation}
t_{\rm QD}(\epsilon) = -i \sqrt{4 \Gamma_L \Gamma_R} \int dt e^{i
  \epsilon t} G_{\rm QD} (t) ,
\label{tQD}
\end{equation}
where $\Gamma_{L,R}$ are the partial width for decay of the level
through the left and right tunneling barrier, respectively (we put
$\hbar =1$).

Electrons in the QD interact with electrons in the QPC.  This
interaction modifies the local potential in the QPC. We use the
standard description of the QPC as a 1D system of noninteracting
electrons. The Hamiltonian $H_N$ of the QPC when exactly $N$
electrons occupy the QD is written in terms of exact scattering
states,
\begin{equation}
H_N = \int {d k \over 2 \pi} k [\psi_L^\dagger (k) \psi_L (k)
+ \psi_R^\dagger (k) \psi_R (k) ],
\label{HN}
\end{equation}
where $\psi_{L,R}$ are the destruction operators for left and right
moving scattering states, respectively. A summation over spin
indices is implied. The presence on the $N+1$ electron 
changes the QPC Hamiltonian to $H_{N+1} = H_N +V$, where
\begin{equation} 
V = V_{LL} +V_{RR} + V_{LR} ,
\label{V}
\end{equation}
with
\begin{eqnarray}
\label{VLL}
V_{LL (RR)} & = & \lambda \int {d k_1 d k_2 \over 2 \pi} \psi_{L
  (R)}^\dagger (k_1) \psi_{L (R)} (k_2) , \\
\label{VLR}
V_{LR} & = & \lambda_{LR} \int {d k_1 d k_2 \over 2 \pi} [ \psi_{L
  (R)}^\dagger (k_1) \psi_{L (R)} (k_2) + \mbox{H.c.}] . 
\end{eqnarray}
The wire with the QPC is connected to two reservoirs between
which the finite bias $e V_{d}$ is applied.

The Green function of the resonant level in the dot interacting with
the wire is of the form
\begin{equation}
G_{\rm QD} (t) = -i \Theta (t) e^{-i E_0 t -\Gamma t}
[P_{N+1} A_-(t) + P_N A_+ (t) ] ,
\label{GQD}
\end{equation}
where $E_0$ is the energy of the level in the absence of the QPC,
and $P_{N (N+1)}$ the probability of the charging state $N (N+1)$ of
the dot, $P_N + P_{N+1} = 1$. The total width is $\Gamma = \Gamma_L
+ \Gamma_R$. The coherence factors $A_{\pm}(t)$ describe the
response of the wire to the addition (removal) of an electron from
the dot,
\begin{eqnarray}
\label{A+}
A_+ (t) & = & \langle e^{iH_N t} e^{-iH_{N+1} t} \rangle_{H_N}
,  \\
\label{A-}
A_- (t) & = & \langle e^{iH_N t} e^{-iH_{N+1} t} \rangle_{H_{N+1}}
.  
\end{eqnarray}
The expectation values are taken with respect to an equilibrium
ensemble in the wire with the Hamiltonian $H_N$ and $H_{N+1}$,
respectively.

An interpretation of dephasing follows directly from these
expressions. They can be understood \cite{Ste90} as the scalar
product of two states of the environment (the QPC) at time $t$.
These states are identical at $t=0$, and thus $A_{\pm}(t=0) =1$.
Then, one state evolves with the Hamiltonian $H_N$, the other state
with $H_{N+1}$. Since the environment is not observed in the
experiment, its coordinate is being integrated over. Quantum
coherence is lost once the states of the environment have become
almost orthogonal.  Then the scalar product becomes almost zero and
the environment has identified the state of the QD.

The operators $V_{LL}$ and $V_{RR}$ mix scattering states that
propagate in the same direction. This only changes the phase of the
transmission amplitude through the QPC. The mixing of scattering
states moving in opposite directions, described by $V_{LR}$,
modifies the transmission coefficient. Note that only the latter
contribution is affected by the voltage drop in the wire. Hence, the
nonequilibrium (finite $V_d$) part of the dephasing rate is entirely
due to this latter contribution.

Aleiner et al.\ \cite{Ale97} obtained $A_{\pm}$ for arbitrary
$\lambda$ using known results for the orthogonality catastrophe,
i.e.\ the response of an equilibrium noninteracting electron system
to a sudden perturbation. The nonequilibrium dephasing can be
calculated in an expansion to second order in $\lambda_{LR}$. The
result has the form
\begin{equation}
A_+ (t) = \left( {i \pi kT \over \xi_0 \sinh \pi k T t} 
\right)^{\alpha + \gamma} e^{-t / t_d + \gamma h(t,kT,eV_d)}.
\label{A+2}
\end{equation}
Here, $\xi_0$ is an high--energy cutoff, the smaller of the Fermi
energy in the wire or the inverse rise time of the perturbation in
the wire. The definitions of the exponents $\alpha$, $\gamma$ and
the crossover function $h$ can be found in Ref.~\cite{Ale97}. It
suffices to note that $h(t)/t \to 0$ for large t. For a symmetric
dot, the dephasing rate is given by
\begin{equation}
{1 \over t_d} = {e V_d \over h} { (\Delta {\cal T}_d)^2 \over 4 {\cal
T}_d   (1-{\cal T}_d)} ,
\label{Gammd}
\end{equation}
which agrees up to a constant factor with the estimate in
Eq.~(\ref{decoh1}).

The calculation of the coherence factor $A_-(t)$ yields $A_-(t)=A_+
(t)^*$. At the transmission peak, $P_N =P_{N+1} = 1/2$, and the
transmission amplitude $t_{\rm QD}$ is the Fourier transform
(\ref{tQD}) of $G_{\rm QD} (t)$. The final result can be
approximated by \cite{Ale97},
\begin{equation}
t_{\rm QD} \simeq {2 \pi \sqrt{\Gamma_L \Gamma_R} \over 4 k T + \pi
  \Gamma_{\rm tot} } \left( kT + \Gamma_{\rm tot} \over \xi_0
\right)^\alpha \left( kT + \Gamma_{\rm tot} + eV_d \over \xi_0
\right)^\gamma ,
\label{tQDfin}
\end{equation}
where the total width $\Gamma_{\rm tot} = \Gamma + \Gamma_d$ is
given in terms of the intrinsic width $\Gamma$ and the width
$\Gamma_d = \hbar / t_d$ introduced by the dephasing.  For the
which--path experiment \cite{Buk98} the exponents $\alpha$, $\gamma$
appear to be rather small so that the last two terms on the right
hand side of Eq.~(\ref{tQDfin}) yield a factor $1$. Moreover,
$\Gamma, \Gamma_d \ll kT$, so that we can expand $t_{\rm QD}$ to
find
\begin{equation}
t_{\rm QD} \simeq {\pi \sqrt{\Gamma_L \Gamma_R} \over 2 kT } \left(
  1 - { e V_d \over k T} {(\Delta {\cal T}_d)^2 \over 32 {\cal T}_d (1-{\cal T}_d) 
} \right).
\label{tQDex}
\end{equation}
The term in brackets on the right--hand--side represents the
reduction of the visibility due to the interaction with the QPC.

\subsubsection{Influence functional}
\label{Sec6.2.3}
A conceptually different approach to the which--path interferometer
was presented by Levinson \cite{Lev97}. He studied an isolated
quantum dot (not connected to leads) with a single level at energy
$E_0$ and Hamiltonian $H_{\rm QD}= E_0 c^\dagger c$.  The QPC and
the interaction between QPC and QD are modeled precisely as in Sec.\ 
\ref{Sec6.2.2}, so that the total Hamiltonian takes the form
\begin{equation}
H = H_{\rm QD} + H_N + c^\dagger c V ,
\label{Htot}
\end{equation}
where $H_N$ and $V$ are defined in Eqs.~(\ref{HN}) and
(\ref{V})--(\ref{VLR}), respectively. Levinson investigated the
effect of the environment on the QD. The charge transmitted through
the QPC creates a fluctuating potential at the QD that modulates the
electron states in the QD. This modulation dephases the states in
the QD.

The coherence of the QD is quantitatively described by the
expectation value $\langle c(t) \rangle$ taken with respect to
the equilibrium ensemble with the Hamiltonian $H$. Here, $c(t) =
e^{i H t} c e^{-i H t}$ is the amplitude in the Heisenberg
representation. The QD has a coherent part if $\langle c(t)
\rangle \neq 0$, while the case $\langle c(t) \rangle = 0$ and
$\langle c^\dagger (t) c(t) \rangle \neq 0$ corresponds to a
totally incoherent quantum dot. The equation of motion for $c(t)$
reads
\begin{equation}
{d c(t) \over d t}  =i ( E_0 + V(t) ) c(t) ,
\label{ct1}
\end{equation}
where $V(t)= e^{i H t} V e^{-i H t}$. According to Eq.\ (\ref{ct1}),
$V(t)$ may be interpreted as a time dependent modulation of the
energy $E_0$.  Solving this equation and averaging, one finds
\begin{equation}
\langle c(t) \rangle = \langle c(0)  e^{-i E_0 t} T_t e^{-i
\int_0^t V(t_1) d t_1 } \rangle ,
\label{ct2}
\end{equation}
with the time ordering operator $T_t$. Note that the time ordered
exponential on the right hand side of this equation is identical
with the product $\exp(i H_N t) \exp(-i H_{N+1} t)$ that enters in
the coherence factors $A_{\pm}$ defined in the previous section.
This proves the equivalence of the present approach with the one
described in Sec.~\ref{Sec6.2.2}. An expectation value of the type
of that on the right hand side of Eq.~(\ref{ct2}) is known as the
Feynman--Vernon influence functional.

The average in Eq.~(\ref{ct2}) is calculated assuming that the
process described by $V(t)$ is a Gaussian process. This amounts to
the approximation $\langle c(t) \rangle = \langle c(0) \rangle $ $\times
 \exp
(-i E_0 t) \exp [ -(1/2) \Phi(t)]$, where $\Phi(t) = \int_0^t d
t^\prime \int_0^t d t^{\prime \prime} K(t^\prime-t^{\prime \prime})$
with
\begin{equation}
K(t) = {1 \over 2} [ \langle V(t) V(0) \rangle + \langle V(0) V(t)
\rangle ] .
\label{Kt}
\end{equation}
The correlator $K(t)$ decays in time on a characteristic time scale
$t_d$. For large $t \gg t_d$, one finds $\langle c(t) \rangle =
\langle c(0) \rangle \exp (-i E_0 t) \exp (-t/t_d)$ where the
dephasing rate is given by
\begin{equation}
{ 1 \over t_d} = {1 \over 2} \int_{-\infty}^\infty d t K(t) .
\label{rateLev}
\end{equation}
The dephasing rate may be found using known results for the current
correlation functions in a QPC. For a symmetric dot one finds the
results \cite{Lev97}
\begin{eqnarray}
\label{Lev1} 
1 / t_d & \simeq {\cal A} e V_d / \hbar & \mbox{ for high bias
$eV_d \gg kT$} , \\
\label{Lev2} 
1 / t_d & \hspace*{0.3cm} \simeq {\cal A} (e V_d)^2 / (\hbar kT)
\hspace*{0.3cm} & \mbox{ for low bias
$eV_d \ll kT$} ,
\end{eqnarray}
with ${\cal A} = (\Delta {\cal T}_d)^2 / [8 \pi {\cal T}_d (1-{\cal
  T}_d)]$. The result for the high bias limit is identical with
Eq.~(\ref{Gammd}) obtained by Aleiner et al.\ \cite{Ale97}.
    
\subsubsection{Master equation}
\label{Sec6.2.4}
Density matrix approaches to dephasing in coupled quantum dots
were devised by Gurvitz \cite{Gur97} and Hackenbroich et al.\ 
\cite{Hac98}. The approaches include the dephasing in the
which--path interferometers as a special case. Gurvitz' work is
reviewed in Sec.~\ref{Sec7}. The derivation given below is
similar in spirit to the approach of Ref.\ \cite{Hac98}.

We consider a quantum dot with a single energy level. The QD is
coupled to a QPC but otherwise isolated from the environment. For 
the Hilbert space of the QD we choose the basis $|a \rangle$, $|b
\rangle$, representing an empty and an occupied dot level,
respectively. We assume that the wire with the QPC supports only a
single transverse mode so that the scattering matrix $S_{\rm QPC}$
through the QPC is a $2 \times 2$ matrix that depends on the
occupation of the QD,
\begin{eqnarray}
\label{event4aa}
S_{\rm QPC} = \left\{ \begin{array}{ll} 
 S_a , & \mbox{if the QD is empty,} \\
 S_b , & \mbox{if the QD is occupied.} 
\end{array}
\right. 
\end{eqnarray}
The relations (\ref{event4aa}) may be combined to the two--particle
scattering matrix
\begin{eqnarray}
S_{\sigma \sigma^\prime} =
\delta_{\sigma \sigma^\prime} [ \delta_{\sigma a} S_a + 
\delta_{\sigma b} S_b ] ,
\label{event4ab}
\end{eqnarray}
where both $\sigma$ and $\sigma^\prime$ can take the values $a,b$. 
Let $\rho_{\rm tot}= \rho \otimes \rho_{\rm QPC}$ be the density
matrix of the total system prior to the passage of an electron
through the QPC.  We choose $\rho_{\rm QPC} = {\rm diag}[1,0]$
representing an incoming particle from one side of the QPC. The
density matrix after scattering is $\rho_{\rm tot}^\prime =
S \rho_{\rm tot} S^\dagger $. The reduced density matrix
$\rho^\prime$ of the QD is obtained by tracing over the QPC variables
\begin{equation}
\rho^\prime = {\rm Tr}_{\rm QPC} [ S \rho_{\rm tot} S^\dagger] .
\label{event4a}
\end{equation}
In the basis of incoming and outgoing scattering states we can
parametrize the scattering matrix
\begin{equation}
S_a = \left( \begin{array}{cc} 
r_a & t_a^\prime \\ t_a  & r_a^\prime
\end{array} \right) =  \left( \begin{array}{cc} 
\cos \theta_a \exp[i \phi_{r a}]& i \sin \theta_a \exp[i 
\phi_{t^\prime a}]
\\ i \sin \theta_a \exp[i \phi_{t a}] & \cos \theta_a \exp[i
  \phi_{r^\prime a}]
\end{array} \right).
\label{event4}
\end{equation}
in terms of one angle $\theta_a$ and four phases $\phi_a$. As a
consequence of the unitarity of $S_a$, the phases are not independent
from each other but obey the relation $\exp[i(\phi_{r a}-\phi_{t
a})]= -\exp[-i( \phi_{r^\prime a}-\phi_{t^\prime a})]$. A
parameterization similar to Eq.\ (\ref{event4}) but in terms of an
angle $\theta_b$ and phases $\phi_b$ may be used for the scattering
matrix $S_b$.  Combining Eqs. (\ref{event4ab}), (\ref{event4}) and
substituting the result into Eq.\ (\ref{event4a}) one finds that the
diagonal elements of $\rho$ do not change upon scattering through the
QPC. The off-diagonal elements change according to
\begin{eqnarray}
\rho_{ab}^\prime & = &  {\cal P} \rho_{ab} ,
\end{eqnarray}
where
\begin{eqnarray}
{\cal P} = r_a^* r_b + t_a^* t_b .
\end{eqnarray}
Note that $\rho_{ba}^\prime = \rho_{ab}^{\prime *}$ since
$\rho^\prime$ is hermitian.

We will assume that the scattering through the QPC takes place on a
time scale much shorter than the relevant time scales of the quantum
dot.  In particular, the scattering time shall be much shorter than
the decoherence time $t_d$. This assumption allows us to use the
Markov approximation and neglect memory effects in the QPC.
Then the density operator after the scattering of $n$ electrons
through the QPC is given by
\begin{eqnarray}
\rho_{ab}(t) = {\cal P}^n e^{i \epsilon n \Delta t} \rho_{ab}(0),
\label{event11}
\end{eqnarray}
where $\Delta t = h /(2 e V_d) $ is the average time between two
scattering events, $t = n \Delta t$, and $E_{a,b}$ are the energies
of the empty and occupied dot level, respectively. The term
involving $\epsilon = (E_b-E_a)/\hbar$ accounts for the free
evolution of the QD. Note that ${\cal P}$ is generally a complex
number. From its modulus we can read off the dephasing rate
\begin{eqnarray}
{1 \over t_d} = -{1 \over \Delta t} \ln |{\cal P} | ,
\label{event12}
\end{eqnarray}
while the phase gives rise to the energy shift
\begin{eqnarray}
\Delta E = { \hbar \over \Delta t} {\rm arg} {\cal P} .
\label{event13}
\end{eqnarray}
Both the dephasing rate and the energy shift can be obtained
explicitly in the weak--coupling limit ${\cal P} \approx 1$. Then
${\cal P}$ may be expanded in $\Delta \theta = \theta_b -\theta_a$,
$\Delta \phi_r = \phi_{rb}-\phi_{ra}$, and $\Delta \phi_t =
\phi_{tb}-\phi_{ta}$. Using the parameterization (\ref{event4}), we
can express $\Delta \theta$ in terms of the transmission
coefficients ${\cal T}_{b,a} = |t_{b,a}|^2$. This yields $(\Delta
\theta)^2 = (\Delta {\cal T}_d)^2 / [4 {\cal T}_d (1-{\cal T}_d)]$
where $\Delta {\cal T}_d = {\cal T}_b - {\cal T}_a$ and ${\cal T}_d
= ({\cal T}_b + {\cal T}_a)/2$.  Substitution into Eqs.\ 
(\ref{event12}), (\ref{event13}) yields the dephasing rate 
\begin{eqnarray}
\label{master4} {1 \over t_d}  =  { e V_d \over h}  
{ (\Delta {\cal T}_d)^2 \over 4 {\cal T}_d (1-{\cal T}_d)} + 
{ e V_d \over h} {\cal
    T}_d (1-{\cal T}_d)  (\Delta \phi)^2  , 
\end{eqnarray}
and the energy shift
\begin{eqnarray}
\Delta E  =  { e V_d \over \pi}  (1-{\cal T}_d)  \Delta
  \phi_r + { e V_d \over \pi} {\cal T}_d  \Delta \phi_t ,
\end{eqnarray}
where $\Delta \phi = \Delta \phi_r -\Delta \phi_t$.  The evolution of
$\rho$ in the weak--coupling case may be cast in the form of a
master equation
\begin{equation} 
{d \rho \over d t} = \left[ -{1 \over t_d} + i {\Delta E \over \hbar}
\right]  \left( \begin{array}{cc} 0 & \rho_{ab} \\ 
      \rho_{ba} & 0
\end{array} \right) .
\label{master5}
\end{equation}
The diagonal elements of $\rho$ do not change in time.  Thus the
occupation probability of the quantum dot does not change in time.
By contrast, the off--diagonal elements decay with the dephasing
rate $1/t_d$. The evolution of $\rho$ thus indeed reflects dephasing
(decoherence) and not energy relaxation.

The dephasing rate (\ref{master4}) reduces to the results derived in
Secs.\ \ref{Sec6.2.2}, \ref{Sec6.2.3} when the phase change $\Delta
\phi$ is negligible. The effect of non--zero $\Delta \phi$ has been
investigated in a paper by Stodolsky \cite{Sto98}. His result for
the dephasing rate ($\propto 1-{\rm Re} {\cal P}$) differs from Eq.\ 
(\ref{event12}) and does not reproduce the dependence $1/ t_d \propto
{\cal T}_d (1-{\cal T}_d)$ observed experimentally \cite{Spr00}.

\subsubsection{Charge relaxation}
\label{Sec6.2.5}
An new approach to the dephasing problem was developed by B\"uttiker
and coworkers \cite{Bue93,Ped98,Bue99,Bue99a}. The authors consider
the charge and potential fluctuations that result from the Coulomb
coupling between mesoscopic conductors. These fluctuations govern
the dephasing process and thus the dephasing rate. Based on earlier
theoretical work on charge relaxation in interacting conductors
\cite{Bue93,Ped98}, B\"uttiker and Martin \cite{Bue99,Bue99a}
calculated the dephasing rate for a QD coupled to a QPC. The method
applies to dephasing both from changes in the transmission
probability and from changes in the transmission phase. The results
obtained in both cases coincide with the formulae given earlier.
B\"uttiker and Martin generalized their results to the case of more
than one channel in the QPC and to a device that includes a phase
randomizing voltage probe between the QPC and the QD (cf.\ the
experiment of Ref.\ \cite{Spr00}). It was found that the dephasing
rate is unaffected by the voltage probe if there is only one channel
in the QPC.  
\newpage

\newpage
\setcounter{equation}{0}
\section{Quantum Zeno effect}
\label{Sec7}
The frequent repetition of a decohering measurement leads to a
striking phenomenon known as the quantum Zeno effect
\cite{Mis77,Mer98}: The suppression of transitions between quantum
states. The standard example is a two--level system with a tunneling
transition between the two levels. Assume that the system at $t=0$
is in prepared in the state $a$. For small times $t$, the
probability to tunnel out of that state is $P_{a \to b}(t)= {1 \over
  \hbar^2} | \langle b| \Omega_0 |a \rangle |^2 t^2$, where
$\Omega_0 /2$ is the tunneling matrix element, and $\omega_0 =
\Omega_0/\hbar$ the tunneling frequency. However, if the interval
$t$ is split into $N$ subintervals each with the length $t/N$, and
the system is measured at the end of each subinterval, the
probability for tunneling is reduced to $N P_{a \to b}(t/N) = (1/N)
P_{a \to b}(t)$. In the limit of arbitrary dense measurements, $N
\to \infty$, the system is frozen in the initial state. Repeated 
measurements can thus completely hinder the natural evolution of a
quantum system. 

\begin{figure}
\centerline{\psfig{file=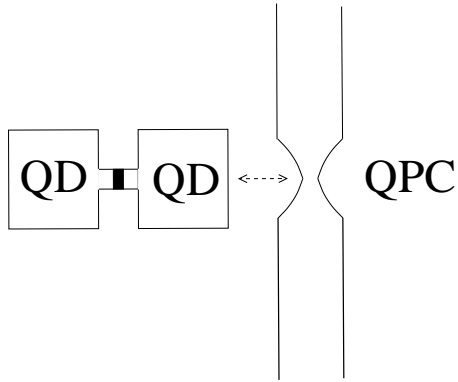,width=6cm,angle=0}}
\vspace*{0.5cm}
\caption[]{Mesoscopic device with two quantum dots and a QPC. 
  Electric charge can tunnel coherently between the dots. The QPC
  measures the charge accumulated in one of the dots.}
\label{isodots}
\end{figure}

Despite considerable theoretical work on the quantum Zeno effect there
is only little experimental proof for it. An experimental test using
an induced hyperfine transition of Be ions \cite{Ita90} has been
reported.  Experiments on optical transitions \cite{Gag92} or atomic
Bragg scattering \cite{Dyr97} have been proposed.  Further
experimental evidence for the existence of the quantum Zeno effect is
clearly desirable. Gurvitz \cite{Gur97} first pointed out that the
quantum Zeno effect may be observed in semiconductor microstructures.
He studied theoretically the tunneling of electric charge between two
weakly coupled quantum dots. A QPC located in the vicinity of one of
the dots served as a non--invasive detector for the charge on that dot
(see Fig.\ \ref{isodots}).  As expected from the quantum Zeno effect,
Gurvitz found that the coupling to the QPC suppresses the tunneling
oscillations between the two dots. New aspects of the problem were
identified by Hackenbroich et al.\ \cite{Hac98}. It was found that an
ac voltage in the QPC leads to parametric resonance and to a strong
reduction of dephasing in the quantum dots. The resonance occurs when
the frequency $\omega$ of the ac signal equals twice the frequency
$\omega_0$ of the internal charge oscillations in the double--dot
system.  The spectral density of the detector current was studied by
Korotkov \cite{Kor99,Kor00}.  When the detector is weakly coupled to
the dots the spectral density has a peak close to the frequency
$\omega_0$ reflecting the oscillation of charge between the two
quantum dots. For strong coupling the peak disappears and the spectral
density develops a Lorentzian shape. In this regime charge transfer
takes place via random jumps rather than by periodic oscillations.

In Sec.~\ref{Sec7.1} we investigate electron tunneling between two
quantum dots in the presence of a QPC detector. The problem is studied
in terms of the reduced density matrix of the coupled dots. The
spectral density of the detector current is analyzed in
Sec.~\ref{Sec7.2}. In Sec.~\ref{Sec7.3} we modify the system by
connecting the dots to external electron reservoirs. We show that
the dc current through the dots is affected by the detector and
provides an indirect signature of the quantum Zeno effect. In Sec.\
\ref{Sec7.4} we analyze the decay of an unstable quantum system, and
investigate whether continuous measurements slow down its decay rate. 

\subsection{Charge oscillations}
\label{Sec7.1}

\subsubsection{Damping}
\label{Sec7.1.1}
Consider electric charge oscillating between two quantum dots in the
presence of a QPC in close vicinity (Fig.~\ref{isodots}). The
coupled dots may be viewed as the mesoscopic realization of a
double--well potential.  We assume that there is only one energy level
in each dot and that the dots are occupied with a single electron. Let
the states $| a \rangle$ and $|b \rangle$ with energies $E_a$ and
$E_b$ represent the electron in the left and right dot, respectively.
We first study the case that the interaction with the QPC is
negligibly small. Then the dynamics of the two--dot system is governed
by the tunneling Hamiltonian $(\Omega_0 /2) ( | a \rangle \langle b| +
| b \rangle \langle a|)$.  The elements $\rho_{ij}$ of the density
matrix satisfy the Bloch equations,
\begin{eqnarray}
\label{Bloch1}
\dot{\rho}_{aa} & = & i {\omega_0 \over 2} ( \rho_{ab}-\rho_{ba} ) ,
\\  \label{Bloch2}
\dot{\rho}_{bb} & = & i {\omega_0 \over 2} ( \rho_{ba}-\rho_{ab} ) ,
\\  \label{Bloch3}
\dot{\rho}_{ab} & = & i \epsilon \rho_{ab} + i { \omega_0 \over 2} (
\rho_{aa}-\rho_{bb} ) ,
\end{eqnarray}
with the frequencies $\epsilon = (E_b-E_a)/ \hbar$ and $\omega_0 =
\Omega_0 / \hbar$. The diagonal elements $\rho_{aa}(t)$ and
$\rho_{bb}(t)$, respectively, are the probabilities to find the
electron in the left and right dot. The off--diagonal elements
satisfy $\rho_{ab}(t)=\rho_{ba}^*(t)$ since $\rho$ is hermitian.
Solving the Bloch equations for the initial conditions $\rho= {\rm
  diag}[1,0]$ one finds
\begin{equation}
\label{rhoaa}
\rho_{aa}(t) = { \omega_0^2 \cos^2 (\omega t /2) + \epsilon^2 \over
\omega_0^2 + \epsilon^2 } ,
\end{equation}
where $\omega = \sqrt{ \omega_0^2 + \epsilon^2}$. An electron
located initially in the left dot oscillates between the dots with
the frequency $\omega$. The amplitude of these oscillations is
$\omega_0^2 /(\omega_0^2 + \epsilon^2)$. Thus, if $\epsilon \gg
\omega_0$, the electron essentially remains localized in the left dot.

Now consider the charge oscillations when the QPC detector is
active. The transmission through the QPC depends on the electron
position in the dot. We first assume that the electron position only
affects the transmission probability and not the transmission phase.
Let ${\cal T}_a$ (${\cal T}_b$) be the transmission probability if
the electron is in the left (right) dot.  Bloch equations for this
case were derived by Gurvitz \cite{Gur97}.  The resulting equations
for the diagonal elements coincide with Eqs.\ (\ref{Bloch1}),
(\ref{Bloch2}). Equation (\ref{Bloch3}) is replaced by
\begin{equation}
\label{Bloch4}
\dot{\rho}_{ab} = i \epsilon \rho_{ab} + i {\omega_0 \over 2} (
\rho_{aa}-\rho_{bb} ) - 2 \kappa_d \rho_{ab} ,
\end{equation}
with the dephasing rate $\kappa_d = eV_d (\Delta {\cal T}_d)^2/[8 h
{\cal T}_d]$. Dephasing causes an exponential damping of the
nondiagonal matrix element. Combination of Eqs.~(\ref{Bloch1}),
(\ref{Bloch2}), and (\ref{Bloch4}) determines the time evolution of
$\rho$. For weak damping $\omega_0 > \kappa_d$ and $\epsilon=0$, one
finds damped oscillations
\begin{eqnarray}
\rho_{aa}(t) -1/2 \sim \exp[- \kappa_d t ] \cos \sqrt{\omega_0^2 -
  \kappa_d^2} t .
\end{eqnarray}
Note that the damping causes a redshift of the frequency $\sqrt
{\omega_0^2 - \kappa_d^2}$ and, hence, slows down the charge
oscillations between the two dots. This suppression of transitions
due to measuring with the QPC is a manifestation of the quantum Zeno
effect. The off--diagonal elements of $\rho$ vanish for large times
in agreement with the standard picture of decoherence, and the
density matrix reduces to the random statistical ensemble $\rho
\rightarrow {\rm diag}[1/2,1/2]$.

Gurvitz' derivation of the rate equation implicitly assumes ${\cal
  T}_d \ll 1$. The generalization to arbitrary value $0< {\cal
  T}_d<1$ was obtained by Hackenbroich et al.\ \cite{Hac98} using a
two-particle scattering approach similar to the calculation of Sec.\
\ref{Sec6.2.4}. The resulting Master equation is identical with 
Eqs.~(\ref{Bloch1}), (\ref{Bloch2}), (\ref{Bloch4}) but yields the
dephasing rate
\begin{equation}
\label{master2}
\kappa_d = { eV_d \over h} { (\Delta {\cal T}_d)^2 \over 8 
{\cal T}_d (1 - {\cal T}_d)}
\end{equation}
that differs from Gurvitz' result by a factor $1/(1-{\cal T}_d)$.
The result (\ref{master2}) holds provided the quantum dots only affect
the transmission probability through the QPC. The result may easily be
generalized to include a change in the transmission phase through the
detector.  One then finds $\kappa_d = 1/(2 t_d)$ where $1/t_d$ is the
dephasing rate (\ref{master4}) of a single quantum dot computed in
Sec.\ \ref{Sec6.2}. [We note that the problem considered here reduces
to the dephasing in a single quantum dot in the limiting case
$\omega_0=0$.]

\begin{figure}
\hspace*{0cm}
\centerline{\psfig{file=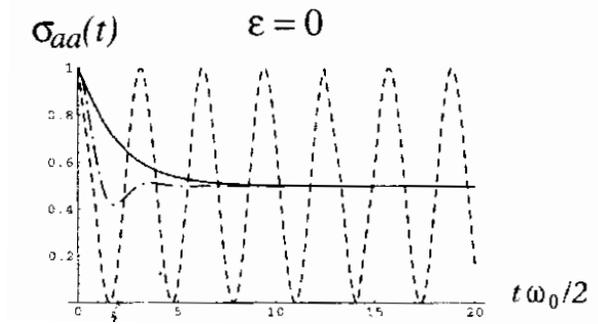,width=8cm,angle=-0.7}}
\caption[]{The occupation probability of one quantum dot as a
  function of time. The curves correspond to different values of the
  dephasing rate: $\kappa_d$ (dashed), $\kappa_d = 4 \omega_0$
  (dot--dashed), and $\kappa_d=16 \omega_0$ (solid). The figure is
  obtained for $\epsilon =0$. Taken from
  Ref.~\protect{\cite{Gur97}}.} \label{gur}
\end{figure}

\subsubsection{Parametric resonance}
\label{Sec7.1.2}
\begin{figure}
\centerline{\psfig{file=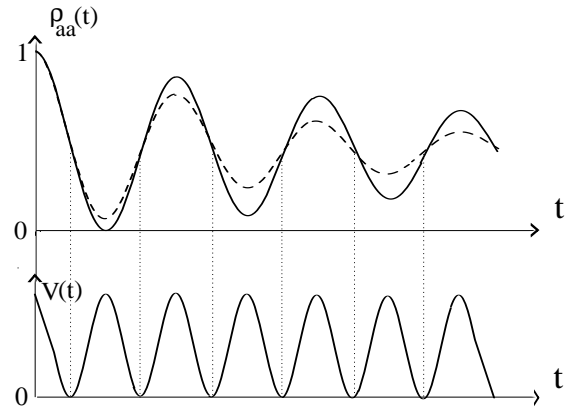,width=7.6cm,angle=0}}
\caption[]{Parametric resonance in the double-dot system coupled to a
  QPC--detector. The upper part shows the oscillations of $\rho_{aa}$
  for constant voltage (dashed curve) and for a time-dependent
  voltage as shown in the lower part (full curve).  Damping of the
  oscillations is reduced by a factor two. Taken from
  Ref.~\protect{\cite{Hac98}}.}
\label{resonance}
\end{figure}

The charge oscillations between the two quantum dots may display a
parametric resonance if the voltage drop $V_d$ has an ac--component
\cite{Hac98}. Consider $V_d(t) = V_0-V_1 \sin \omega_1 t$ with
$V_0,V_1 \ge 0$ and $V_1 \le V_0$.  Substitution into the equation of
motion for $\rho_{aa}$ yields a damped harmonic oscillation with an
{\em oscillatory} damping constant. Using a simple ansatz for
$\rho_{aa}(t)$ and neglecting terms of order ${\cal O} (V_d^2)$, one
is led to an equation of the Mathieu type which is known \cite{Lan69}
to display parametric resonance close to the frequencies $\omega_1 = 2
\omega /n$ where n is a positive integer.  Parametric resonance is
most pronounced for $\omega_1 \approx 2 \omega$. The damping near the
resonance is strongly reduced, $\kappa_d = (\Delta {\cal T}_d)^2 /[8 h
{\cal T}_d (1-{\cal T}_d)] e (V_0-V_1/2)$. The resulting time
evolution of $\rho_{aa}$ near resonance is illustrated in
Fig.~\ref{resonance} and compared with the case where $V_d$ is
time--independent. 

\subsection{Detector current}
\label{Sec7.2}
While the reduced density matrix $\rho$ of the two quantum dots is a
convenient tool for the theoretical investigation of the quantum
Zeno effect, one cannot directly measure $\rho$ in experiments.
However, information about the dynamics in the quantum dots may be
obtained from the current passing through the detector. The power
spectrum of the detector current was studied by Hackenbroich et al.\ 
\cite{Hac98} and by Korotkov \cite{Kor99,Kor00}.

Korotkov's approach \cite{Kor99,Kor00} is based on the description
of decoherence in terms of a stochastic wave function. This approach
is well-known in quantum optics \cite{Ple98}. Variants of the
approach are known as quantum trajectory or quantum jump approach.
The method describes the evolution of the quantum system under
investigation (the two quantum dots) conditioned on a particular
measurement of the detector system (the QPC). For the conditional
evolution of the quantum dot density matrix given the detector
current $I(t)$ one finds \cite{Kor99,Kor00}
\begin{eqnarray}
\label{Kor1}
\dot{\rho}_{aa} & = & i {\omega_0 \over 2} (\rho_{ab} - \rho_{ba})
 - {2 \Delta I \over
  S_0} \rho_{aa} \rho_{bb} [I(t)-I_0] , \\
\label{Kor2}
\dot{\rho}_{ab} & = & i \epsilon \rho_{ab} + 
i {\omega_0 \over 2} (\rho_{aa} - \rho_{bb})  
 +{\Delta I \over S_0} (\rho_{aa}-\rho_{bb})[I(t)-I_0] 
\rho_{ab} - \gamma \rho_{ab} ,
\end{eqnarray}
where $S_0$ is the low--frequency spectral power of the detector
shot noise, and $I_0 = (I_a + I_b)/2$, $\Delta I = I_b -I_a$ where
$I_a$ and $I_b$ are the average currents when the electron localized
on the left and right dot, respectively. The damping rate $\gamma$
accounts for dephasing due to interaction with some environment not
included in the detector. Eqs.\ (\ref{Kor1}), (\ref{Kor2}) are
supplemented by the equation
\begin{eqnarray}
I(t)-I_0 = \Delta I (\rho_{aa}-\rho_{bb}) /2 + \xi(t)
\label{Kor3}
\end{eqnarray}
for the detector current. Here, $\xi(t)$ describes a Gaussian random
process with zero average and spectral density $S_0$. Equations
(\ref{Kor1})-(\ref{Kor3}) describe the evolution of the system for a
single realization of the detector current. Averaging over $\xi$
reduces these equations to the deterministic Master equation for the
evolution of the ensemble average derived in Sec.\ \ref{Sec7.1}. 

\begin{figure}
\centerline{\psfig{file=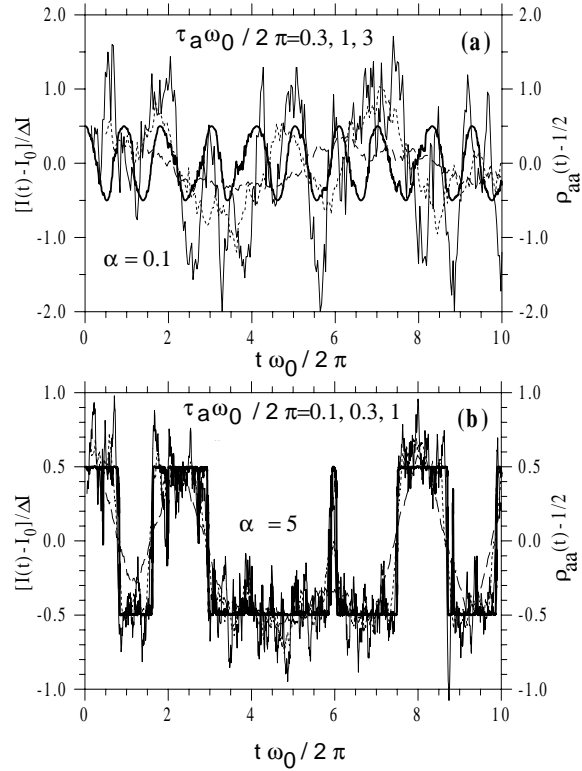,width=7.6cm,angle=0}}
\caption[]{Evolution of $\rho_{aa}$ (thick line) and the
  corresponding detector current $I(t)$ (thin slid, dotted, dashed
  line) averaged using rectangular time windows of size $\tau_a$.
  (a) Weak--coupling $\alpha=0.1$, (b) strong--coupling $\alpha=5.0$. 
  Taken from Ref.~\protect{\cite{Kor00}}.}
\label{figKor1}
\end{figure}

\begin{figure}
\centerline{\psfig{file=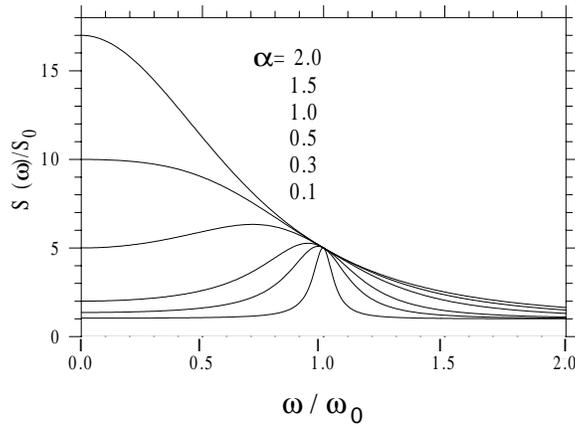,width=7.6cm,angle=0}}
\caption[]{The detector current spectral density $S(\omega)$ for
  different coupling strength $\alpha$. Taken from
  Ref.~\protect{\cite{Kor00}}.}
\label{figKor2}
\end{figure}

The evolution of $I(t)$ and $\rho_{aa}$ are illustrated in Fig.\ 
\ref{figKor1} for two different values of the coupling $\alpha
\equiv \kappa_d / \omega_0$. Note that the results hold for a
particular realization of the random process $\xi(t)$. Figure
\ref{figKor1}(a) is obtained for $\alpha=0.1$ representing a weakly
coupled detector. Charge can oscillate several times between the two
dots while generating only weak modulations of the detector current.
Superimposed on the modulations is the detector shot noise. Fig.\ 
\ref{figKor1}(a) shows the detector current averaged over different
time intervals. For small intervals the signal is noisy, while for
long intervals individual oscillations cannot be resolved. As a
result, charge oscillations are difficult to observe in the
weak--coupling limit.  The situation is different in the strong
coupling case $\alpha >1$ (Fig.\ \ref{figKor1}(b)). The strong
influence of the detector suppresses the quantum oscillations, so
that the charge performs random jumps between two localized states.
In this case, the properly averaged detector current follows the
evolution of the quantum dots.

The detector current may be further characterized by its spectral
power. For a symmetric double--dot ($\epsilon=0$), the spectral
power can be obtained analytically \cite{Kor00} yielding
\begin{eqnarray}
S(\omega) = S_0 + 4 S_0 { \alpha^2 \omega_0^4 \over (\omega^2
  -\omega_0^2)^2 + \alpha^2 \omega^2 \omega_0^2} .
\label{Kor4}
\end{eqnarray}
Figure \ref{figKor2} shows $S(\omega)$ for different values of the
coupling strength $\alpha$. For weak coupling, the spectral power
displays a peak close to the frequency $\omega_0$. The peak height
equals four times the noise background. The full width at half
maximum is given by $2 \kappa_d$ and the peak is centered at the
shifted frequency $\sqrt{\omega_0^2 -\kappa_d^2}$ \cite{Hac98}. The
frequency shift with respect to the frequency $\omega_0$ is a
consequence of the quantum Zeno effect. The peak in $S(\omega)$
gradually disappears with increasing coupling. In the strong
coupling limit, $S(\omega)$ has a Lorentzian shape known from the
classical theory of telegraph noise \cite{Kor00}.

\subsection{Current through the quantum dots}
\label{Sec7.3}
Due to high--frequency noise, a frequency resolved measurement of
the QPC current is difficult to perform. To avoid the problems of
high--frequency measurements several authors \cite{Gur97,Hac98}
investigated manifestations of the quantum Zeno effect in dc
transport. The dc current through two quantum dots with a QPC in the
vicinity was studied both for the dots in series and with the dots in
a parallel circuit. The current displays a characteristic dependence
on the detector efficiency which provides an indirect signature of the
quantum Zeno effect.

\subsubsection{Dots in a series}
\label{Sec7.3.1}
Consider the transmission through a double--well structure consisting
of two coupled quantum dot in a series. The dots are modeled as in
Sec.~\ref{Sec7.1} in terms of two levels with energies $E_a$, $E_b$,
and coupling matrix element $\Omega_0 /2$. Dot $a$ is coupled to an
electron reservoir $a$ and dot $b$ to a reservoir $b$.  The coupling
strength is described by the partial widths $\Gamma_{a(b)}$ for the
decay of level $a(b)$ into the reservoir $a(b)$. The Fermi energies in
the reservoirs are chosen such that $E_F^b \ll E_a, E_b \ll E_F^a$.
The current $I_s$ through the double--dot system can be calculated 
\cite{Gur97} using rate equations and is given by
\begin{equation}
\label{Is}
I_s ={e \over \hbar} { (\Gamma_b + \hbar \kappa_d ) \Omega_0^2 
\over 4 (E_a-E_b)^2
  + (\Gamma_b +\hbar \kappa_d)^2 + \Omega_0^2 (\Gamma_b + \hbar 
  \kappa_d) \left( {2 \over \Gamma_b} + {1 \over \Gamma_a} \right)}
.
\end{equation}
For small energy difference $|E_a-E_b|$ the current decreases with
$\kappa_d$. However, for large $|E_a-E_b|$ the current increases
with $\kappa_d$. This may be understood \cite{Gur97} as resulting
from the delocalization of the electron due to a continuous
measurement of the charge in one of the dots..

\subsubsection{Parallel quantum dots}
\label{Sec7.3.2}
\begin{figure}
\centerline{\psfig{file=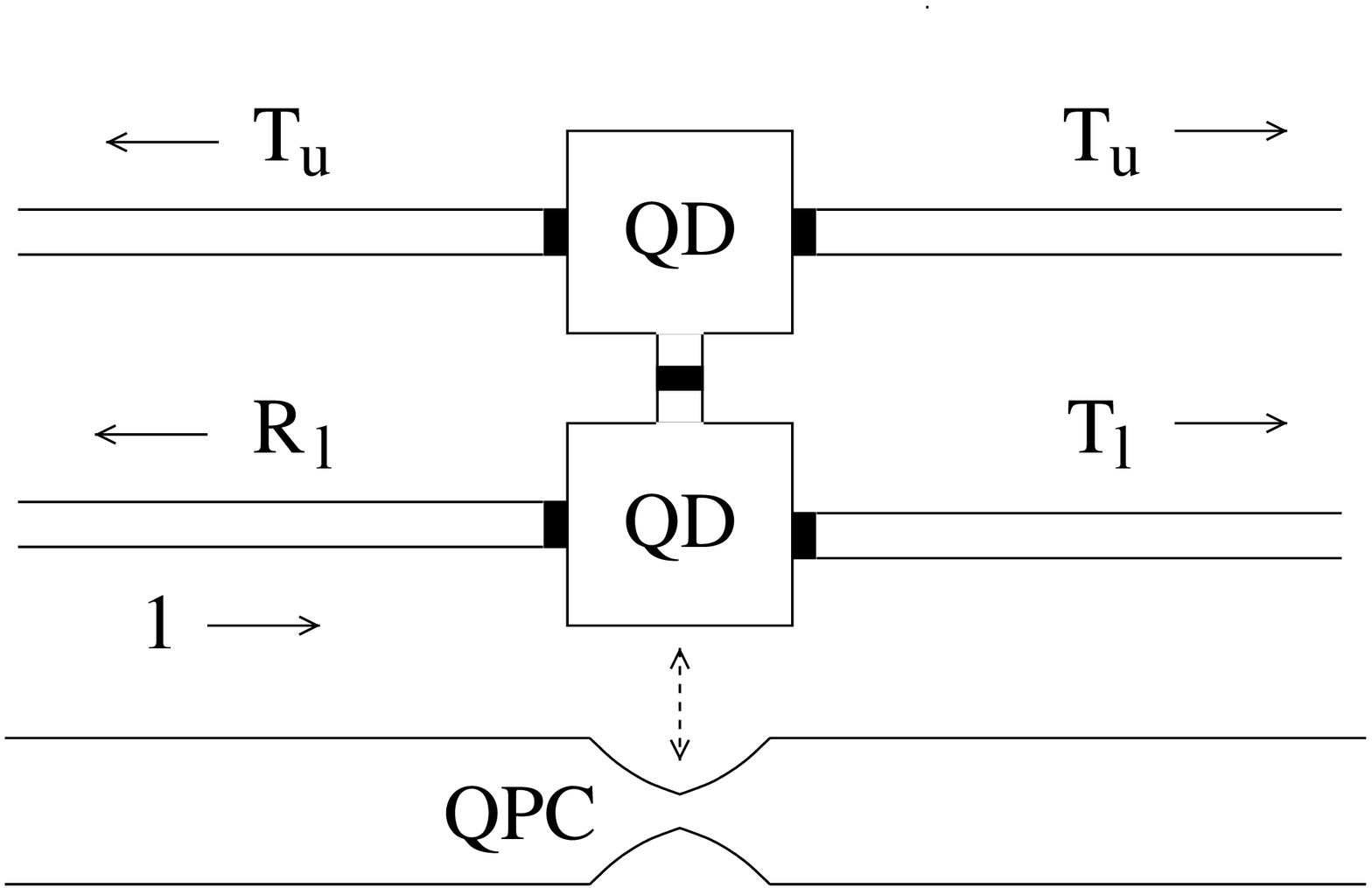,width=7cm,angle=0}}
\vspace*{0.5cm}
\caption[]{Two coupled quantum dots form a double well potential 
  and interact with a QPC measuring the position of an excess
  electron in the two dots. A current is driven through the dots via
  four external leads. The transmission and reflection coefficients
  of the leads are denoted by $T_u$, $T_l$, $R_{l}$.}
\label{newddot}
\end{figure}

We now discuss the quantum Zeno effect in a set--up for dc transport
across two parallel quantum dots. The arrangement, proposed by
Hackenbroich et al.\ \cite{Hac98}, is shown in Fig.~\ref{newddot}.
Two dots are coupled via a tunneling barrier.  Each dot is connected
with two external leads that allow for the transport of current to
and out of the dot.  With current flowing into the lower dot, we
calculate the branching ratio of the current transmitted through the
upper and the lower dot, respectively. 

We assume that both dots are in the Coulomb blockade regime, and
that $k T$ is much smaller than the single--particle level spacing
in the dots.  Then we need to consider only a single energy level in
each dot. Both levels are degenerate with energy $E_0 - i/2 \Gamma$.
The width $\Gamma$ is due to the coupling to the leads. A QPC
measures the charge in the lower quantum dot. The measurement
modifies the transmission through the two--dot system. This
modification is found by calculating the two--particle scattering
matrix for an electron passing through the QPC and another
electron oscillating in the two--dot system. 

We model the QPC in terms of plane waves with energy $\epsilon_k$,
mean density $\rho_F$, and Hamiltonian
\begin{eqnarray}  H_{\rm QPC}+ V \! = \sum_{k} \epsilon^{}_k 
b_k^\dagger 
b_{k}^{} + \! \sum_{k,k^\prime} (W_{k k^\prime}
+  V_{k k^\prime} 
d^\dagger_l d^{}_l) b_k^\dagger b_{k^\prime}^{}  \ .
\label{Ham}
\end{eqnarray}
Here, $V$ describes the interaction between the QPC and the lower QD.
The potential $W_{k k^\prime}$ mimics the constriction in the wire
with the QPC while $d^{\dagger}_l$ ($b_k^{\dagger}$) denotes the
creation operator for the state on the lower QD (for the QPC plane
wave states).

In each of the leads, we consider only one transverse channel
labeled by $c={\sigma,\mu}$, with $\sigma = l,u$ denoting the lower
(l) and upper (u) lead, and $\mu=+,-$ the left and right lead. The
two--particle scattering matrix can be obtained from the
Lippmann--Schwinger equation yielding the scattering amplitudes
\begin{equation}
  S_{c c^\prime,k k^\prime} = 
  \delta_{c c^\prime} \delta_{k k^\prime}-
  2 \pi i \gamma_{c} \gamma_{c^\prime} G_{\sigma \sigma^\prime, k k^\prime} .
\label{scatt2}
\end{equation}
Here, $\gamma_{c}$ is the matrix element for tunneling from channel
$c$ to the adjacent dot. The two--particle Green function $G$ for the
joint transition between dots and QPC is given in terms of its inverse
\begin{equation} 
\label{Green4}
(G^{-1})_{\sigma \sigma^\prime \!,k k^\prime \!} \!=\!  \left(\!
  \begin{array}{cc} \!\!G_0^{-1} \delta_{k k^\prime}\!-\!W_{k
      k^\prime}\!&  \Omega_0 /2\; \delta_{k k^\prime}\\ 
\Omega_0/ 2\; \delta_{k k^\prime} & \!\!G_0^{-1} \delta_{k
k^\prime}\!-\!W_{k k^\prime}\!-\!V_{k k^\prime} \end{array}\! \! 
\right)\!.
\end{equation}
On the right hand side, we have explicitly displayed the matrix in
$\sigma$--space. Here $G_0^{-1}= E-E_0 -\epsilon_k + i \Gamma/2$ is
the inverse propagator of the single Breit--Wigner resonance, and $E$
is the sum energy of the two incoming particles. We have assumed that
all $\gamma_{c} \equiv \gamma$ are identical and used the relation
$\Gamma =4 \pi \gamma \gamma^*$. For $V = 0$, we can diagonalize the
matrix $\epsilon_k \delta_{k k^{\prime}} + W_{k k^{\prime}}$ by a
unitary transformation in $k$--space, and $G$ reduces to 
the product of the unit matrix in $k$--space and the two $k$--dependent
coupled Breit--Wigner resonances for the double--dot system. Then, all
scattering processes are elastic, and the branching ratio for the
transmission through the upper and lower lead is ${\cal T}^{(0)}_u
/{\cal T}^{(0)}_l= \Omega_0^2 / \Gamma^2$.

The full two--particle scattering matrix (\ref{scatt2}) allows for
{\em energy exchange} between dots and detector: In contrast to the
sum energy of the two incoming particles, the energy of electrons in
the QPC is not conserved in the scattering process. Such inelastic
processes are essential to ensure the unitarity of the S--matrix.
Physically, the energy exchange allows for a position measurement of
the dot electron without violation of the Heisenberg uncertainty
relation.

To calculate the transmission and reflection coefficients through the
double--dot system, we restrict ourselves to constant
scattering potentials $W_{k k^\prime} \equiv W$ and $V_{k k^\prime}
\equiv V$. We expand $G$ in Eq.~(\ref{Green4}) in powers of $V$ and
resum the resulting series. We obtain two contributions to $G$. The
first is independent of $V$ and describes independent elastic
scattering through the QPC and the dots. This term obviously does not 
contribute to the quantum Zeno effect. The second contribution ${\tilde
G}$ involves energy exchange $\Omega=\epsilon_k -\epsilon_{k^\prime}$
between dots and QPC. For fixed incident energy $E=E_0 + \epsilon_k$,
\begin{eqnarray} {\tilde G}_{\sigma
\sigma^\prime,k k^\prime}  \! & =\! & A(\Omega)  
\left( \begin{array}{cc} \Omega_0^2 &-i  \Omega_0 \Gamma \\
\!\!-\Omega_0 (2\Omega + i \Gamma) & i \Gamma (2 \Omega+i\Gamma )\!\!
\end{array} \right) \ ,
\label{inter}
\end{eqnarray}
with the amplitude
\begin{equation}
A(\Omega)= -{  4V \over  F_W F_{W+V} (\Gamma^2 + \Omega_0^2) 
( (2\Omega + i\Gamma)^2 - \Omega_0^2 ) }.
\end{equation}
Here, $F_W=1+2 \pi i W \rho_{F}$. The transmission and reflection
coefficients are calculated by adding the two--particle scattering
probabilities and tracing over the degrees of freedom of the QPC.
The resulting expressions are further simplified by using a
weak--coupling expansion to second order in $V$. In this limit
the application of a drain source voltage $V_d$ across the QPC is
equivalent to the simultaneous scattering of  $2 e V_d \rho_F$
particles in different {\em longitudinal} QPC modes. The total effect
of these particles is obtained by multiplying the result for one
QPC--particle with the number of longitudinal modes.

The $V$--dependent correction to the branching ratio arises both
from coherent (elastic) and incoherent (inelastic) scattering. We
find that measurements with the QPC--detector have a twofold effect:
(i) They suppress tunneling from the feeding lead into the lower dot
and (ii) they suppress tunneling from the lower into the upper dot.
Observation (i) is reflected in the {\em increase} of the reflection
coefficient, while (ii) follows from the {\em decrease} of the
branching ratio
\begin{eqnarray}
  {{\cal T}_{u}\over {\cal T}_{l}}= {{\cal T}^{(0)}_{u}\over
{\cal T}^{(0)}_{l}} \left[ 1
  - {eV_d \over \pi \Gamma}{(\Delta {\cal T}_d)^2 \over {4 \cal T}_d (1-
    {\cal T}_d)}\right]\ .
\label{branching}
\end{eqnarray}
Here, the ratio ${\cal T}^{(0)}_{u} / {\cal T}^{(0)}_{l}$ was given
above, and $\Delta {\cal T}_d$ and ${\cal T}_d$ were defined in
Sec.~\ref{Sec6}. Both effects (i), (ii) have an obvious
interpretation as manifestations of the quantum Zeno effect. The second
term in the square bracket is up to a factor $\Gamma / (2 \hbar)$
equal to the dephasing rate found for the isolated double--dot system.

\subsection{Decay rate}
\label{Sec7.4}

In early work on the quantum Zeno effect it was argued \cite{Mis77}
that an unstable quantum system slows down its decay under frequent
or continuous observations. However, despite of further theoretical
work \cite{Pan99} the suppression of decay remained controversial.
Indeed, a simple argument suggests that the decay rate should not be
influenced by observation: Consider the exponential decay from a
discrete initial state into a continuum of final states. The
probability for decay during a small time $t$ is linear in $t$ and
given by $P(t)= \Gamma t/ \hbar$. For $N$ measurements the decay
probability is $N P(t/N)=P(t)$ and hence identical to the decay
probability in the absence of any measurement.

\begin{figure}
\centerline{\psfig{file=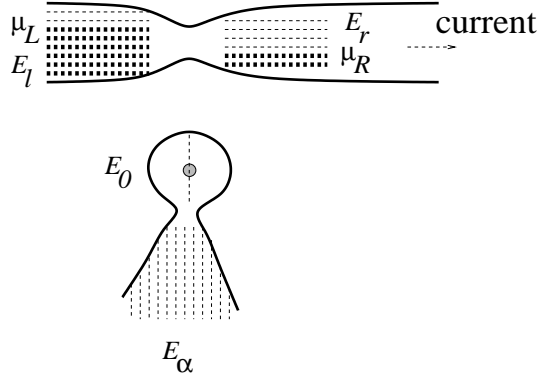,width=7cm,angle=0}}
\vspace*{0.5cm}
\caption[]{Schematic illustration of a QPC--detector near a quantum
  dot that is coupled to a reservoir. The energy level $E_0$ of the
  dot is occupied by an electron which can decay into the states
  $E_\alpha$ of the reservoir. Taken from
  Ref.~\protect{\cite{Elaa00}}.}
\label{figEla}
\end{figure}

Elattari and Gurvitz \cite{Elaa00,Elab00} addressed the problem for
the system shown in Fig.\ \ref{figEla}. A quantum dot with a single
energy level $E_0$ is coupled to an electron reservoir. The states
$E_\alpha$ in the reservoir are dense. The dot is placed near a QPC
connected to two separate reservoirs filled up to the Fermi energies
$\mu_L$ and $\mu_R$, respectively. The transmission probability
through the QPC depends on the occupation of the dot level. Thus,
the QPC continuously monitors the dot occupation.  The dynamics of
the entire system is determined by the Schr\"odinger equation $i
\hbar \partial_t | \Psi(t) \rangle = H | \Psi(t) \rangle$, where the
Hamiltonian $H$ comprises the quantum dot, the QPC, their mutual
interaction and the electron reservoirs. As the initial state $|
\Psi(0) \rangle$ we choose an occupied quantum dot level and
electron reservoirs filled up the Fermi levels $\mu_L$ and $\mu_R$.

The time evolution of the system can be expressed in the Bloch--type
equations \cite{Elaa00}
\begin{eqnarray}
\label{dec1}
\dot{\sigma}_{00} & = & - {\Gamma \over \hbar} \sigma_{00} , \\
\label{dec2}
\dot{\sigma}_{\alpha \alpha} & = & i {\Omega_\alpha \over \hbar} 
(\sigma_{0 \alpha} -\sigma_{\alpha 0}) , \\
\label{dec3}
\dot{\sigma}_{\alpha 0} & = & i {E_0-E_\alpha \over \hbar}
\sigma_{\alpha 0} - i {\Omega_\alpha \over \hbar} \sigma_{00} 
- {\Gamma + \hbar \kappa_d \over 2 \hbar} \sigma_{\alpha 0}.
\end{eqnarray}
Here, $\sigma_{00}$ is the probability for the electron to occupy
the dot level, $\sigma_{\alpha \alpha}$ is the probability for a
transition into level $\alpha$ in the continuum, and $\kappa_d$ is
the decoherence rate induced by the detector. The coupling matrix
element between the dot level and the level $\alpha$ is denoted by
$\Omega_\alpha$, and $\Gamma= 2 \pi |\Omega_\alpha|^2 \rho$ is the
total width for decay into the continuum, where $\rho$ is the
density of states in the reservoir coupled to the quantum dot.  From
Eq.\ (\ref{dec1}) one immediately finds
\begin{eqnarray}
\sigma_{00}(t)= \exp(-\Gamma t/\hbar ).
\end{eqnarray}
Hence, continuous monitoring of the unstable system does not slow
down its exponential decay. In contrast, the energy distribution
of the electron in the continuum, $P(E_\alpha) \equiv \sigma_{\alpha
  \alpha} (t \to \infty)$ is affected by the detector. From the
solution of Eqs.\ (\ref{dec1})-(\ref{dec3}) in the limit $t \to
\infty$, one finds a Lorentzian distribution centered around
$E_\alpha=E_0$,
\begin{eqnarray}
P(E_\alpha) \propto {\Gamma + \hbar \kappa_d \over (E_0 - E_\alpha)^2
+ (\Gamma + \hbar \kappa_d)^2/4} .
\end{eqnarray}
The measurement results in a broadening of the linewidth from $\Gamma$
to $\Gamma + \hbar \kappa_d$. The broadening does not affect the
decay rate $\Gamma$.

\newpage
\setcounter{equation}{0}
\section{Conclusion}
\label{Sec9}
Semiconductor quantum dots allow to measure and control current at
the single--electron level. Multiple quantum dots, wires and quantum
point contacts can be integrated on a single chip into more
complicated structures. Such integrated microstructures not only
offer control over the number of transmitted electrons, they also
allow to measure and manipulate the quantum state of these
electrons.

We demonstrated that devices employing a quantum dot in an
Aharonov--Bohm ring allow to measure the phase of the transmission
amplitude through the quantum dot. Controlled dephasing of quantum
dot states can be achieved by means of a quantum point contact in
close proximity to the dot. It is a formidable task to extend these
experiments to quantum dots in the Kondo regime.  Theoretical
predictions for the phase shift in this regime have been made
\cite{Ger00}.  Novel experiments testing the dephasing of charge
tunneling between coupled quantum dots have been proposed
\cite{Gur97,Hac98}.  The research in this field allows for the first
time to test fundamental principles of quantum mechanics, such as
the particle--wave duality, with solid state devices. Questions that
previously could only be addressed in atomic physics or quantum
optics may now be amenable \cite{Hac98d} to solid state physics.
Further question may be investigated, e.g.\ exploring the connection
of dephasing and the Fermi statistics \cite{Bue99b}.  These
considerations promise interesting future research on electron
coherence in a wide variety of mesoscopic systems.


\newpage
\centerline{\bf Acknowledgments}

It is a pleasure to thank Hans A. Weidenm\"uller for introducing me
to the subject of this work and for the fruitful work done together
during the course of the last years.  I am grateful to Reinhard
Baltin, Yuval Gefen, Dieter Heiss, and Bernd Rosenow for good
collaboration.  I thank Moti Heiblum and the members of his group at
the Weizmann institute, in particularly Eyal Buks, Ralph Schuster,
and David Sprinzak for sharing their experimental results with me,
for providing figures and for interesting discussions.  I enjoyed
insightful discussions with friends and colleagues, among others,
with Y.\ Alhassid, C.\ W.\ J.\ Beenakker, F.\ Haake, Y.\ Imry, E.\ 
Narimanov, A.\ D.\ Stone, and S.\ Tomsovic. Support of the Deutsche
Forschungsgemeinschaft via the SFB 237 "Unordnung und gro\ss e
Fluktuationen" is gratefully acknowledged.

\bibliography{draft}

\end{document}